\documentclass[twocolumn]{aastex63}
\newcommand{\Look}{``look-elsewhere effect''}
\newcommand{\p}{$\eta$}
\newcommand{\M}{$\mathcal{M}$}
\newcommand{\PPone}{{\mathbf{p}}_1}
\newcommand{\PPtwo}{{\mathbf{p}}_2}    
\newcommand{\PPthree}{{\mathbf{p}}_3}
\newcommand{\PPfour}{{\mathbf{p}}_4}
\newcommand{\PPfive}{{\mathbf{p}}_5}
\newcommand{\PPsix}{{\mathbf{p}}_6}
\newcommand{\AAone}{{\mathbf{A}}_1}
\newcommand{\AAtwo}{{\mathbf{A}}_2}
\newcommand{\AAthree}{{\mathbf{A}}_3}
\newcommand{\AAfour}{{\mathbf{A}}_4}
\newcommand{\AAfive}{{\mathbf{A}}_5}
\newcommand{\AAsix}{{\mathbf{A}}_6}
\newcommand{\PC}{Correct-$p$}
\newcommand{\PH}{Half-$p$}
\newcommand{\PD}{Double-$p$}
\newcommand{\PI}{Interference-$p'$}
\newcommand{\Vv}{Hp}  
\newcommand{\Dan}{Dp} 
\newcommand{\oo}{^{\mathrm{o}}}
\newcommand{\paperone}{{{\color{black} Paper~I}}}
\usepackage{skull}    

\newcommand{\Adataone}{{First$226^{\mathrm{y}}$-data}}
\newcommand{\Adatatwo}{{Last$9^{\mathrm{y}}$-data}}
\newcommand{\Bdataone}{{First$185^{\mathrm{y}}$-data}}
\newcommand{\Bdatatwo}{{Last$50^{\mathrm{y}}$-data}}

\newcommand{\RModel}[1]{{{model}$_{#1}$}}

\newcommand{\LK}{Lp}
\newcommand{\AD}{Ad}
\newcommand{\SP}{Sp}
\newcommand{\HD}{If}
\newcommand{\RD}{Um}
\newcommand{\REJ}{$Q_F\!\!<\!\!10^{-16}$}
\newcommand{\Dd}{{\mathrm{[d]}}}
\newcommand{\Webcolour}{\color{blue}}
\newcommand{\Link}[2]{\href{#1}{\Webcolour \bf #2}}

\received{\today}
\revised{..., 2021}
\accepted{..., 2021}
\submitjournal{ApJ}
\shorttitle{New companion candidates of Algol}
\shortauthors{Jetsu}
\begin{document}
\title{Say hello to Algol's new companion candidates}
\correspondingauthor{Lauri Jetsu}
\email{lauri.jetsu@helsinki.fi}
\author[0000-0002-0786-7307]{Lauri Jetsu}
\affiliation{Department of Physics\\
P.O. Box 64, FI-00014 University of Helsinki \\
Finland}
\begin{abstract}
  Constant orbital period
  ephemerides of eclipsing
  binaries give
  the computed eclipse epochs (C).
  These ephemerides based on
  the old data
  can not accurately predict
  the observed future eclipse epochs (O).
  Predictability can be improved by 
  removing linear or quadratic trends from
  the O-C data. 
  Additional companions
  in an eclipsing binary system 
  cause light-time travel effects that
  are observed as strictly 
  periodic O-C changes. 
  Recently,
    \citet{Haj19} estimated that
    the probability for detecting the
    periods of two
    new companions from
  the O-C data is only 0.00005.
  We apply the new Discrete
  Chi-square Method (DCM)
  to 236 years of O-C 
  data of the eclipsing
  binary Algol ($\beta$ Persei).
  We detect the tentative signals
  of at least five companion
  candidates
  having periods between
  1.863 and 
  219.0 years.
  The weakest one of these five
  signals does not reveal
  a ``new'' companion candidate,
  because its
  $680.4 \pm 0.4$ days
  signal period 
  differs only $1.4 \sigma$
  from the 
  well-known 
  $679.85 \pm 0.04$ days
  orbital period of Algol~C.
  We detect these same signals also
  from the first 226.2 years of data,
  and they give an excellent
  prediction for
  the last 9.2 years of our data. 
  The orbital planes of Algol~C and the new
  companion candidates are
  probably co-planar, because
  no changes have been observed in Algol's
  eclipses.
  The 2.867 days orbital period
  has been constant since it was
  determined by 
  \citet{Goo83}.
\end{abstract}

\section{Introduction} \label{SectIntro}

The oldest preserved historical
document of the discovery
of a variable star
is the Ancient Egyptian papyrus Cairo 86637,
where naked eye observations
of Algol's eclipses
have been recorded into
the Calendar of Lucky and Unlucky days
\citep{Por08,Jet13,Jet15,Por18}.
Montanari re-discovered its variability in
the year 1669.
\citet{Goo83} determined the orbital
period $P_{\mathrm{orb}}=2.^{\mathrm{d}}867$
of this eclipsing binary (EB).
The close orbit eclipsing stars are
Algol~A (B8~V) and Algol~B (K2~IV).
\citet{Cur08} discovered 
the $1.^{\mathrm{y}}863$
wide orbit third companion
Algol~C (K2~IV).
Direct interferometric images of these
three members have been obtained
\citep[e.g.][]{Zav10,Bar12}.

Periodic long-term changes occur
in the observed (O) minus the
computed (C) primary eclipse epochs
of EBs.
The most probable causes are
a third body \citep[e.g.][]{Li18},
a magnetic activity cycle \citep[e.g.][]{App92}
or 
an apsidal motion \citep[e.g.][]{Bor05}.
\citet{Haj19} searched for third bodies
in a large sample of 80~000 EBs.
They detected 992 triple systems
from the O-C data, and only
four candidates that may have a fourth body.
Their fourth body detection
rate was $4/80~000=0.00005$.
Recently,
\citet[][hereafter \paperone]{Jet20} formulated
the new Discrete Chi-Square Method (DCM).
He applied DCM to the O-C data of XZ~And, and
detected the periods of
a third and a fourth body.

In Algol,
  the mass transfer from
  the less massive Algol~B $(0.8 m_{\odot})$
  to the more massive Algol~A $(3.7 m_{\odot})$
  should cause a long-term $P_{\mathrm{orb}}$ period
  increase
  \citep{Kwe58}, which
  should have been observed as
  quadratic long-term O-C changes
  \citep{Kis98}.
  Long-term
  $P_{\mathrm{orb}}$ increase
  or quadratic O-C changes 
  have not been observed in Algol
  since its period was determined
  238 years ago.
However,
 its orbital period modulation
  does cause
  negative and positive
  O-C changes.
  The short-term low amplitude O-C changes
  follow
  $1.^{\mathrm{y}}863$ orbital motion
  cycle of Algol~C,
  while the 
  high amplitude O-C changes follow
  $30^{\mathrm{y}}$ and $200^{\mathrm{y}}$
  quasi-periodic activity cycles
  \citep[][]{App92}.
  The physical origin of period changes
  is not fully understood,
  because Algol's puzzling O-C diagram
  contains unknown signals and trends
  \citep[e.g.][]{Fri70,App92}.
  We apply DCM to Algol's O-C data,
  because this method can detect many signals
  superimposed on unknown trends.

%
%
\citet{Kim18} note that their TIDAK database
  O-C ephemerides ``cannot be used for the
  prediction of future times of
    the primary or secondary minima.''
    These ephemerides are determined
    by eliminating linear or quadratic
    trends from the available O-C data
    \citep{Kre01}.
    They usually need to be
    re-determined when new data
    are obtained.
    Although the O-C changes caused by a third body
    are strictly periodic,
    the predictions usually
    fail to separate 
    aperiodic trends from
    periodic signals
    \citep[e.g.][]{Bou14,Loh15,Son19}.
    Furthermore,
    the detection rate of third bodies
    from O-C data is extremely low \citep[e.g.][]{Haj19}.
    Against this background,
    it is totally unexpected that we can
    detect numerous periods in Algol's O-C data,
    as well as predict its O-C changes.

\section{Data} \label{SectData}

The epochs
  of the observed light curve minima
  give the {\it observed} (O) values.
  We obtained the $n=2238$ observed eclipse
  epochs of Algol
  from the 2018 version of TIDAK database
  \citep{Kim18}.
  These eclipse epochs have been
  determined by hundreds of astronomers
  during the past two centuries.
    The nights when these eclipses could
    be observed were known beforehand.
    Every eclipse lasted eight hours.
    Both dimming and brightening took four hours.
    The probability for a negative or positive
    mid eclipse epoch error was the same,
    because the eclipse light curve was symmetric.
    It is therefore probable that the observational
    errors follow a Gaussian distribution,
    the epoch values contain no observational trends, and
    the observational errors are not heteroskedastic.
  Naturally, the accuracy of these data
  improves towards modern times,
  because the observational
  techniques have improved.
  We study only the primary minimum epochs
    when the dimmer Algol~B eclipses the brighter Algol~A.
    Therefore, we reject 
    all fourteen secondary minima,
because they occur $P_{\mathrm{orb}}/2$ after
the primary minima. 
We analyse only 
the remaining $n=2224$ primary minima
between
November 12th, 1782 and
October 18th, 2018.
These data
are given in Table \ref{TableData}
($\Delta T = 86171^{\mathrm{d}}=  236^{\mathrm{y}}$).
We obtain the {\it computed} (C) epoch values
  from the TIDAK database ephemeris 
\begin{eqnarray}
\mathrm{HJD~}2445641.5135 + 2.86730431{\mathrm{E.}}
\label{EqEphe}
\end{eqnarray}
This ephemeris predicts that all Algol's primary
eclipses occur at multiples
$\mathrm{HJD~}2445641.5135+E \times P_{\mathrm{orb}}$,
where $P_{\mathrm{orb}}= 2.^{\mathrm{d}}86730431$ is the
orbital period of Algol and $E$ is an integer number.
This constant orbital period
ephemeris ``model'' is quite accurate,
because all O-C values are between
$-0.^{\mathrm{d}}24$ and $+0.^{\mathrm{d}}15$ during 236 years.

Out of all 2224 estimates, only 197 
    have an error estimate, and none of the 1236 first ones.
    However, this does not mean that these values without
    error estimates are unreliable or inconsistent.
  The error estimates are available for only about
  10\% of data. These are all new observations
  after the year 1921. The range of these known errors
is between $0.^{\mathrm{d}}0002$ and $0.^{\mathrm{d}}013$.
The most accurate TIDAK database O-C values
have four decimals.
Since the errors 
are not known for over 90\% of observations,
we use arbitrary errors $\sigma_i=0.^{\mathrm{d}}00010$
for all O-C values.
These arbitrary numerical values do
not influence our results, because we
use the same weight $w_i=\sigma_i^{-2}$ for every observation,
and we compute the DCM test statistic $z$ from the
sum of squared residuals (Eq. \ref{Eqz}).
We will also show that a weighted DCM search,
  where the O-C data accuracy improves towards
  modern times, does not alter our results
(Sect. \ref{SectExperiments}).

We also analyse shorter
  subsamples of all data (Table \ref{TableSamples}).
  In Sect. \ref{SectAdata}, we apply
  DCM to the first 226.2 years of all data
  (\Adataone). This gives us
  a prediction for the last
  9.2 years of all data (\Adatatwo).
  In Sect. \ref{SectAdata},
  the same DCM procedures are also
  applied to the first 185.5 years of all data (\Bdataone),
  and the last 50 years of all data (\Bdatatwo).

\section{DCM-method}
\label{SectMethod}

The Discrete Chi-Square Method (DCM) 
notations for the data are
$y_i=y(t_i) \pm \sigma_i$,
where $t_i$ are the observing
times and $\sigma_i$ are the errors
$(i=1,2, ..., n)$. 
The time span of data is $\Delta T=t_n-t_1$.
The mid point of data is
  $t_{\mathrm{mid}}=t_1+\Delta T/2$.

We analyse these
data with DCM,
which can detect many signals superimposed
on arbitrary trends.
Detailed instructions for
using the DCM python
code were given
in the
appendix of
\paperone.
In this current study,
we provide all necessary
information
for reproducing our DCM analysis
of Algol data.\footnote{All necessary files
  for reproducing our results are
  published in
  \Link{https://zenodo.org}{Zenodo
    database: doi {10.5281/zenodo.5082125}}}
DCM model is
\begin{eqnarray}
  g(t) = g(t,K_1,K_2,K_3)
   =  h(t) + p(t). 
\label{Eqmodel}
\end{eqnarray}
It is a sum of periodic and aperiodic functions
\begin{eqnarray}
h(t)   & = & h(t,K_1,K_2) = \sum_{i=1}^{K_1} h_i(t) \label{Eqharmonicone} \\
h_i(t) & = & \sum_{j=1}^{K_2} 
B_{i,j} \cos{(2 \pi j f_i t)} + C_{i,j} \sin{(2 \pi j f_i t)}
\label{Eqharmonictwo} \\
p(t)   & = & p(t,K_3) = \sum_{k=0}^{K_3} p_k(t) \\
p_k(t) & =  & M_k \left[
                    {{2(t-t_{\mathrm{mid}}}) \over {\Delta T}}
                    \right]^k.
\label{Eqpolynomial}
\end{eqnarray}
The periodic $h(t)$ function is a sum of 
$K_1$ harmonic $h_i(t)$
signals having 
frequencies $f_i$.
The signal order is $K_2$. 
These signals
are superimposed
on the aperiodic $K_3$ 
order polynomial trend $p(t)$.

  In the original DCM version,
  the $p(t)$ terms were
  $M_k[(2t)/\Delta T]^k$ (\paperone: Eq. 5),
  and the first time point $t_1$ was subtracted
  from all time points $t_i$ before modelling.
  For odd and even $k$ values,
  every $M_k[(2t)/\Delta T]^k$ term could only
  increase or decrease
  monotonically
  during 
  the whole $\Delta T$ interval,
  because the argument $(2t)/\Delta T$
  was always positive.
  In our new formulation (Eq. \ref{Eqpolynomial}),
  the even $k=2,4,...$ terms
  $M_k[2(t-t_{\mathrm{mid}})/\Delta T]^k$ can now both
  increase and decrease during the whole $\Delta T$ interval.
  This increases the flexibility of the model.
  Furthermore, it is no longer necessary to subtract the
  first observing time $t_1$ before the modelling.
  Note that the $2(t-t_{\mathrm{mid}})/\Delta T$ argument
  equals $-1$ at $t_1$, and $+1$ at $t_n$.
  Hence, the scale of polynomial $M_k$
  coefficients (Eq. \ref{Eqpolynomial}) is
  comparable to the scale of
  trigonometric $B_{i,j}$ and $C_{i,j}$
  coefficients (Eq. \ref{Eqharmonictwo}).
  This change of $p(t)$ trend terms
    from $p_k(t)=M_k[(2t)/\Delta T]^k$
    to $p_k(t)=M_k[2(t-t_{\mathrm{mid}})/\Delta T]^k$
    does not change the detected signal periods
    in any of the analysed O-C samples.

Our abbreviation ``\RModel{K_1,K_2,K_3}''
refers to a $g(t)$ model having orders
$K_1$, $K_2$ and $K_3$.
The free parameters are 
$\bar{\beta}=$ $[\beta_1, \beta_2, ..., $
$\beta_{\eta}] = $
$[B_{1,1},C_{1,1},f_1, ..., $ 
$B_{K_1,K_2}, $ 
$C_{K_1,K_2}, f_{K_1},$ 
$M_0, ..., M_{K_3}]$,
where 
\p$= K_1 \times (2K_2+1) + K_3+1$
is the number of free parameters.
We divide the free parameters $\bar{\beta}$ into
two groups  $\bar{\beta}_{I}$ and $\bar{\beta}_{II}$.
The first group of
free parameters are
the frequencies $\bar{\beta}_{I}=[f_1, ..., f_{K_1}]$.
These frequencies make the $g(t)$ model non-linear,
because all free parameters are not
eliminated from 
all partial derivatives $\partial g / \partial \beta_i$.
If the $\bar{\beta}_{I}$ frequencies are fixed to constant
known tested numerical values, the model becomes linear,
because all partial derivatives
$\partial g / \partial \beta_i$ no longer
contain any free parameters.
In this case, the solution for the 
remaining second group of free parameters,
$\bar{\beta}_{II}=[B_{1,1}C_{1,1}, ..., $
$B_{K_1,K_2}, C_{K_1,K_2},$ $M_0, ..., M_{K_3}]$,
is unambiguous.
We refer to this type of models and their
free parameter solutions,
when we use the concepts
``linear model'' and ``unambiguous result''.

DCM model residuals are
\begin{eqnarray}
  \epsilon_i=y(t_i)-g(t_i)= y_i-g_i.
  \label{EqResiduals}
\end{eqnarray}
For every 
  combination $\bar{\beta}_{I}=[f_1, f_2, ..., f_{K_1}]$
  of tested frequencies,
  we compute the DCM test statistic
  \begin{eqnarray}
    z=z(f_1, f_2, ..., f_{K_1})=\sqrt{R/n}
    \label{Eqz}
  \end{eqnarray}
  from the
sum of squared residuals $R=\sum_{i=1}^{n}\epsilon_i^2$ of a non-weighted
linear model least squares fit.
We use this non-weighted test statistic,
because the errors for the data are unknown.

The global periodogram minimum is at
\begin{eqnarray}
  z_{\mathrm{min}}=
  z(f_{\mathrm{1,best}},f_{\mathrm{2,best}},...,
  f_{\mathrm{K_1,best}}),
\end{eqnarray}
where
$f_{\mathrm{1,best}},f_{\mathrm{2,best}},...,
f_{\mathrm{K_1,best}}$ are the 
frequencies of the best DCM model for the data.
Every scalar value of this
$z$ periodogram is computed from $K_1$ frequency values.
For example, the $K_1=2$ periodogram could be plotted
like a map, where $f_1$ and $f_2$ are the coordinates,
and $z=z(f_1,f_2)$ represents the height.
However, a graphical presentation for $K_1 \ge 3$
is impossible, because it requires more than three
dimensions.
In \paperone,
we solved this
problem by
presenting only the
following one-dimensional
slices of the full periodograms
\begin{eqnarray}
  z_1(f_1) & = & z(f_1,f_{\mathrm{2,best}}, ...,f_{\mathrm{K_1,best}}) \nonumber \\
  z_2(f_2) & = & z(f_{\mathrm{1,best}},f_2, f_{\mathrm{3,best}}, ...,f_{\mathrm{K_1,best}}) \nonumber \\
  z_3(f_3) & = & z(f_{\mathrm{1,best}},f_{\mathrm{2,best}},f_3, f_{\mathrm{4,best}},...,f_{\mathrm{K_1,best}}) ~\label{EqSlices} \\
  z_4(f_4) & = & z(f_{\mathrm{1,best}},f_{\mathrm{2,best}},f_{\mathrm{3,best}},f_4, f_{\mathrm{5,best}},f_{\mathrm{K_1,best}}) \nonumber \\
  z_5(f_5) & = & z(f_{\mathrm{1,best}},f_{\mathrm{2,best}},f_{\mathrm{3,best}},f_{\mathrm{4,best}},f_5,f_{\mathrm{K_1,best}}) \nonumber \\
  z_6(f_6) & = & z(f_{\mathrm{1,best}},f_{\mathrm{2,best}},f_{\mathrm{3,best}},f_{\mathrm{4,best}},f_{\mathrm{5,best}},f_6). \nonumber 
\end{eqnarray}
In the above $K_1=2$ map analogy,
$z_1(f_1)$ would represent the height at $f_1$ coordinate
when moving along the constant line $f_2=f_{\mathrm{2,best}}$
that crosses the global minimum $z_{\mathrm{min}}$.

DCM determines the following
$h_i(t)$ signal
parameters
\begin{itemize}

\item[] $P_i = 1/f_i = $ Period
\item[] $A_i = $ Peak to peak amplitude
\item[] $t_{\mathrm{i,min,1}} = $ Deeper primary minimum epoch 
\item[] $t_{\mathrm{i,min,2}} = $ Secondary minimum epoch (if present)
\item[] $t_{\mathrm{i,max,1}} = $ Higher primary maximum epoch
\item[] $t_{\mathrm{i,max,2}} = $ Secondary maximum epoch (if present),
\end{itemize}
and the $M_k$ parameters 
of the $p(t)$ trend.
For us,
the most interesting
parameters
are 
the signal periods $P_i$ and
the signal  amplitudes $A_i$,
and the $p(t)$ trend coefficient $M_1$.

We determine the DCM model
parameter errors
with
the bootstrap procedure
\citep{Efr86,Efr94}.
During each bootstrap round,
we select
a random sample $\bar{\epsilon}^*$
from
the residuals $\bar{\epsilon}$
of the DCM model (Eq. \ref{EqResiduals}).
Each $\epsilon_i$ 
can be chosen
as many times as
the random selection 
happens to favour it.
This gives the {\it artificial} 
bootstrap data sample
\begin{eqnarray}
y_i^*=g_i+\epsilon_i^*.
\nonumber
\end{eqnarray} 
DCM model
for each 
$\bar{y}^*$ sample gives
one estimate for
every model parameter.
For each particular 
model parameter,
its error estimate is
the standard deviation
of all estimates obtained
from all $\bar{y}^*$
bootstrap samples.
We have already used
this same
bootstrap procedure
in our TSPA-
and CPS-methods
\citep{Jet99,Leh11}.
Finally, we note that
our bootstrap procedure can not
assess the bias in the $y_i$ input data,
which first contaminates 
the $\epsilon_i$ values,
and then also the $\epsilon_i^{\star}$
and $y_i^{\star}$ values.

We use the Fisher-test to compare
any pair $g_1(t)$ and $g_2(t)$ of
simple and complex models.
Their number of free parameters $(\eta_1 < \eta_2)$, 
and 
their sums of squared residuals $(R_1,R_2)$
give the test statistic 
\begin{eqnarray}
F_R =
\left(
{
{R_1}
\over
{R_2}
}
-1
\right)
\left(
{
{n-\eta_2-1}
\over
{\eta_2-\eta_1}
}
\right).
\label{EqFR}
\end{eqnarray}
Our null hypothesis is
\begin{itemize}

\item[] $H_{\mathrm{0}}$: {\it ``The complex model $g_2(t)$ 
does not provide
a significantly better fit to the data
than the simple
model $g_{\mathrm{1}}(t)$.'' }

\end{itemize}
\noindent
Under $H_0$, the test statistic
$F_R$ has an $F$ distribution with 
$(\nu_1,\nu_2)$ degrees of freedom,
where $\nu_1=\eta_2-\eta_1$ and $\nu_2=n-\eta_2$
\citep{Dra98}. 
The probability for 
$F_R$  reaching values higher than $F$
is called the critical level $Q_{F} = P(F_R \ge F)$.
We reject the $H_0$ hypothesis,
if
\begin{eqnarray}
  Q_F < \gamma_F=0.001,
\label{EqFisher}
\end{eqnarray}
where 
$\gamma_F$ is the pre-assigned significance level.
It represents the probability of falsely rejecting
$H_0$ when it is in fact true.
The $H_0$ rejection means that
we rate the complex $g_2(t)$ model 
better than the simple $g_1(t)$ model.

The $Q_F$ critical level becomes smaller when $F_R$
increases. In other words, the $H_0$ hypothesis
rejection probability increases for larger $F_R$ values.
The basic idea of the Fisher-test is simple.
The sum of complex model residuals $R_2$ decreases
when the $\eta_2$
number of free parameters increases.
When the complex model has more $\eta_2$
free parameters,
the first $(R_1/R_2-1)$ term 
increases $F_R$ (Eq. \ref{EqFR}),
but at the same time the second
$(n-\eta_2-1)/(\eta_2-\eta_1)$ penalty 
term decreases $F_R$.
In conclusion, this second penalty
term prevents overfitting. 
  
The key ideas of DCM are 

  \begin{enumerate}

  \item  \label{IdeaOne}
    The {\it non-linear} DCM model $g(t)$ of Eq. \ref{Eqmodel}
    becomes {\it linear} when the frequencies
    $f_1, ..., f_{K_1}$ are fixed to their tested
    numerical values. These
    linear
    models
    give {\it unambiguous} results.

  \item \label{IdeaTwo}
    DCM tests a dense grid of all possible
    frequency combinations
    $f_1 \!> \!f_2 \!>\!... \!>\!f_{K_1}$. For every 
    frequency combination, the
    {\it linear} model least squares fit
    gives the test statistic 

    \subitem $z=\sqrt{\chi^2/n}$ if
    errors $\sigma_i$ are known

    \subitem $z=\sqrt{R/n }$ if errors $\sigma_i$ are
    unknown, 
   
  \item[] where $\chi^2=\sum_i^n \epsilon_i^2/\sigma_i^2$,
    $R=\sum_i^n \epsilon_i^2$ and $\epsilon_i=y_i-g_i$ are
    the model residuals.

  \item  \label{IdeaThree}
    The $f_1 > f_2 > ... >f_{K_1}$
    grid combination of the best DCM model
    minimizes the $z$
    test statistic.

  \item  \label{IdeaFour}
    The bootstrap method gives
    the error estimates for all model
    parameters.

  \item  \label{IdeaFive}
    All different 
    $K_1$, $K_2$ and $K_3$
    order nested models
    are compared using the Fisher-test,
    which reveals the best one of all models
    \citep{Dra98,All04}.

  \end{enumerate}
  
  In short, DCM applies 
  the following  robust and well tested
  statistical approaches:
  Linear least squares fits (Idea \ref{IdeaOne}),
  $\chi^2$ and $R$ test statistic (Idea \ref{IdeaTwo}),
  Dense tested frequency grids (Idea  \ref{IdeaThree}),
  Bootstrap utilizing residuals (Idea \ref{IdeaFour})
  and
  Fisher-test comparison of nested models (Idea \ref{IdeaFive}).
    
  The caveats of DCM are

  \begin{enumerate}
    
  \item 
      DCM is designed for
      periodicity detection,
      but it
      gives no {\it direct} significance estimates
      for these detected periodicities.
      In this sense,
      DCM resembles 
      our former TSPA- and CPS-methods
      \citep{Jet99,Leh11}.
      DCM utilizes {\it indirect}
      Fisher-test significance estimates
      to identify the best model among
      all tested models,
      but it gives no
      significance estimates for the detected
      periodicities of this best model.
      We will later discuss our indirect
      significance estimates,
      especially in connection with
      the look-elsewhere effect
      (Sect. \ref{SectLook}). 
    
  \item The best
      frequency combination can
    be missed if the tested grid is too sparse
    (Idea \ref{IdeaThree}).
    However, an adequately dense tested frequency
    grid eliminates the possibility for
      this kind of an error.
    The caveat is that denser grids
    require more computation time. 

    For example, all three signal
    $z_1$,
    $z_2$ and
    $z_3$
    periodograms for the original data
    are continuous and display no abrupt jumps,
    because the periodogram values for all
    close tested frequencies correlate
    (see Fig. \ref{1hjd14R321Sz}).
    Since the frequencies of the minima
    of all these periodograms
    are accurately determined,
    there is no need to test
    an even denser grid (i.e. more trials),
    because this would not alter
    the final result of the non-linear
    iteration (\paperone: Eq. 18). 
    In other words, the detected
      period values would no longer change,
      if we increased the number of of tested frequencies
      (\paperone: $n_L$ and $n_S$ trials).
      Since DCM gives no {\it direct}
      significance estimates for
      the detected periods (Caveat 1),
      there is no need to
      determine
      the number of independent trials,
      like for example
      the number of independent tested frequencies 
      \citep[e.g.][their Eq. A.1]{Jet00A}.

  \item
    If the grid of each tested
    $f_1> f_2 > ... > f_{K_1}$ frequency contains
    $n_f$ values,
    the total number of tested frequency combinations
    is proportional to $\propto n_f ^{K_1}$.
    For example,
    it took about one month for an ordinary PC
    to compute the four signal DCM \RModel{4,2,1} 
    search,
    and to analyse its twenty
    bootstrap samples  (Table \ref{Table1hjd14R}, model \M=4). 
   
  \item 
    Some DCM models are unstable
    because they are simply wrong
    models for the data.
    For example, a wrong $p(t)$
  trend order $K_3$,
  or a search for too many $K_1$
  signals,
  can cause such instability. 
    In this
      paper, we denote such unstable
      models with ``\RD''.
  We denote the two
  signatures of such unstable models
  with

  \begin{itemize}

  \item[] ``\AD'' = Dispersing amplitudes =
    Amplitudes and/or amplitude errors disperse.

  \item[] ``\HD'' = Intersecting frequencies =
    At least two model frequencies are too
    close to each other.
    
  \end{itemize}

\item[]  We give list all
    our symbols in
    Table \ref{TableSymbols}.
    Both of the above instabilities were defined
    in \paperone ~(Sect. 4.3), where a typical example of the
    wildly oscillating signals was also shown
    in Fig. 6 of \paperone.

  \end{enumerate}

 \noindent
 DCM tests all reasonable alternative
 linear models
 for the data, and determines the
 {\it unambiguous results}
  for the best values of their free parameters.
  This brute numerical approach finds
  the best model among all alternative models.
  DCM ``works like winning a lottery by buying
  all lottery tickets'' (\paperone).

  
  \section{Third body O-C changes
    \label{SectOCmodel}}

  The light-time
  travel effect (LTTE) caused by a third body is
\begin{eqnarray}
  \mathrm{(O-C)} & = &
  \! {{K} \over {\sqrt{(1\!-\!e^2 \cos^2{\omega})}} } \times
  \label{EqIrwOne} \\
                 &   &
  \left[
{
{1\!-\!e^2}
\over
{1\!+\!e \cos{\nu(t)}}
}
\sin{(\nu(t) \!+\! \omega)}
\!+\!
e \sin{\omega}
\right]
\nonumber \\
  & = &
        {
        {a \sin{i}}
        \over
        {173.15}
        } \times  \nonumber \\
  &  & 
\left[
{
{1\!-\!e^2}
\over
{1\!+\!e \cos{\nu(t)}}
}
\sin{(\nu(t) \!+\! \omega)}
\!+\!
e \sin{\omega}
        \right] \nonumber
\end{eqnarray}
where 
\begin{eqnarray}
K={{a~ \sin{i} \sqrt{1-e^2 \cos^2{\omega}}} \over {173.15}}
\label{EqIrwTwo}
\end{eqnarray}
\citep{Irw52}.
This relation gives EB
orbit around the common centre
of mass of all three stars.
The orbit parameters are
the semimajor axis $([a]={\mathrm{AU}})$,
the orbital plane inclination $([i]={\mathrm{rad}})$,
the eccentricity of orbit $(e)$,
the longitude of periastron $([\omega]=\mathrm{rad})$,
the true anomaly $([\nu]=\mathrm{rad})$
and
the amplitude of light-time travel effect 
\begin{eqnarray}
  K=A/2,
\label{EqIrwThree}
\end{eqnarray}
which is half of the  
peak to peak amplitude $A$ of the
observed O-C changes $([A]={\mathrm{d}})$.

We compute the true anomaly from the Fourier expansion
\begin{eqnarray}
  \nu(t) & =     & M(t) \!+\! (2e \!-\! {{1}\over {4}} e^3) \sin{[M(t)]}
                   \label{EqIrwFour} \\
       & +     & {{5}\over{4}} e^2 \sin{[2M(t)]} 
                 \!+\! {{13}\over{12}} e^3 \sin{[3M(t)]} \!+\! O(e^4),
                 \nonumber
\end{eqnarray}
where 
\begin{eqnarray}
  M(t) = {{2\pi(t-t_p)}\over{p}},
  \label{EqIrwFive}
\end{eqnarray}
is the mean anomaly \citep{Mul95,Roy05}.
The other parameters are
the EB orbit pericentre epoch $([t_p]={\mathrm{HJD}})$, 
the third body orbital period $([p]={\mathrm{d}})$
and 
the omitted fourth
order terms $([O(e^4)]={\mathrm{rad}})$. 

If the orbit is circular $(e=0)$,
the third body mass $m_3$ can be solved from
the  mass function 
\begin{eqnarray}
  f(m_3)= {{m_3 \sin{i}^3} \over {(m_1+m_2+m_3)^2}}={{[173.15 (A/2)]^3}\over{p^2}},
\label{EqMass}
\end{eqnarray}
where $m_1$ and $m_2$ are the masses of EB
\citep[][]{Wol99,Zas07,Man16,Ekr21}.
The semi-major axis of the third body orbit is
\begin{eqnarray}
a_3= a {{(m_1+m_2)}\over{m_3}},
\label{EqSemi}
\end{eqnarray}
where $a=173.15 (A/2) /\sin i$.

For circular third body orbit,
the suitable O-C curve DCM model order is $K_2=1$,
the pure sinusoid
(Eq. \ref{EqIrwOne}: $e=0$).
For an eccentric $e>0$ third body orbit,
the O-C curve is not a pure sinusoid,
and
the suitable DCM model order 
is $K_2=2$ \citep[][]{Hof06}.

\section{Results \label{SectResults}}

Here, we present separately
the DCM period search
results for all data (Sect. \ref{SectAlldata}),
\Adataone ~(Sect. \ref{SectAdata})
and
\Bdataone ~(Sect. \ref{SectBdata}).
We also make some additional experiments
(Sect. \ref{SectExperiments}).

\subsection{All data \label{SectAlldata}}

\subsubsection{All data: Trend \label{SectTrend}}

  In Table \ref{TableAllFisher},
  the Fisher-test is used to compare the
  results for all data in
  twelve separate DCM
period searches between
$P_{\mathrm{min}}=6000^{\mathrm{d}}$
and
$P_{\mathrm{max}}=80000^{\mathrm{d}}$.
These models have one, two or three signals
$(K_1=1, 2$ or 3).
The third body orbits can be eccentric
$(K_2=2 \equiv e>0)$.
The alternative tested $p(t)$ trends are
$K_3=0, 1, 2$ or 3.
Table \ref{TableAllFisher}
contains many  notations ``--'',
because it makes no sense to compare the
same pair of models twice, nor to
compare the model to itself.
The total number of compared pairs is
$(12 \times 11)/2$.
For example, the Fisher-test
comparison of the one signal \M=1 and \M=2 models
gives a large test statistic value $F=2821$.
The critical level $Q_F$ of this $F$ value falls below
the computational\footnote{This is the computational $Q_F$
  estimate accuracy for
  f.cdf subroutine in scipy.optimize python library,}
accuracy of $10^{-16}$ (Table \ref{TableAllFisher}: \REJ).
This means that 
the linear $K_3=1$ trend \RModel{1,2,1} is
absolutely certainly a better model than
the constant $K_3=0$ trend \RModel{1,2,0}.
The upward arrow ``$\uparrow$''
indicates this result.
Note that Table \ref{TableAllFisher} contains numerous
``\REJ'' cases, where the identification of the
better model is absolutely certain.

All column \M=10 arrows point upwards $(\uparrow)$,
and all line \M=10 arrows point leftwards $(\leftarrow)$
in Table \ref{TableAllFisher}.
Hence, this stable \M=10 model is better than
all other eleven alternative models.
This best DCM \RModel{3,2,1} for all data is a
sum of $K_1=3$ signals having an order $K_2=2$,
and  a linear $K_3=1$ trend.
We use this $K_3=1$ linear trend
in all analysis of original data.
The meaning of this linear trend is discussed
later (Sect. \ref{SectStability}, Eq. \ref{EqConstantP}).
We will also show that all data
contains {\it only three}
$K_2=2$ order signals between 8000 and 80000 days
(Sect. \ref{SectAllDataEccentric}).

Four of the twelve models are unstable ``\RD''
(Table \ref{TableAllFisher}: \M= 3, 5, 8 and 9).
There are three models,
  where the detected
  period exceeds $\Delta T$ time span of data
  (Table \ref{TableAllFisher}: \M= 2, 3 and 7).
  They are denoted with the symbol
  \begin{itemize}
  \item[] ``\LK'' =  Leaking period = At least
    one detected period exceeds $\Delta T$ time span of data.
    \end{itemize}
  
\subsubsection{All data: Eccentric orbits \label{SectAllDataEccentric}}

In Table \ref{TableAllFisher},
  we compared $(12 \times 11)/2$ pairs of
  models against each other.
  The better model in each pair was
  identified with the Fisher test:
  the complex model above ``$\uparrow$'',
  or the simple model on the left ``$\leftarrow$''.

  The structure of our next Table \ref{Table1hjd14R}
  is more complicated, because
  we squeeze {\it all}
  DCM eccentric orbit search
  results for all data into this single table.
  We search for periods between 8000 and 80000 days.
  The left side of this table
  gives the detected periods and amplitudes.
  The right side gives the Fisher-test comparison results.
  For example, 
  the one signal \M=1 model
  period and amplitude
  are
  $P_1=88183^{\mathrm{d}}\pm816^{\mathrm{d}}$
  and
  $A_1=0.^{\mathrm{d}}313 \pm0.^{\mathrm{d}}004$.
  The next six 
  ``--'' notations for this \M=1 model
  mean that it has no other periods
  $P_2$, $P_3$ or $P_4$, nor amplitudes $A_2$, $A_3$ or $A_4$.
  Fisher-test comparison
  between this one signal \RModel{1,2,1} ~(\M=1)
  and the two signal \RModel{2,2,1} ~(\M=2) gives
  an extreme test statistic
  value $F=183$.
  The critical level \REJ ~confirms that 
  \M=2 model is
  certainly the better one in this pair of models.
  Comparison of \M=1 model
  to \M=3 and \M=4 models gives the same result.
  
  For the next \M=2, 3 and 4 models,
  the number of detected periods and amplitudes
  increases one by one.
  The number of Fisher-tests decreases one by one,
  because it is unnecessary to test the same
  pair of models twice ``--'', nor to compare any model
  to itself ``--''.

The periods and amplitudes for one,
two and three signal models
are consistent (Table \ref{Table1hjd14R}: \M=1-3).
When we detect a new signal,
we re-detect the same
old earlier signal periods and amplitudes
for models having less signals.
The one signal \M=1 model shows a leaking period ``\LK'',
because the $P_1=88183^{\mathrm{d}}$ period exceeds the
$\Delta T = 86171^{\mathrm{d}}$ time span of data.
The two and the three signal \M=2 and \M=3
models are stable, but the \M=4 model is not ``\RD''.

 The one-dimensional
  $z_1(f_1)$,
  $z_2(f_2)$,
  $z_3(f_3)$
  and
  $z_4(f_4)$
   periodogram
  slices (Eq. \ref{EqSlices}) of \M=4 ~model
are shown in
Fig. \ref{1hjd14R421Sz}.
The transparent
diamonds denote locations of
the red $z_1(f_1)$,
the blue $z_2(f_2)$
the green $z_3(f_3)$
and
the yellow $z_4(f_4)$
periodogram minima.
These minima are
clearly separated.

The four signal \M=4 model is unstable, because it
suffers from the amplitude dispersion ``\AD'' effect.
The periodograms of this model do not betray this
effect (Fig. \ref{1hjd14R421Sz}),
but the exceedingly high amplitude
green $h_3(t)$ and yellow $h_4(t)$ signals
do (Fig. \ref{1hjd14R421Sgdet}).
The errors of both $A_3$ and $A_4$
amplitudes are large.
The $P_4=55172^{\mathrm{d}}$ period is about two times
longer than the $P_3=26846^{\mathrm{d}}$ period.
DCM exploits the anti-phase sum of these two
dispersing high
amplitude signals for modelling all data.

The stable three signal \M=3 model is a better
model for all data
than the failing unstable
``\RD''
four signal \M=4 model.
Fisher-test reveals
with an absolute certainty of \REJ
~that this three signal
\M=3 model is also better than the \M=1 model
or the \M=2 model
(Table \ref{Table1hjd14R}: two times ``$\uparrow$'' in Col 8).

Model \M=3 periodogram minima
are also clearly separated 
(Fig. \ref{1hjd14R321Sz}, lower panel).
When all three periodograms are plotted
in the same scale,
the two $z_1(f_1)$ and $z_2(f_2)$ periodogram
minima appear to be
shallower than the $z_3(f_3)$ periodogram minimum,
because the 
high amplitude $h_3(t)$  signal
dominates in this three signal \M=3 model
(Fig. \ref{1hjd14R321Sz}, upper panel).
This $79999^{\mathrm{d}}$ period $h_3(t)$ signal has a much bigger
impact on the sum of squared residuals $R$
than the two
lower amplitude
$20358^{\mathrm{d}}$ period $h_1(t)$,
and
$24742^{\mathrm{d}}$ period $h_2(t)$ signals.
This three signal \M=3 model is shown in Fig. \ref{1hjd14R321Sgdet}.
The level of residuals, denoted by blue dots,
is stable
and there are no trends.
Each $h_j(t_i)$ signal
\begin{eqnarray}
y_{i,j}=y_i-[g(t_i)-h_j(t_i)]
\label{EqSignals}
\end{eqnarray}
is also shown
separately (Fig. \ref{1hjd14R321SSignals}). 
The red $h_1(t)$ and the blue  $h_2(t)$ curves show
two minima and two maxima, but the green large
amplitude $h_3(t)$
curve shows only one minimum and one maximum.

It takes about one month for an ordinary
PC to compute the results for
the four signal \M=4 model,
as well as to analyse at least twenty bootstrap samples
(Table \ref{Table1hjd14R}: \RModel{4,2,1}).
The computation of five signal model
would take several months.
``Fortunately'',
there is no fourth or fifth signal between
8000 and 80000 days in all data,
because the \M=4 model is unstable ``\RD''.
The three signal \M=3 model
is the best model for all data.
Therefore, we can search for additional
periods shorter than 8000 days
from the \M=3 model residuals.

Since the  \M=3 model residuals contain no trends,
we analyse them by using
$K_3=0$ models
having a constant $p(t)$ level.
The period search between
$500^{\mathrm{d}}$ and  $8000^{\mathrm{d}}$
gives two new periods
$680.^{\mathrm{d}}4$ and $7290^{\mathrm{d}}$ 
(Table \ref{Table1hjd14R}, model \M=6).
In the three signal \M=7 model, the periods
$P_2=7124^{\mathrm{d}}$
and 
$P_3=7698^{\mathrm{d}}$
give
$[P_2^{-1}-P_3^{-1}]^{-1}=95541^{\mathrm{d}}\pm13902^{\mathrm{d}}$,
which 
is equal to the time span $\Delta T =  86171^{\mathrm{d}}$
of all data
(Table \ref{Table1hjd14R}: \M=7).
In other words, the difference between the real
$P_2=7124^{\mathrm{d}}$
and the spurious 
$P_3=7698^{\mathrm{d}}$ period is one
round during $\Delta T$.
Our symbol for this type of
spurious periods is
\begin{itemize}
\item[] ``\SP'' = Spurious period = Unreal periods caused
  by data time span and real
  periodicity.
\end{itemize}
Therefore, we reject the \M=7 model, and the best
model for residuals is the \M=6 model.
In this analysis of residuals,
DCM again consistently re-detects
the same periods and amplitudes of earlier models
having less signals.
Model \M=6  periodograms, and the model itself,
are shown in Figs.
\ref{1hjd58R220Sz}
and 
\ref{1hjd58R220Sgdet}.
The two last $680.^{\mathrm{d}}4$ and $7290^{\mathrm{d}}$ 
signals detected from the residuals
are shown in Fig. \ref{1hjd58R220SSignals}.
As expected of a real O-C signal,
both curves have only one minimum and one maximum.
These two signals are 44.8 and 41.0 times 
weaker than the strongest first
detected
$79999^{\mathrm{d}}$ signal.

  For the original data,
  DCM detects {\it simultaneously}
  the three signals signals and the trend
  of \M=3 model.
  For the residuals,
  the same applies to
  the two signals and the trend of
  \M=6 model.
  In this sense, DCM differs from the ``pre-whitening''
  technique, which requires that the 
  trend must be determined and removed
  {\it before even one signal at the time}
  can be detected
  \citep[e.g.][]{Rei13}.
  This ``pre-whitening'
  technique, which applies the Discrete Fourier Transform
(DFT), was compared to DCM in \paperone ~(Sect. 6).

\begin{figure*}[ht!]
\plotone{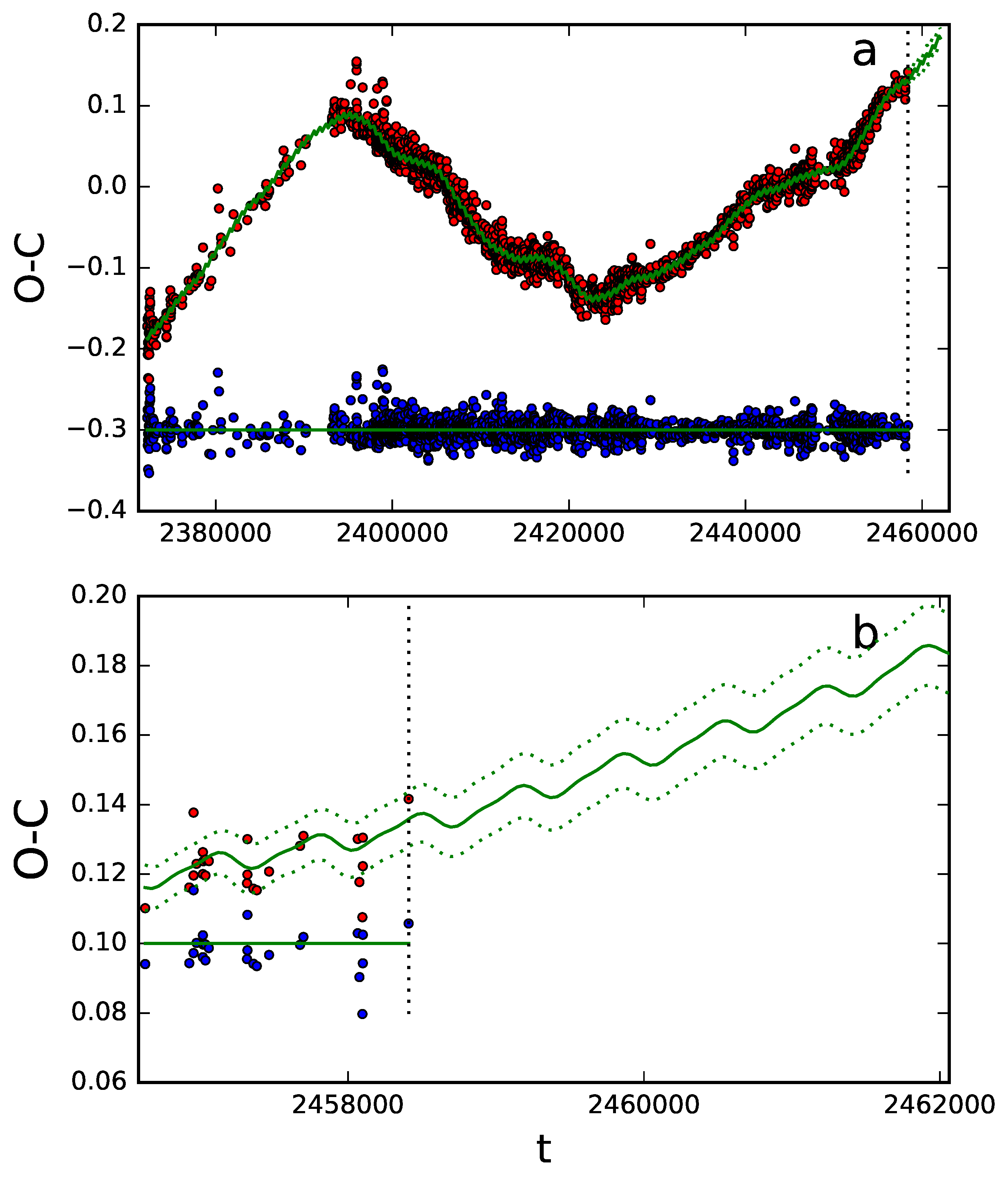}
\caption{All data eccentric orbit analysis
  (Sect. \ref{SectAllDataEccentric}).
  (a) Best five signal \M=3+6 model,
  sum of three signal model \M=3 for original data
  and two signal \M=6 model for
  residuals (green continuous curve),
  is overplotted on all O-C data (red dots).
  Residuals (blue dots) are offset to level -0.3
  (horizontal line).
  Notice the tiny flickering caused by 127 rotations of Algol~C 
  around Algol~AB.
  Vertical dotted
  line marks last observation (Oct 18th, 2018),
  where prediction for next ten years begins.
  (b) Past five years of data and residuals
  (red and blue circles).
  Residuals are offset to level +0.10 (horizontal line).
  Continuous and dotted green lines denote \M=3+6
  model and its $\pm 3 \sigma$ error limits.
  Units are $[t]=$ HJD and $[O\!-\!C]=$ d.
  \label{Prediction1}}
\end{figure*}

We conclude that
DCM detects five signals from all
data $(n=2224)$. 
The full model for all Algol's O-C data 
is the sum of the \M=3 model for the original data,
and the \M=6 model for the residuals
(Table \ref{Table1hjd14R}).
Our notation for this sum
\begin{eqnarray}
{\mathrm{model}}_{3,2,1}+{\mathrm{model}}_{2,2,0}
\nonumber
\end{eqnarray}
 of two models
  in Table \ref{Table1hjd14R}
is simply the
 ``\M=3+6 model''.
 This model is denoted
 with the green continuous line
in Figs. \ref{Prediction1}ab.
Its standard deviation of residuals is
$0.^{\mathrm{d}}011$.
We also give a ten year
prediction for Algol's O-C changes
after our last observation on Oct 18th, 2018
(Fig. \ref{Prediction1}b).

\subsubsection{All data: Circular  orbits \label{SectAllDataCircular}}


  In our appendix, we show that
  if an eccentric
  orbit O-C curve has a period $p$,
  then this curve is a sum
  of two circular orbit O-C
  curves having periods
  $p$ and $p/2$.
  For this reason, the DCM period search
  results obtained for circular orbits
  in this section can be used to
  check the eccentric orbit
  results presented earlier
  in Sect. \ref{SectAllDataEccentric},
  and vice versa (Table \ref{TableCompare}).

  For third body circular orbit,
  the correct DCM model $h_i(t)$ signal  order is $K_2=1$
  (Eq. \ref{EqIrwOne}: $e=0$).
  We fix the $p(t)$ trend to $K_3=1$,
  and search for the correct $K_1$ number of
  circular orbit sinusoidal
  signals in all data.
  Two alternative approaches are tested.
  We will show that both
  approaches give the same results.
  
In the first alternative approach,
we search for one,
two, three and four sinusoidal circular orbit signals
having periods between 8000 and 80000 days
in all data
(Table \ref{Table1hjd14C}).
The one signal \M=1 model is stable.
The two, three and four signal
\M=2, \M=3 and \M=4 models are unstable ``\RD'',
because they all suffer from dispersing amplitudes
``\AD''. 
The largest periods   (``\LK'')
in these three models
exceed
the all data time span $\Delta T=86171^{\mathrm{d}}$.

From the \M=4 model residuals, we detect
the fifth sinusoidal
signal period $10175^{\mathrm{d}}$ (Table \ref{Table1hjd14C}: \M=5).
The next \M=6 model is unstable ``\RD'',
and it is also rejected
with the Fisher-test criterion (Eq. \ref{EqFisher}).

DCM detects signatures of five sinusoidal
signals having periods longer
than 8000 days.
Therefore,
we search for shorter
periods from the \M=5 model residuals.
This reveals three additional
sinusoidal \M=9 model signals
(Table \ref{Table1hjd14C}).
The next
four signal model \M=10
is rejected with the Fisher-test
  criterion (Eq. \ref{EqFisher}).

In our first alternative approach,
the best circular orbit model is
the \M=4+5+9 model
(Table \ref{Table1hjd14C}).

Our typical number of tested  periods 
is $n_{\mathrm{L}}=80$ in the long search,
and $n_{\mathrm{S}}=40$ in the short search.
We use these dense grids to eliminate
the ``trial factor'' error 
(Sect. \ref{SectMethod}: Caveat 2).
Computation time is proportional to
$\propto n_{\mathrm{L}}^{K_1}$ and
$\propto n_{\mathrm{S}}^{K_1}$.
For larger number of signals,
these dense tested grids of ours
take a long time to compute.
For example, the computation of
four signal model for all data,
and its twenty bootstrap samples,
takes about one month for an ordinary PC.

In the second alternative approach we
also search for circular orbit periods
between 8000 and 80000 days.
However, we reduce the computation time dramatically
by testing only
$n_{\mathrm{L}}=30$ and $n_{\mathrm{S}}=8$
frequencies.
In this case, an ordinary PC can
perform the six signal DCM
search in about
one week.
Unlike in the first alternative
approach,
we do not need to search for the
fifth and sixth signal from
the four signal model residuals.
We can perform
the five and the six signal
DCM search {\it directly} to all original data.
The four, five and six signal
circular orbit model
results for all original
data are given in Table \ref{Table1hjd16C}.
All \M=1, 2 and 3  models suffer from
amplitude dispersion ``\AD'', as well as
from 
leaking periods ``\LK'',
because their largest detected periods 
exceed $\Delta T$.
Model \M=3 also suffers from
intersecting frequencies ``\HD''.
We reject it
with the Fisher-test criterion
(Eq. \ref{EqFisher}).
The best circular orbit
model for all original data is the
five sinusoidal signal \M=2 model.
The \M=2
model periodogram
is shown in Fig. \ref{1hjd16R511Sz}.
The periodogram minimum of the largest
$P_5=120740^{\mathrm{d}}$ period
is real,
because the violet  $z_5(f_5)$ curve in the lower panel
turns upwards at smaller tested frequencies
(i.e. periods larger than $\Delta T$).
The \M=2 model itself is shown in Fig. \ref{1hjd16R511Sgdet}.

From model \M=2 residuals,
we find two periods shorter than 8000 days
(Table \ref{Table1hjd16C}: \M=5).
We reject model \M=6,
because the periods
$P_1=7034^{\mathrm{d}}\pm148^{\mathrm{d}}$ and
$P_2=7478^{\mathrm{d}}\pm82^{\mathrm{d}}$
give $(P_1^{-1}-P_2^{-1})^{-1}=118469^{\mathrm{d}}\pm46755^{\mathrm{d}}$.
Hence, the spurious ``\SP'' period $P_1$ 
is connected to the real period $P_2$
and the time
span $\Delta T=86171^{\mathrm{d}}$ of all data.

The second alternative
approach best circular orbit model is the \M=2+5 model
(Table \ref{Table1hjd16C}).

We compare the results of 
our two alternative approach
circular orbit DCM analyses in
Table \ref{TableCompareCircular}.
All results are consistent.
The periods and amplitudes agree
within their error limits.
We detect the same five
longer sinusoidal signal
periods from the original data,
and the same two shorter period
sinusoids from the residuals.
We get these consistent
results even after dramatically
reducing the number of tested frequencies.
Hence, these two analyses
not suffer from
the ``trial factor'' effect
(Sect. \ref{SectMethod}: Caveat 2).
The dispersing amplitudes ``\AD'' or
the leaking periods ``\LK'' do not
either mislead this analysis.

\subsection{\Adataone   \label{SectAdata}}

The eccentric orbit DCM search results
for subsample
\Adataone ~are given
in Table \ref{Table2hjd14R}.
The one signal \M=1 model 
and two signal \M=2 model suffer from
leaking periods ``\LK''.
The stable three signal \M=3 model is the best
one for the original data,
because the four signal \M=4 model is unstable ``\RD''.

For the \M=3 model residuals,
the best model is \M=6 model.
We reject model \M=7,
because the relation
$[P_2^{-1}-P_3^{-1}]^{-1}=81963^{\mathrm{d}}\pm13594^{\mathrm{d}}$
reveals that the third $P_3=7757^{\mathrm{d}}$ period
is a spurious  ``\SP''
period connected
to the real period $P_2=7078^{\mathrm{d}}$
and the time span $\Delta T=82602^{\mathrm{d}}$
of data.

\begin{figure*}[ht!]
\plotone{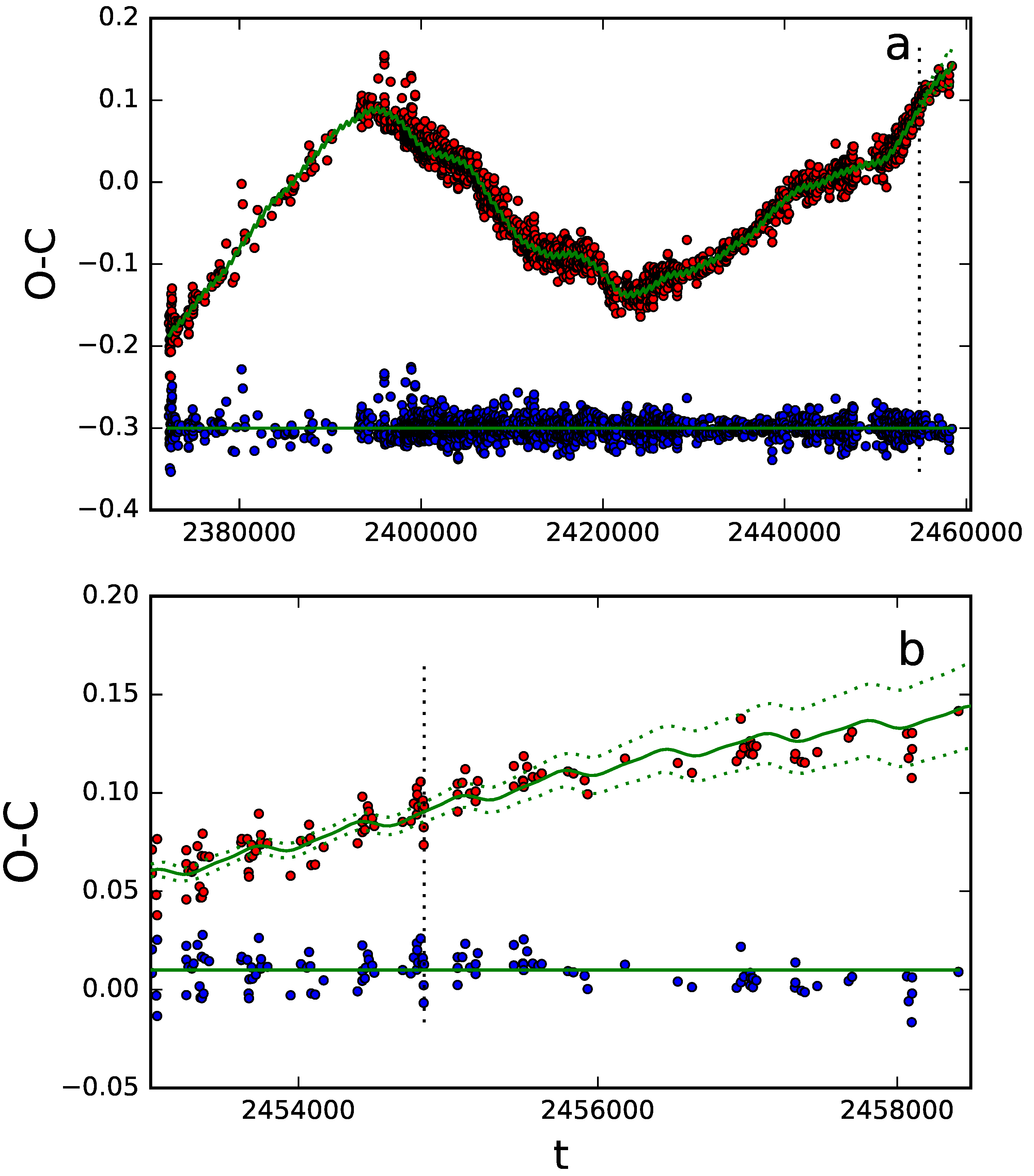}
\caption{\Adataone ~eccentric
    orbit analysis (Sect. \ref{SectAdata}).
  (a) Model \M=3+6 (Table \ref{Table2hjd14R}).
  Otherwise as in Fig. \ref{Prediction1}a.
  (b) Prediction for
  \Adatatwo.
Otherwise as in Fig. \ref{Prediction1}b.
  \label{Prediction2}}
\end{figure*}

The best model for
\Adataone ~is the \M=3+6 model
(Table \ref{Table2hjd14R}).
This model is shown in Fig. \ref{Prediction2}.
It gives 
an excellent prediction for
the next nine years of
\Adatatwo ~(Fig. \ref{Prediction2}b).
The standard deviation of prediction
residuals is only $0.^{\mathrm{d}}0078$ $(n=50)$.
It is smaller than the standard deviation
$0.^{\mathrm{d}}011$ of 
the ~predictive \M=3+6 model residuals
$(n=2174)$.
However,
the larger errors of the older observations
can explain this contradiction.
The main conclusion is that
our nine years
prediction succeeds.

\subsection{\Bdataone   \label{SectBdata}}

The eccentric orbit 
DCM search results for 
the shortest subsample
\Bdataone ~are
given in Table \ref{Table3hjd14R}.
The one signal \M=1 model is stable.
The two and three signal
\M=2 and \M=3 models are unstable
(Table \ref{Table3hjd14R}: ``\RD'').
The best model for \Bdataone
~is the stable four signal \M=4 model.

For the \M=4 model residuals,
the stable \M=6 model is the best one,
because the \M=7 model is unstable ``\RD''.

\begin{figure*}[ht!]
\plotone{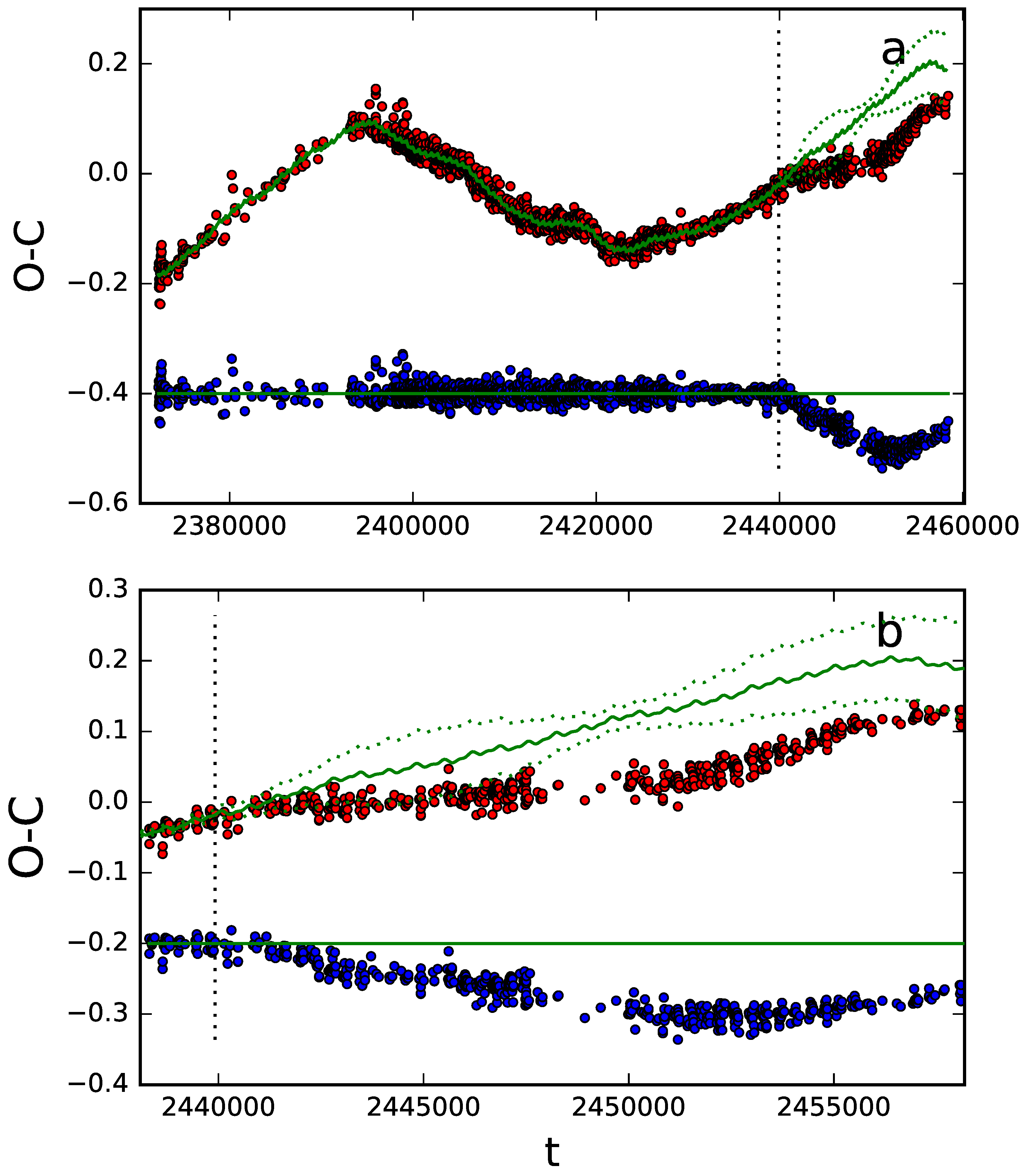}
\caption{\Bdataone ~eccentric
  orbit analysis (Sect. \ref{SectBdata}).
  (a) Model \M=4+6 (Table \ref{Table3hjd14R}).
  Otherwise as in Fig. \ref{Prediction1}a.
(b) Prediction for \Bdatatwo.
Otherwise as in Fig. \ref{Prediction1}b.
  \label{Prediction3}}
\end{figure*}

The best \M=4+6 model for \Bdataone
~is shown in Fig. \ref{Prediction3}.
Our fifty years prediction succeeds
only for the first few years (Fig. \ref{Prediction3}b).
However, this is no surprise, because
the time span of predictive data 
is only $\Delta T=67680^{\mathrm{d}}=185^{\mathrm{y}}$.
For this reason,
the longest and the
strongest detected predictive signal
period is $P_4=62992^{\mathrm{d}}=172^{\mathrm{y}}$
(Table \ref{Table3hjd14R}: \M=4).
This high amplitude signal
determines the long-term
prediction trend for 
\Bdatatwo.
We have already shown that
the correct period for
this long-term trend would be
$219^{\mathrm{y}}$
(Table \ref{Table1hjd14R}: \M=3, Table \ref{Table2hjd14R}: \M=3).
The short $185^{\mathrm{y}}$ time span
of \Bdataone ~prevents the 
detection this correct $219^{\mathrm{y}}$ period.
The {\it correct} $219^{\mathrm{y}}$ signal trend
turns upwards slower than
the {\it wrong}  $172^{\mathrm{y}}$ signal trend.
This is the simple reason for the failure of
our fifty years prediction
for \Bdataone.

The \Bdatatwo ~prediction
error for shows
a peculiarity that seems
to defy the laws of statistics.
First, the $\pm 3 \sigma$ prediction error increases,
as one would expect
(Fig. \ref{Prediction3}b: green dotted lines).
Surprisingly,
this prediction error then begins to decrease,
and the prediction becomes very
accurate close to HJD~2450000.
After this,
the prediction error begins to
increase again.
This peculiarity certainly
requires an explanation.

The reason of this peculiarity 
could already be inferred from the black 
interference curve in Fig. \ref{FigOCthree}
(lowest right panel:  $P_1=24771$).
The scatter of $g(t)$ interference curve is
not the same at all phases.
In this particular case,
this scatter increases close to the maxima,
but it decreases close to the minima.
The largest and the smallest scatter coincides
with the phases when the
first time derivative fulfills $\dot{g}(t)=0$.

However,
the above mentioned effects in
Fig. \ref{FigOCthree} are caused by
interference of only two signals,
while the peculiar error limit effect
in Fig. \ref{Prediction3}
occurs in the \M=4+6 model sum of  
six signals.
We show 
this model for twenty bootstrap samples
in  Fig. \ref{AllSolutions} (red dotted curves).
The scatter of these curves increases
when the predictive data ends
at the dotted black vertical line.
However, all dotted red 
curves converge close to the
vertical continuous black
line at HJD~2450000.
After this line,
they diverge again.
Before this line,
the data shows an increasing trend,
but the positive slope is decreasing
(Fig. \ref{Prediction3}a: red circles).
A suitable model would be
$\dot{g}(t)>0$ and $\ddot{g}(t)<0$.
After this line, this slope is still positive,
but it is increasing. Now the suitable
model would be
$\dot{g}(t)>0$ and $\ddot{g}(t)>0$.
This means that there is a
turning point $\dot{g}(t)=0$
close this HJD~2450000 epoch,
where the $\ddot{g}(t)$ sign changes from
negative to positive.
The second derivative sign change
of {\it any} function forces this function
to change its direction twice.
This \M=4+6 model turning point forces the bootstrap
model solutions to converge.
This simple effect explains
why the prediction error increases,
decreases, and again increases
(Fig. \ref{Prediction3}: green dotted lines).

Our turning point hypothesis would explain the 
gap in O-C data close to HJD~2450000
(Fig. \ref{AllSolutions}: vertical continuous line).
There are no such gaps in Algol's modern O-C data,
not even during the two World Wars.
TIDAK database contains
  only four O-C values
    between HJD~2448288
    and HJD~2449988 ($\equiv$ 4.6 years).
    Even today, one of these four
    is still marked ``unpublished'' 
    (1997, Drozdz: HJD~2449317.4171).
Close to the above mentioned turning point,
the O-C data did no longer
support the well established
expected long-term $\dot{g}(t)>0$ and $\ddot{g}(t)<0$ trend.
Perhaps for this reason, the
contradictory new data was not published at that time.
Only when the new  $\dot{g}(t)>0$ and $\ddot{g}(t)>0$ trend
was securely established,
the continuous flow of supporting O-C
observations began again.

We conclude that, except for the first few years,
our \Bdatatwo ~prediction fails.
However, our turning point epoch prediction
HJD~2450000
is excellent.

\subsection{Additional experiments
  \label{SectExperiments}}

 We divide all
   original data into two parts.
   Both halves are too short for
the detection of the long 219 years
period. This hampers their
period analysis.
 In the first low accuracy half,
 we detect only one signal
 of about 137 years.
  From the more
 accurate second half,
 we detect four signals
 of 1.86
 30.9,
 39.7 and 103.3 years.
 The shortest one is equal
 to the orbital period of Algol~C.

 We also test two alternatives, where the
weights of observations increase linearly.
In two alternative experiments,
the weights are doubled or quadrupled
during the time span of all data.
In both cases,
the five strongest signals detected from
the weighted data
are identical to those
detected from non-weighted data
(Table \ref{Table1hjd14R}, \M=3+6 model).

\subsection{Signals identified in all data
  \label{SectAllDataSignals}}

The eccentric orbit analysis
indicates that all data
contains five signals (Table \ref{Table1hjd14R}, \M=3+6).
Here, we argue that
the correct number of signals may
also be six.
We use bold letters 
$\PPone$,
$\PPtwo$,
$\PPthree$,
$\PPfour$,
$\PPfive$ and
$\PPsix$ for the periods
of these signals (Table \ref{TableOrbits}).
This notation helps the readers to separate
these six periods from the numerous other
$P_1,P_2,..., P_6, p, p_1, p_2, p_3$ and $p'$ periods.
We use the tentative names
  Algol~C,
  Algol~D,
  Algol~E,
  Algol~F,
  Algol~G
  and Algol~H
  for the objects
  possibly connected to these
  periods.
The corresponding peak to peak
amplitudes are
$\AAone$,
$\AAtwo$,
$\AAthree$,
$\AAfour$,
$\AAfive$ and
$\AAsix$.

Our six signal argument relies on two tables.
The first table compares the
eccentric and circular orbit analysis
periods for all data  (Table \ref{TableCompare}).
The second table compares the 
periods detected in three different samples:
All data,
\Adataone ~and \Bdataone
~(Table \ref{TableThreeSamples}).

In our Appendix,
we apply
DCM to simulated O-C data (Eq. \ref{EqIrwOne}).
We show that the following
four different effects are encountered when 
the O-C data contains one period $p$,
or two periods $p_1$ and $p_2$.

\begin{itemize}

  \item[] ``\PC'': DCM detects the correct period $p$.
  \item[] ``\PH'': DCM detects the spurious period $p/2$.
  \item[] ``\PD'': DCM detects the spurious period $2p$.
  \item[] ``\PI'': DCM detects the spurious period $p'$
    caused by $p_1$ and $p_2$ interference (Eq. \ref{EqSpurious}).

\end{itemize}

\noindent
The ``\PH'' and ``\PD'' effects can mislead
  DCM analysis of low eccentricity O-C curves, which
  resemble pure sinusoids.

There is only one minimum and one maximum
in the real O-C curve caused by the
LTTE of a single third body.
  This third body can approach and recede
  only once during one orbital period $p$.
  Hence, the O-C ``$p'$ interference''
  curves having two minima and two maxima
  can not be caused by one body alone,
  but they may indicate the presence of
  more than one body.

  In the next
  Sects. \ref{SectSignalOne}-\ref{SectSignalSix},
  we illustrate
  one
  $\PPone$, $\PPtwo$,
$\PPthree$,
$\PPfour$,
$\PPfive$ and
$\PPsix$ 
  signal at the time,
  how the above mentioned 
  four effects can explain
  all eccentric and
  all circular orbit DCM period search results.
  
  \subsubsection{Signal $\PPsix=79999^{\mathrm{d}}
    =219^{\mathrm{y}}.0$
\label{SectSignalOne}}

The circular orbit signal
period $P_{\mathrm{c,7}}=120740^{\mathrm{d}} \pm 41002^{\mathrm{d}}$ 
differs about $\pm 1 \sigma$
from the eccentric orbit period
$\PPsix=P_{e,5}=79999^{\mathrm{d}} \pm 1216^{\mathrm{d}}$
(Table \ref{TableCompare}).
Hence,
the circular and
eccentric orbit analyses give
the same correct $\PPsix$ period (``\PC'' effect).

This $\PPsix$ period is two times longer
than the next circular orbit period
$P_{c,6}=42422^{\mathrm{d}} \pm 640^{\mathrm{d}}$
(``\PH'' effect).
The $\PPsix=219^{\mathrm{y}}$ signal curve
shows only one minimum
and one maximum
(Fig. \ref{1hjd14R321SSignals}: lowest panel green curves),
because the two strongest circular orbit
$P_{c,7}$ and $P_{c,6}$
signals are ``in phase''.
These results confirm that
DCM succeeds in detecting the $p$ and $p/2$ regularities
illustrated in Fig. \ref{FigOCone}
and Table \ref{TableOCtwo}. 

DCM detects the $\PPsix=219^{\mathrm{y}}$ signal
in all data
and 
\Adataone ~(Table \ref{TableThreeSamples}).
The too short
\Bdataone ~time span 
 prevents the detection of
 the $\PPsix$ period.
 Therefore,
 the largest detected
$P_4=62992^{\mathrm{d}}$ $\pm 2499^{\mathrm{d}}$ period
differs more than $\pm 3 \sigma$ from $\PPsix$.

We use an amplitude estimate
$\AAsix=A_{e,5}=0.^{\mathrm{d}}287 \pm 0.^{\mathrm{d}}005$
for this $\PPsix=219^{\mathrm{y}}$ signal
(Table \ref{TableCompare}).

\subsubsection{Signals $\PPfive=24247^{\mathrm{d}}=
  66.^{\mathrm{y}}4$  and
  $\PPfour=12294^{\mathrm{d}}=33.^{\mathrm{y}}7$
\label{SectSignalTwo}}

The connection between the eccentric orbit
$\PPfive=P_{e,4}=24742^{\mathrm{d}}\!\pm \!141^{\mathrm{d}}$ signal
and the circular orbit
$P_{c,5}=24747^{\mathrm{d}}\!\pm\! 872^{\mathrm{d}}$ signal
is definitely the ``\PC'' effect (Table \ref{TableCompare}).
The ``\PH'' effect certainly
connects this $\PPfive$ signal also to
circular orbit
$P_{c,4}=12294^{\mathrm{d}}\!\pm\! 109^{\mathrm{d}}$
signal.

However, two questions need to be answered.
Why does the $\PPfive=66^{\mathrm{y}}.4$
signal show two minima
and two maxima (Fig. \ref{1hjd14R321SSignals}:
mid-panel blue curves)?
This is impossible for
any single third body eccentric orbit.
Why are the $A_{\mathrm{c,5}}$ and $A_{\mathrm{c,4}}$ amplitudes
of the two circular orbit
$P_{\mathrm{c,5}}$ and $P_{\mathrm{c,4}}$ signals
practically equal (Table \ref{TableCompare})?

The easiest answer to both 
questions would be that
the $\PPfive=66.^{\mathrm{y}}4$ and
$\PPfour=33.^{\mathrm{y}}7$
signals represent
two separate independent signals,
which are ``off-phase''.
Their ``\PI'' effect could induce
the two unequal minima and
two unequal maxima of the
blue O-C curve (Fig. \ref{1hjd14R321SSignals}),
which
resembles
the black interference curve in Fig. \ref{FigOCthree}.
In this case,
the circular orbit
$P_{\mathrm{c,4}}=12294^{\mathrm{d}}\pm 109^{\mathrm{d}}$ signal
could represent a real fourth
independent $\PPfour=33.^{\mathrm{y}}7$ signal.

The $\PPfive=66.^{\mathrm{y}}4$ signal
is detected in all data and
\Adataone ~(Table \ref{TableThreeSamples}).
This $\PPfive$ signal
is not detected in the shortest
\Bdataone ~sample,
but the $\PPfour=33.^{\mathrm{y}}7$ signal is.
We conclude that the $\PPfive=66.^{\mathrm{y}}4$
and $\PPfour=33.^{\mathrm{y}}7$
signals are most probably two independent real signals.

The amplitudes of the circular orbit
$P_{\mathrm{c,5}}$ and $P_{\mathrm{c,4}}$ signals 
give our 
$\AAfive=A_{c,5}=0.^{\mathrm{d}}018\pm 0.^{\mathrm{d}}002$
and 
$\AAfour=A_{c,4}=0.^{\mathrm{d}}018\pm 0.^{\mathrm{d}}001$
amplitude
estimates for the $\PPfive$ and $\PPfour$ signals
(Table \ref{TableCompare}).

Here,
we have
shown that the eccentric orbit
$\PPfive=66.^{\mathrm{y}}4$ signal
  may arise from the  ``\PI'' effect of two circular orbit
  $\PPfive=66.^{\mathrm{y}}4$ and
  $\PPfour=33.^{\mathrm{y}}7$ sinusoids.
  Later,
  we will present an alternative explanation
  (Fig. \ref{Configurations}: Configurations 2 and 3).

  \subsubsection{Signal $\PPthree=10144^{\mathrm{d}}
    =27.^{\mathrm{y}}8$
  \label{SectSignalFour}}

None of the eccentric orbit periods is close
to the circular orbit period
$P_{c,3}=10144^{\mathrm{d}}\pm 30^{\mathrm{d}}
=27.^{\mathrm{y}}8 \pm 0.^{\mathrm{y}}1$ 
(Table \ref{TableCompare}).
However,
the ``\PD'' effect certainly connects this $P_{c,3}$
period to the eccentric orbit period
$P_{e,3}=20358^{\mathrm{d}}\pm 128^{\mathrm{d}}$.
This $P_{e,3}$ signal
shows two maxima and two minima
(Fig. \ref{1hjd14R321SSignals}: lower panel red curves).
These two equal maxima and two equal minima 
are symmetric.
This kind of symmetry is detected
in our simulations of
low eccentricity spurious double sinusoids
(Table \ref{TableOCone}:  ``\Dan''$\equiv$''\PD'' effect).
Therefore,
the $P_{c,3}$ period probably represents a real signal
$\PPthree=27.^{\mathrm{y}}8$.

The eccentric orbit
$P_{e,3}=20358^{\mathrm{d}}$ signal
is detected in
all data,
\Adataone ~and
  \Bdataone ~(Table \ref{TableThreeSamples}).
This means that DCM detects
the $\PPthree \approx P_{e,3}/2$ signal
in all these three different samples.

Our amplitude estimate
for this $\PPthree=27.^{\mathrm{y}}8$
signal is
$\AAthree=A_{c,3}=0.^{\mathrm{d}}0097 \pm 0.^{\mathrm{d}}0004$
(Table \ref{TableCompare}).

In this section,
we have shown that the eccentric orbit
$P_{e,3}=56.^{\mathrm{y}}0$ signal
probably represents the ``double wave''
of  the $\PPthree=27.^{\mathrm{y}}8$
signal.
We will later present an alternative explanation
(Fig. \ref{Configurations}: Configuration 3).

\subsubsection{Signal $\PPtwo=7269^{\mathrm{d}}
  =20.^{\mathrm{y}}0$
\label{SectSignalFive}}

The eccentric orbit
$\PPtwo=P_{e,2}=7269^{\mathrm{d}}\pm29^{\mathrm{d}}$
signal and the circular orbit
$P_{c,2}=7395^{\mathrm{d}}\pm 37^{\mathrm{d}}$
signal are certainly connected
(Table \ref{TableCompare}: ``\PC'' effect).

Like any real third body O-C curve,
this  $\PPtwo=20.^{\mathrm{y}}0$
signal shows only
one minimum and one maximum
(Fig. \ref{1hjd58R220SSignals}: lower panel blue curves). 
DCM detects this $\PPtwo=20.^{\mathrm{y}}0$ signal
in all data and
\Adataone ~(Table \ref{TableThreeSamples}).
In shortest
\Bdataone ~sample,
this $\PPtwo$ period may be connected
to its double period
$P_3=15429^{\mathrm{d}}\pm 222^{\mathrm{d}}$ 
(Table \ref{TableThreeSamples}: ``\PD'' effect).

Our amplitude estimate
for this $\PPtwo=20.^{\mathrm{y}}0$ signal is
$\AAtwo=A_{e,2}=0.^{\mathrm{d}}007 \pm 0.^{\mathrm{d}}001$
(Table \ref{TableCompare}).

\subsubsection{Signal $\PPone=680.4^{\mathrm{d}}
  =1.^{\mathrm{y}}86$
\label{SectSignalSix}}

The eccentric orbit and circular orbit
DCM searches give
the same $\PPone=680.^{\mathrm{d}}4 \pm 0.^{\mathrm{d}}4=
1.^{\mathrm{y}}863\pm0.^{\mathrm{y}}001$ signal
(Table \ref{TableCompare}: ``\PC'' effect).

DCM detects this $\PPone=
1.^{\mathrm{y}}863$
signal from all three samples
(Table \ref{TableThreeSamples}).
Like any real O-C curve, this signal
shows only one minimum and one maximum
(Fig. \ref{1hjd58R220SSignals}: higher panel red curves).

We use
$\AAone\!=\!A_{e,1}\!=
\!0.^{\mathrm{d}}0064\! \pm \! 0.^{\mathrm{d}}0007$
(Table \ref{TableCompare}).
This signal is discussed later in
greater detail (Sect. \ref{SectAlgolC}). 

\subsubsection{Two weakest signals}

DCM detects indications of two additional weaker signals
$P_{c,2}=2986^{\mathrm{d}}\pm 39^{\mathrm{d}}$
(Table \ref{TableCompareCircular})
and
$P_{2}=3387^{\mathrm{d}}\pm 17^{\mathrm{d}}$
(Table \ref{TableThreeSamples}).
They could be separate signals, because 
their $\pm 3\sigma$ 
error limits do not overlap.
They are 0.80 and 0.45 weaker than the
weakest detected $\PPone=
1.^{\mathrm{y}}863$ signal.
We can not confirm whether these two
weakest signals are real or spurious.

  \begin{deluxetable*}{cccccccccccccccl}
    \tablecaption{Third body circular orbits
        (Sect. \ref{SectAllDataSignals}).
     Periods $(\PPone, ... \PPsix)$
     and
     amplitudes $(\AAone, ..., \AAsix)$ used to compute
     third
     mass $m_3$ and
     semi-major axis $a_3$ estimates
     (Eqs. \ref{EqMass} and \ref{EqSemi}).
     Inclination alternatives are
     $i=90^{\mathrm{o}}$, 
     $60^{\mathrm{o}}$ and $30^{\mathrm{o}}$.
     Last column gives our tentative object names.
     We emphasize that our
       approximate
       $m_3$ and $a_3$ estimates are based
       on four assumptions.
       (1) All six signals are caused by LTTE of wide orbit candidates.
       (2) Correct hierarchial system alternative
       is Configuration 1 (Fig. \ref{Configurations}).
       (3) All orbits are circular.
       (4) Every candidate can be treated as a "third body".
       In other words, effects of other
       candidates inside 
      ``third body'' orbit can be ignored in
   Eqs. \ref{EqMass} and \ref{EqSemi}.
     \label{TableOrbits}
   }
\tablewidth{700pt}
   \addtolength{\tabcolsep}{-0.10cm}
\tabletypesize{\scriptsize}
\tablehead{ 
  \colhead{}   & \colhead{Periods}   & \colhead{} & \colhead{} & \colhead{}  & \colhead{Amplitudes} &
  \colhead{} & \colhead{$m_3^{i=90}$}   & \colhead{$a_3^{i=90}$} & &
  \colhead{$m_3^{i=60}$}   & \colhead{$a_3^{i=60}$} & &
  \colhead{$m_3^{i=30}$}   & \colhead{$a_3^{i=30}$} & \\
  \cline{1-3}
  \cline{5-6}
  \cline{8-9}
  \cline{11-12}
  \cline{14-15}  
 & [d] & [y] & & &  [d] & & $[m_{\odot}]$ & [AU] && $[m_{\odot}]$ & [AU] && $[m_{\odot}]$ & [AU] & Name}
\startdata
       $\PPsix$ &$    79999   \pm     1216   $&$    219.0  \pm      3.3  $&&
       $\AAsix$ &$     0.287  \pm      0.005 $&&
$   2.50  \pm    0.02 $&$   44.7   \pm    0.4  $&&
$   3.03  \pm    0.03 $&$   42.6   \pm    0.3  $&&
$   6.94  \pm    0.08 $&$   32.2   \pm    0.2  $&
   Algol H \\
      $\PPfive$ &$    24246   \pm      872   $&$     66.4  \pm      2.4  $&&
      $\AAfive$ &$     0.018  \pm     0.002  $&&
$   0.27  \pm    0.02 $&$   26.1   \pm    0.5  $&&
$   0.31  \pm    0.02 $&$   25.9   \pm    0.5  $&&
$   0.56  \pm    0.04 $&$   25.0   \pm    0.5  $&
   Algol G \\
   $\PPfour$ &$    12294   \pm      109      $&$     33.7     \pm      0.3    $&&
      $\AAfour$ &$     0.018  \pm     0.001  $&&
$   0.43  \pm    0.03 $&$   16.19  \pm    0.04 $&&
$   0.50  \pm    0.04 $&$   16.06  \pm    0.01 $&&
$   0.91  \pm    0.07 $&$   15.2   \pm    0.05 $&
   Algol F \\
     $\PPthree$ &$    10145   \pm       30   $&$     27.78    \pm      0.08  $&&
     $\AAthree$ &$     0.0097 \pm     0.0004 $&&
$   0.26 \pm    0.01 $&$   14.596 \pm    0.002$&&
$   0.30  \pm    0.01 $&$   14.50  \pm    0.01 $&&
$   0.54  \pm    0.02 $&$   14.053 \pm    0.005$&
   Algol E \\
       $\PPtwo$ &$     7290  \pm       29   $&$     19.96  \pm      0.08  $&&
       $\AAtwo$ &$     0.007 \pm     0.001  $&&
$   0.24  \pm    0.03 $&$   11.72  \pm    0.07 $&&
$   0.28  \pm    0.04 $&$   11.67  \pm    0.08 $&&
$   0.49  \pm    0.07 $&$   11.34  \pm    0.12 $&
   Algol D \\
       $\PPone$ &$      680.4 \pm        0.4 $&$      1.863 \pm      0.001 $&&
     $\AAone$   &$     0.0064 \pm     0.0007 $&&
$   1.2   \pm    0.1  $&$    2.14  \pm    0.04 $&&
$   1.4   \pm    0.2  $&$    2.09  \pm    0.04 $&&
$   2.8   \pm    0.4  $&$    1.82  \pm    0.07 $&
   Algol C \\
   \enddata
   \addtolength{\tabcolsep}{+0.10cm}
 \end{deluxetable*}

\section{Discussion
\label{SectDiscussion}}

\cite{App92} mechanism can not explain
the numerous
strictly periodic O-C signals
of Algol, because quasi-periodic
activity cycles are never regular.
Apsidal motion follows only one period.
LTTE of Algol's companion candidates could
cause these numerous strictly periodic cycles.
Assuming circular orbits,
we use $m_1=3.7 m_{\odot}$ and $m_2=0.8 m_{\odot}$ \citep[][]{Zav10}
to compute 
the $m_3$ mass
and
the $a_3$ semi-major axis
estimates for these tentative companion
candidates (Table \ref{TableOrbits}).
These approximate mass and semi-major
  axis estimates are obtained
  by assuming that 
  each candidate is a "third" component.
  The effects of other
candidates inside the orbit
of the ``third'' component
are ignored in
   Eqs. \ref{EqMass} and \ref{EqSemi}.

\subsection{Hierarchial structure \label{SectHierarchial}}

We call the eclipsing Algol~A and Algol~B pair the
central eclipsing binary (cEB).
Algol~C is called a wide orbit star (WOS),
as well as all other new tentative companion candidates.
We use the same hierarchial system diagrams
as \citet{Tok21}.

Our first hierarchial system
diagram shows the circular orbit 
$i=90^{\mathrm{o}}$ inclination
case  of Table \ref{TableOrbits}
(Fig. \ref{Configurations}: Configuration 1).
The eight members in this configuration
are cEB and six WOSs.
The orbital periods WOS candidates
are between 1.863 and 219.0 years.
The most massive $(m_3=2.50 m_{\odot})$
companion candidate Algol~H is
also the most distant one $(a_3=44.7{\mathrm{AU}})$.
The four other WOS candidates are low mass stars
$(0.23 m_{\odot} \le m_3 \le 0.43 m_{\odot})$.
The
closest $m_3^{i=90}=1.16 m_{\odot}$ companion
candidate has
an orbital period $\PPone=680.^{\mathrm{d}}4\pm 0.^{\mathrm{d}}4$,
which is close to the known orbital period
$P_{\mathrm{orb}}=679.^{\mathrm{d}}85 \pm 0.^{\mathrm{d}}04$
of Algol~C \citep{Zav10}.
We will discuss this probable detection of Algol~C later
in Sect. \ref{SectAlgolC}.

Our second hierarchial system
diagram shows one alternative for Configuration 1
(Fig. \ref{Configurations}: Configuration 2). 
The seven members are cEB and five WOSs.
We have already
shown that the sum of ``off-phase'' sinusoidal
$\PPfive=66.^{\mathrm{y}}4$ and
$\PPfour \approx \PPfive/2=33.^{\mathrm{y}}7$
signals can cause
the
$\PPfive=66.^{\mathrm{y}}4$
period double wave 
(Sect. \ref{SectSignalTwo}).
However, a single  $\PPfive=66.^{\mathrm{y}}4$
long-period binary 
can cause a similar effect,
if the masses of its members are unequal. 
These unequal masses could also explain
the two unequal maxima and minima of the
blue O-C curve in Fig. \ref{1hjd14R321SSignals}.
The red lines in our Configuration 2 diagram show
this hypothetical long-period $\PPfive=66.^{\mathrm{y}}4$
binary having
an orbital period $\PPsix=219.^{\mathrm{y}}0$
around the barycentre of the
whole system
(Fig. \ref{Configurations}). 

Our third hierarchial system
diagram is a minor modification of Configuration 2
(Fig. \ref{Configurations}: Configuration 3). 
The seven members are, again, cEB and five WOSs.
Now we take the five periods
of \M=3+6 model as such.
Signal $66.^{\mathrm{y}}4$ is not separated
into two signals (Sect. \ref{SectSignalTwo}).
We use the full $P_{e,3}=55.^{\mathrm{y}}8$ signal period,
not the half of this period (Sect. \ref{SectSignalFour}).
This $P_{e,3}=55.^{\mathrm{y}}8$ signal
could also represent a long-period binary,
where the masses of both components
are approximately equal.
In Configuration 3,
the two long-period 
$\PPfive=66.^{\mathrm{y}}4$
and
$P_{e,3}=55.^{\mathrm{y}}8$
binaries
orbit each other
during $\PPsix=219.^{\mathrm{y}}0$.
This may be the most stable one
of our three configuration alternatives,
because
the cEB and the remaining
two inner orbit WOSs would only
weakly perturb the two hypothetical
long-period binaries, and vice versa.
This type of quintuple binary systems
have been discovered
\citep[e.g.][their Fig. 2 of V994 Her]{Zas13}.

\subsection{Detectability \label{SectDetection}}

In binaries,
the radial velocity observations can
reveal the presence of a third body,
like in the discovery of Algol~C \citep{Cur08}.
For nearby hierarchial systems,
combined astrometric orbit and radial velocity
observations can be used to solve their detailed
structure \citep[e.g.][]{Tok21}.
When \citet{Haj19} searched for WOSs
from the O-C data of 80~000 EBs, they detected 
992 systems having one WOS,
but only four systems possibly
had two WOSs.
Our DCM analysis of Algol's O-C data
suggests the presence of five or six WOSs.
These O-C data can not reveal a lot about
the structure this hierarchial system,
not even the exact number of stars
(Fig. \ref{Configurations}: Configurations 1, 2 or 3).
However, we can give some ideas that may help
in the detection of Algol's WOS candidates.
Here, we assume that the candidate orbits are circular
and their orbital plane inclinations are $i_3=90^{\mathrm{o}}$
(Table \ref{TableOrbits}).
In this Configuration 1,
the observed maximum and minimum radial velocities of WOSs
candidates are
\begin{eqnarray}
  v_{\mathrm{max}} & = & v_0 + {{2 \pi a_3}\over{p_3}}
                         \label{EqVradMax} \\
  v_{\mathrm{min}} & = & v_0 - {{2 \pi a_3}\over{p_3}},
                         \label{EqVradMin}
\end{eqnarray}
where
$v_0=4.0$ km/s is Algol's radial velocity \citep{Wil53}.


The angular distance between Algol
and its WOSs changes constantly.
We compute these angular distance
changes in Algol's
cEB frame of rest.
At the O-C curve minima and maxima,
the largest distance changes 
are
\begin{eqnarray}
  \Delta a_{\mathrm{max}}(\Delta t)= 2 a_3 \sin{(\pi \Delta t/p_3)}
  \label{EqMoveMax}
\end{eqnarray}
during a time interval $\Delta t \le p_3/2$.
For longer time intervals,
we use $\Delta t=p_3/2$ which
gives $ \Delta a_{\mathrm{max}}=2a_3$.
The smallest 
\begin{eqnarray}
  \Delta a_{\mathrm{min}}(\Delta t)=a_3 [1-\cos{(\pi \Delta t/p_3)}]
  \label{EqMoveMin}
\end{eqnarray}
distance changes
coincide with the
O-C curve mean level.
This relation holds for $t_0 \le p_3$.
For longer time intervals,
we use $\Delta t=p_3$ which gives
$\Delta a_{\mathrm{min}}=2a_3$.

The proper motion of Algol is $\mu_0=2.49$ mas/y \citep{Van07}.
The minimum and maximum proper motion of each candidate is
\begin{eqnarray}
  \mu_{\mathrm{min}} & = & \mu_0 - \mu_c \label{EqMuMin}\\
  \mu_{\mathrm{max}} & = & \mu_0 + \mu_c, \label{EqMuMax}
\end{eqnarray}
where $\mu_c=\Delta a_{\mathrm{max}}(\Delta t=1^{\mathrm{y}})$ is
the maximum proper motion during
one year.
Note that $\mu_{\mathrm{min}}=0$ for every candidate,
because their $\mu_c > \mu_0$.

We emphasize that our
$\Delta a_{\mathrm{min}}$ and $\Delta a_{\mathrm{max}}$
estimates refer to the candidate
distance changes with respect to cEB,
while our $\mu_{\mathrm{min}}$ and
$\mu_{\mathrm{max}}$ estimates refer
to the proper motion of all members in the sky.

All parameters of
Eqs. \ref{EqVradMax}-\ref{EqMuMax}
are given
in Table \ref{TableDetection}.
The two estimates for $\Delta a_{\mathrm{min}}$ and
$\Delta a_{\mathrm{max}}$ are computed for
observations spanning 5 or 20 years.
This information is useful for future
searches of our Algol's member candidates.

In December 2020,
the latest third Gaia data release (DR3)
confirmed no certain detections
$\pm4"$ around Algol,
and only one certain $\pm 40"$ detection.
%
%
%
%
%
In their analysis of Gaia DR3 data,
\citet{Tor20} note that ``most problems come from the bright
sources and the strange image profiles.''
They rejected 8159.3 million bright sources,
158.0 million very bright sources and
4066.7 million odd window profiles.
Algol is definitely ``too bright''.
Its brightness profile
is constantly changing
due to the movement of the known members 
Algol~A, Algol~B and Algol~C,
let alone due to
the primary and secondary eclipses.
Therefore, Gaia could
not have measured 
the positions and movements of
the objects in our Table \ref{TableDetection}. 

Algol~H candidate would be easiest to detect,
because its distance from the cEB is the largest.
This most massive candidate is very probably
also the brightest candidate.
At the moment, its O-C curve
is close to the mean level
(Fig. \ref{1hjd14R321SSignals}:
left  hand lowest panel green $h_3(t)$ curve).
Hence, Algol~H would be close to its projected
maximum $a_3=1569$ mas distance from cEB.
The cEB is receding from us
because its O-C values are
increasing for the next fifty years.
Currently, Algol~H would be
approaching us at its minimum
radial velocity $v_{\mathrm{min}}=-2$ km/s
(Table \ref{TableDetection}).
The distance changes between cEB and 
Algol~H would be small,
only 
$\Delta a_{\mathrm{min}}=4$ or 64~mas
during the next 5 or 20 years.

Direct interferometric images have been obtained
of Algol~A, Algol~B and Algol~C
\citep[e.g.][]{Zav10,Bar12}.
If Algol~B
and
Algol~C really are less massive
than our distant Algol~H candidate,
why did the earlier interferometric
imaging not reveal
the presence of this massive
candidate?
Firstly, this Algol~H candidate
is about 20 times further away from
the cEB than Algol~C, which means
that the area of interferometric
imaging should have been
about $20\times20=400$ larger.
Secondly, this Algol~H candidate could be
a long-period binary,
where both members are much less massive
and much dimmer
than a single  $2.50 m_{\odot}$ star
(Fig. \ref{Configurations}: Configurations 2 and 3).
One or two members of this long-period binary
could be an evolved object, like a white dwarf.
Thirdly, \citet{Zav10} and \citet{Bar12}
applied a three star model.
Algol~H contribution to their modelled total
flux would have remained constant, because 
its position did not change
during their observations
(Table \ref{TableDetection}).
We conclude that using an over 400 times
larger imaging area,
and a model of at least four stars,
may lead to the interferometric detection
of this distant Algol~H candidate.
The detection of the other
four less massive candidates
with this technique is much more challenging
(Table \ref{TableOrbits}: $i=90^{\mathrm{o}}$,
$0.23m_{\odot} \le m_3 \le 0.43 m_{\odot}$).
However,
the $\Delta a_{\mathrm{min}}$ and
$\Delta a_{\mathrm{max}}$ values
of these less massive candidates
show that their movements are easier to
detect even during shorter periods of observations
(Table \ref{TableDetection}).

\citet{Pow21} studied the sextuple-eclipsing
binary system
TIC 168789840 with the speckle interferometry
technique.
They could resolve this hierarchial system
of three eclipsing binaries.
Their estimate for the outer period in
this hierarchial system was
about 2000 years.
Algol is about twenty times closer to us than
TIC 168789840 $(d \approx 570{\mathrm{pc}})$.
The orbital period of 
our Algol~H candidate is about 200 years. 
Hence, it might be possible to
detect Algol~H with the speckle interferometry.

\subsection{Algol C detection  \label{SectAlgolC}}

DCM detects the weakest
$\PPone=680.^{\mathrm{d}}4\pm 0.^{\mathrm{d}}4$ signal
in all three samples:
all data,
\Adataone ~and \Bdataone.
This $\PPone$ signal is 44.8 times weaker
than the strongest $\PPsix$ signal
(Table \ref{TableOrbits}).
DCM detects this weakest $\PPone$ signal 
although it is buried under the interference
of five stronger
$\PPtwo$, $\PPthree$, $\PPfour$
$\PPfive$ and $\PPsix$ signals,
and a linear $p(t)$ trend.
The period of this $\PPone$ signal
differs only $1.4\sigma$
from the known orbital period 
$P_{\mathrm{orb}}=679.^{\mathrm{d}}85 \pm 0.^{\mathrm{d}}04$
  of Algol~C \citep{Zav10}.
This indicates that all other five 
detected stronger
signals are real periodicities,
but it does not irrefutably prove this idea.
Our O-C data contains 127 rounds of Algol~C
around Algol~AB, and this orbit is known
to be stable \citep{Zav10,Bar12,Jet13}.
Our lower limit for the mass of Algol~C 
(Table \ref{TableOrbits}: $i=90^{\mathrm{o}}$ and  $1.2 m_{\odot}$)
is smaller
than the interferometric estimates by
\citet[][$i=83.^{\mathrm{o}}7\pm 0.^{\mathrm{o}}1$
and $1.5\pm 0.1 m_{\odot}$]{Zav10}
and
\citet[][$i=83.^{\mathrm{o}}66\pm 0.^{\mathrm{o}}03$ and
$1.76\pm 0.15 m_{\odot}$]{Bar12}.
This indicates that not even DCM can retrieve
the full amplitude of this
weak Algol~C signal when it is buried under
five stronger signals and a linear trend.

\subsection{Stability
\label{SectStability}}

All detected signals are
  strictly periodic,
  because they are also detected in
  the 9.2 years shorter
  subsample \Adataone.
  Except for the $\PPtwo$ and $\PPfive$ signals,
  the other four signals are also detected
  in the fifty years shorter subsample
  \Bdataone.
  This apparent absence of these two
$\PPtwo$ and $\PPfive$
  signals
  in \Bdataone
  ~could be explained by the ``\PH'' and ``\PD''
  effects (Table \ref{TableThreeSamples}).
  However, strict periodicity alone does
  not prove that Algol's hierarchial system
  is stable.
   
  The perturbations
    of WOS can cause periodic
cEB orbital plane changes \citep[][Eq. 27]{Sod75}.
Such long-term orbital plane changes
with respect to the line of sight
may even stop the eclipses completely,
or at least reduce the depth of eclipses,
like in the case of AY~Mus
\citep{Sod74}.
However, the cEB orbital plane
is stable for
$\Psi=0^{\mathrm{o}}$ or $90^{\mathrm{o}}$,
where $\Psi$ is the angle between cEB and WOS
orbital planes.
This is the case for Algol~C,
the only currently known WOS of Algol
\citep[][$\Psi=90.^{\mathrm{o}}20\pm0.^{\mathrm{o}}32$]{Bar12}.
No changes have been observed in the eclipses
of Algol in modern times, and these events
were most probably also observed over
three thousand year ago \citep{Jet13}.
This is possible only if all WOSs have
$\Psi=0^{\mathrm{o}}$ or $90^{\mathrm{o}}$.
If the orbital planes of all WOS are co-planar,
then all WOSs must have $\Psi=90^{\mathrm{o}}$,
because this is the known case for Algol~C.
If all WOS orbit were not co-planar,
this would certainly
reduce the stability of this system,
and perhaps also weaken or
stop the observed eclipses.

The mass transfer from the less massive Algol B to
the more massive Algol A should increase the orbital
period \citep[][Eq. 5]{Kwe58}.
The numerous published mass transfer rate estimates
range from $10^{-13}m_{\odot}\mathrm{yr}^{-1}$
to $10^{-7}m_{\odot}\mathrm{yr}^{-1}$
\citep[][Sect. 4]{Jet13}.
However, no regular long-term
period increase has been observed since
\citet{Goo83} discovered Algol's periodicity.
All WOSs can also perturb the cEB by other physical mechanisms,
like the  Kozai effect \citep{Koz62},
or the combination of Kozai cycle and
tidal friction \citep{Fab07}.
Against this background,
our linear $K_3=1$ trend result for $p(t)$ 
is surprising (Sect. \ref{SectTrend}).
For 236 years,
Algol's orbital period has been constant
\begin{eqnarray}
  P_{\mathrm{orb}}=\left(
  {{1}\over {P_0}}
  -
  {{2M_1}\over{\Delta T}} \right)^{-1}=
  2.^{\mathrm{d}}86732870,
  \label{EqConstantP}
\end{eqnarray}
where $P_0\!=\!2.^{\mathrm{d}}86730431$ (Eq. \ref{EqEphe})
and $M_1\!=\!0.1278$ is $p(t)$ coefficient
for \M=3 model in Table \ref{Table1hjd14R}.
This causes the linear O-C change
of $0.^{\mathrm{d}}256$
in Fig. \ref{1hjd14R321Sgdet}
(upper panel: dotted line).
It also means that
LTTE effects alone can explain
all observed O-C changes.
No additional effects,
like the quadratic $K_3=2$ trend
caused by mass transfer,
are needed to explain these O-C data.

In the future,
  long-term integrations
  may confirm
  the dynamical stability of 
  this system.
  Currently,
  even the exact number of WOS
  candidates remains unknown,
  because three different 
  hierachial system
  diagrams can explain the
  detected WOS periods 
  (Fig. \ref{Configurations}:
  Configurations 1, 2 and 3).
  For any WOS period $p_3$, the correct
  $m_3, e_3, a_3, i_3, \omega_3$ and $\Psi_3$
  initial value
  combinations for the long-term
  integrations 
  are also unknown.
  Therefore,
  our O-C data can not give
  an unambiguous solution
  for this stability problem.
  Whether or not this system is stable,
  we can determine
  the $p_3$ periods that are observed today.

\subsection{Predictability
\label{SectPredictability}}

We admit that an unambiguous
  identification 
  of all individual signals
  from the interference
  sum of numerous signals
  is not always possible.
  One example is
  the $\PPfive$ and $\PPfour$ signal
  identification 
  in Sect. \ref{SectSignalTwo}.
  However, this whole identification
  problem is irrelevant
  from the predictability point of view.
  The sum of {\it identified} signals is equal
  to the sum of {\it unidentified} signals.
  Both alternatives give the same prediction.

  The linear and quadratic EB ephemerides
  can not predict the exact epochs of future
  eclipses \citep[e.g.][]{Kre01,Kim18}.
  For O-C changes caused by a third body,
these predictions also usually fail,
like
in \citet[][their Fig. 7]{Bou14},
\citet{Loh15} or
\citet[e.g.][their Fig. 1]{Son19}.
Different O-C subsets can give different periods,
but this does not mean that there is something
wrong with the period search
methods themselves, like DCM.
Our 9.2 years O-C prediction for Algol
is based on 
\Adataone
~(Fig. \ref{Prediction2}).
Strict periodicity
can explain why this
prediction succeeds.
Predictability is impossible
without strict periodicity.
This prediction would fail,
if even one of
our detected signals were not
strictly periodic,
or if the $K_3=1$ linear $p(t)$ trend were wrong.

Our next fifty years prediction is based on
\Bdataone.
Except for the first few years,
this long-term
prediction fails (Fig. \ref{Prediction3}).
The reason for this failure is simple.
The longest $172^{\mathrm{y}}$ period
detected from \Bdataone
~is not correct.
The short $\Delta T=$ 185 years time span
of this sample
prevents the detection of the correct 
signal period $219^{\mathrm{y}}$.
This correct signal
can be detected only
from all data and \Adataone.
  Together with the $0.^{\rm{d}}26$ trend $p(t)$,
  this highest $0.^{\mathrm{d}}29$ amplitude 
  dominating $219^{\mathrm{y}}$ signal
  determines
  all long-term O-C predictions.
  The insignificant
  long-term trend
  contribution of
  all other weaker signals
  is always less than
  $\pm 0.^{\mathrm{d}}03$,
  because the sum of their amplitudes 
  is $0.^{\mathrm{d}}06$. 
  Although our fifty years
  prediction for the O-C level fails
  (Fig. \ref{Prediction3}),
we get an excellent prediction for
the turning point epoch at
HJD~2450000
(Fig. \ref{AllSolutions}).

New O-C data  after October 2018
can already  be used to test our
prediction for the next ten years
(Fig. \ref{Prediction1}b). 
These predictions
should improve in the future, when 
all orbital period estimates become more accurate.
Predictability should ultimately prove
that all these signals are orbital periods.
At the moment, we can not prove this.
In the history of Astronomy,
the seasons of the year
posed a similar problem.
Their one year periodicity
was detected easily, but 
the reasons for it
were understood much later:
the orbit of the Earth around the Sun, 
and the tilted axis of Earth.
However,
it was possible to predict the seasons
without understanding their origin.
Our detected
periods of Algol are certainly there,
and for some reason or another
they can
be used to {\it predict}. 

\subsection{Look-elsewhere effect \label{SectLook}}

We test over thirty models having
free parameters between $\eta=6$ and 22
(Tables \ref{TableAllFisher}-\ref{Table3hjd14R}).
The total number of free parameters is even higher
when the model for the original data is
added to the model for the residuals.
For example, the best \M=3+6 model
for all data has $\eta=17+11=28$ free parameters
(Fig. \ref{Prediction1}).
Our search for the correct model
over a vast parameter space
increases the probability for finding
spurious apparently significant signals.
This is called the \Look ~
\citep[e.g.][]{Mil81,Bay20}.
There are statistical methods that can
account for the \Look,
and give {\it direct} significance
estimates $S$ for the periods of models
having different degrees of freedom 
\citep[e.g.][their Eq. 3.12]{Bay20}.

DCM applies Fisher-test to compare
the significance of all pairs of simple and complex models.
Fisher-test identifies the best model among all tested
models (Eqs. \ref{EqFR} and \ref{EqFisher}). 
This approach does not
account for the \Look,
because it gives no {\it direct}
significance estimate $S$ for the periodicities of
this best model.
Nevertheless, we can present several arguments
indicating that the \Look
~has no significant impact on our results.

\begin{enumerate}
  
\item  We apply the robust Fisher-test to
compare any complex model having more signals
than any simple model.
We use
the pre-assigned
significance level $\gamma_F=0.001$ to reject the simple model
(Eq. \ref{EqFisher}).
This prevents over-fitting, because
the probability that this best model selection fails
is always smaller than one out of one thousand.
In many cases,
the extreme \REJ ~critical levels confirm
that the complex model is absolutely certainly
better than the simple model.
This confirms that the data contain more signals
than those present in the simple model.
Our {\it indirect} $Q_F$ significance
estimates confirm the presence of additional complex model
periodicities, but
they do not give us {\it direct} $S$ significance
estimates for these periodicities.
Regardless of the \Look,
Fisher-test can confirm that the five
signal \M=3+6 ~model
is the best model for all data.

\item The $z$ periodogram values of close tested frequencies
correlate and display no sudden jumps
(see Sect. \ref{SectMethod}: Caveat 2).
At some tested frequency grid density level,
this means that
the detected period values no longer depend on 
the number of tested periods (\paperone: $n_L$ and $n_S$).
These unambiguous best
period values are obtained from linear models.
Increasing the number of tested periods does
not change the values of these detected periods.
Hence, the tested frequency grid density is
not a trial factor effect (\Look)
that can change the five period
values of our best \M=3+6 model.

\item For all O-C data, we use Fisher-test to
  compare constant, linear, quadratic and cubic
  $p(t)$ trends for one, two and three signal models
  (Table \ref{TableAllFisher}).
  The linear $K_3=1$ trend is the best one. 
  This means that if the O-C data had been computed
  with the period $2.^{\mathrm{d}}86732870$ (Eq. \ref{EqConstantP}),
  the best trend would have been the constant $K_3=0$ trend.
  After exploring numerous trend and signal combination
  alternatives in the vast free parameter space,
  we arrive at this simplest alternative:
  no trend at all in the O-C data!
  Although the \Look ~is certainly present,
  DCM detects this simplest trend alternative
  for our five signal \M=3+6 model.
  
\item Our \M=3+6 model prediction
    is excellent (Fig. \ref{Prediction2}).
  This indicates that the \Look,
  or any other spurious effect,
  does not mislead DCM periodicity detection.

\end{enumerate}

\subsection{Uncertainties}

The time span of our data is
``only''  236 years.
Our biggest uncertainty is
therefore the longest detected
219 years periodicity.
It has been claimed that
the Discrete Fourier Transform
can sometimes detect clear signal
periods slightly longer than the
$\Delta T$ time span of data,
``but with
poor resolution'' \citep{Hor86}. 
The detection of periods close to
$\Delta T$ depends strongly on the
signal-to-noise ratio of the data. 
Such detections may 
not always succeed in our case,
because we detect the 172 years period
from the shortest sample of 185 years.
This period is shorter than 
time span of this particular sample.
We do not detect this  ``old'' 172 years period
from the longer samples of 226
and 236 years,
but we do detect the ``new'' 219 years period.
New additional O-C data may, or may not, confirm that
this 219 years period of ours is correct.

The direct discovery of Algol~H would solve the
above problem for good.
\citet{Egg48} analysed Algol's O-C data.
He arrived at an orbital period of 188.4 years 
for this hypothetical distant companion.
\citet{Irw52} estimated its orbital elements.
We argue that this distant 
Algol~H candidate may be currently found
about 1.6 arc seconds away
from the cEB, the eclipsing pair Algol A and Algol~B.

As for other uncertainties,
we can not determine the exact number of
stars in this hierarchial system,
but this does not prevent us from presenting an
excellent 9.2 year prediction based on the first
226 years of O-C data (Fig. \ref{Prediction2}).
We admit that our longer
fifty years prediction fails,
because our 172 years period detected in
the shortest 185 year sample is wrong
(Fig. \ref{Prediction3}).
However, our turning point
in this same prediction would
explain the four years gap in the published O-C data
around the year 1995 (Fig. \ref{AllSolutions}).
It will be interesting to see how
well we can predict the future
O-C data after October 2018 (Fig. \ref{Prediction1}b).

\section{Conclusions
  \label{SectConclusions}}

The ephemerides of
  eclipsing binaries
  can be improved
  by removing linear or
  quadratic trends
  from the observed (O)
  minus computed (C)
  eclipse epochs
  \citep[e.g.][]{Kre01,Kim18}.
  However, even such
  improved ephemerides
  can not predict
  the exact epochs of
  future eclipses.
  The light-time travel effect
  of a third body
  causes
  strictly periodic predictable
  O-C changes \citep{Irw52}.
  The typical  third and fourth body
  detection rates 
  from O-C data are low,
  only
  992/80~000 and 4/80~000, respectively
  \citep{Haj19}.
   Eclipse epoch predictions based
  on linear or quadratic trends,
  and light-time travel effects,
  usually fail
  because aperiodic trends 
  mislead the detection of periodic signals
  \citep[e.g.][]{Bou14,Loh15,Son19}.

  Considering this general background,
  it is unprecedented that
  our new Discrete Chi-square Method
  can detect 
  five 
  strictly periodic signals
  from 236 years of Algol's O-C data
  (Fig. \ref{Prediction1}a).
  These tentative companion candidate
  orbital periods are
  between 1.863 and 219.0 years.
  One of these periods
  is definitely not a surprise,
  because our 
  $680.4\pm 0.4$ days
  period estimate
  for this weakest detected
  signal differs only $1.4\sigma$ from 
  the well-known 
  $679.85 \pm 0.04$ days
  orbital period of Algol~C.
  From our O-C data alone,
  we can not determine
  the exact number of
  companions in Algol's
  hierarchial system,
  or the stability of this system. 

  From the shorter 226.2 years subsample,
  we detect these same five above mentioned
  strictly periodic signals.
  They give an excellent prediction for the
  last 9.2 years of our O-C data
  (Fig. \ref{Prediction2}b).
  Although it is
  impossible to detect the
  longest 219 year period from our
  shortest analysed subsample of
  185 years,
  we can still predict
  the O-C data
  turning point epoch in the year 1995
  (Fig. \ref{Prediction3}b).
  This unexpected turning point
  event
  could
  explain the odd publication gap 
  in the otherwise continuous
  modern 
  O-C data of Algol.
  
  We detect the linear O-C trend,
  which
  confirms that
  Algol's orbital period
  has not changed
  since it was discovered
  by \citet{Goo83}.
  The orbital planes of
  Algol~C and the new
  other wide
  orbit star candidates are
  probably co-planar,
  because
  Algol's eclipses were
  observed already in
  Ancient Egypt
  \citep{Jet13,Jet15,Por18}.

  In the bigger picture,
  the predictions for 
  complex {\it non-linear}
  models rarely succeed.
  We give a prediction
  for the next decade of Algol's
  O-C changes
  after October 18th, 2018
  (Fig. \ref{Prediction1}b).
  These future O-C changes
  may prove that the abstract
  Discrete Chi-square Method
  approach works for complex
  {\it non-linear
  models}, and that
Algol's data merely allowed
us to check this.

\acknowledgments
We thank Dr. Chun-Hwey Kim for
sending us the TI\-DAK database O-C data of Algol. 
We also thank Dr. Sara Beck,
Dr. Lindsay Ward,
Dr. Gerard Samolyk, Dr. Stella Kafka and
Dr. Nancy Morrison, who helped us in finding
the O-C data of Algol from the
Lichten\-knecker Database of the BAV.
This work has made use of NASA's 
Astrophysics Data System (ADS) services
and the data from the European Space Agency (ESA)
mission Gaia.
We thank Linux Specialist Markus Minkkinen
and Dr. Sebastian Porceddu from
Center for Information Technology
(University of Helsinki).
Their computer support
during the Covid-19 pandemia crisis
enabled us 
to complete this work.


\begin{thebibliography}{}
\expandafter\ifx\csname natexlab\endcsname\relax\def\natexlab#1{#1}\fi

\bibitem[{Allen(2004)}]{All04}
Allen, M. 2004, Understanding Regression Analysis (Springer US), 113--117

\bibitem[{{Applegate}(1992)}]{App92}
{Applegate}, J.~H. 1992, \apj, 385, 621

\bibitem[{{Baron} {et~al.}(2012){Baron}, {Monnier}, {Pedretti}, {Zhao},
  {Schaefer}, {Parks}, {Che}, {Thureau}, {ten Brummelaar}, {McAlister},
  {Ridgway}, {Farrington}, {Sturmann}, {Sturmann}, \& {Turner}}]{Bar12}
{Baron}, F., {Monnier}, J.~D., {Pedretti}, E., {et~al.} 2012, \apj, 752, 20

\bibitem[{{Bayer} \& {Seljak}(2020)}]{Bay20}
{Bayer}, A.~E., \& {Seljak}, U. 2020, \jcap, 2020, 009

\bibitem[{{Borkovits} {et~al.}(2005){Borkovits}, {Forg{\'a}cs-Dajka}, \&
  {Reg{\'a}ly}}]{Bor05}
{Borkovits}, T., {Forg{\'a}cs-Dajka}, E., \& {Reg{\'a}ly}, Z. 2005, in
  Astronomical Society of the Pacific Conference Series, Vol. 333, Tidal
  Evolution and Oscillations in Binary Stars, ed. A.~{Claret},
  A.~{Gim{\'e}nez}, \& J.~P. {Zahn}, 128

\bibitem[{{Bours} {et~al.}(2014){Bours}, {Marsh}, {Breedt}, {Copperwheat},
  {Dhillon}, {Leckngam}, {Littlefair}, {Parsons}, \& {Prasit}}]{Bou14}
{Bours}, M.~C.~P., {Marsh}, T.~R., {Breedt}, E., {et~al.} 2014, \mnras, 445,
  1924

\bibitem[{{Curtiss}(1908)}]{Cur08}
{Curtiss}, R.~H. 1908, \apj, 28, 150

\bibitem[{Draper \& Smith(1998)}]{Dra98}
Draper, N.~R., \& Smith, H. 1998, Applied Regression Analysis (John Wiley {\&}
  Sons, Inc.), doi:10.1002/9781118625590

\bibitem[{{Efron} \& {Tibshirani}(1986)}]{Efr86}
{Efron}, B., \& {Tibshirani}, R. 1986, Statistical Science, 1, 54

\bibitem[{Efron \& Tibshirani(1994)}]{Efr94}
Efron, B., \& Tibshirani, R. 1994, An Introduction to the Bootstrap, Chapman \&
  Hall/CRC Monographs on Statistics \& Applied Probability (Taylor \& Francis)

\bibitem[{{Eggen}(1948)}]{Egg48}
{Eggen}, O.~J. 1948, \apj, 108, 1

\bibitem[{{Esmer} {et~al.}(2021){Esmer}, {Ba{\c{s}}t{\"u}rk}, {Hinse}, {Selam},
  \& {Correia}}]{Ekr21}
{Esmer}, E.~M., {Ba{\c{s}}t{\"u}rk}, {\"O}., {Hinse}, T.~C., {Selam}, S.~O., \&
  {Correia}, A. C.~M. 2021, arXiv e-prints, arXiv:2103.00062

\bibitem[{{Fabrycky} \& {Tremaine}(2007)}]{Fab07}
{Fabrycky}, D., \& {Tremaine}, S. 2007, \apj, 669, 1298

\bibitem[{{Frieboes-Conde} {et~al.}(1970){Frieboes-Conde}, {Herczeg}, \&
  {H{\o}g}}]{Fri70}
{Frieboes-Conde}, H., {Herczeg}, T., \& {H{\o}g}, E. 1970, \aap, 4, 78

\bibitem[{{Goodricke}(1783)}]{Goo83}
{Goodricke}, J. 1783, Philosophical Transactions of the Royal Society of London
  Series I, 73, 474

\bibitem[{{Hajdu} {et~al.}(2019){Hajdu}, {Borkovits}, {Forg{\'a}cs-Dajka},
  {Sztakovics}, {Marschalk{\'o}}, \& {Kutrov{\'a}tz}}]{Haj19}
{Hajdu}, T., {Borkovits}, T., {Forg{\'a}cs-Dajka}, E., {et~al.} 2019, \mnras,
  485, 2562

\bibitem[{{Hoffman} {et~al.}(2006){Hoffman}, {Harrison}, {McNamara},
  {Vestrand}, {Holtzman}, \& {Barker}}]{Hof06}
{Hoffman}, D.~I., {Harrison}, T.~E., {McNamara}, B.~J., {et~al.} 2006, \aj,
  132, 2260

\bibitem[{{Horne} \& {Baliunas}(1986)}]{Hor86}
{Horne}, J.~H., \& {Baliunas}, S.~L. 1986, \apj, 302, 757

\bibitem[{{Irwin}(1952)}]{Irw52}
{Irwin}, J.~B. 1952, \apj, 116, 211

\bibitem[{{Jetsu}(2020)}]{Jet20}
{Jetsu}, L. 2020, The Open Journal of Astrophysics, 3, 4

\bibitem[{{Jetsu} \& {Pelt}(1999)}]{Jet99}
{Jetsu}, L., \& {Pelt}, J. 1999, \aaps, 139, 629

\bibitem[{{Jetsu} \& {Pelt}(2000)}]{Jet00A}
---. 2000, \aap, 353, 409

\bibitem[{{Jetsu} \& {Por\-ced\-du}(2015)}]{Jet15}
{Jetsu}, L., \& {Por\-ced\-du}, S. 2015, PLoS ONE, 10(12), e0144140

\bibitem[{{Jetsu} {et~al.}(2013){Jetsu}, {Porceddu}, {Lyytinen}, {Kajatkari},
  {Lehtinen}, {Markkanen}, \& {Toivari-Viitala}}]{Jet13}
{Jetsu}, L., {Porceddu}, S., {Lyytinen}, J., {et~al.} 2013, \apj, 773, 1

\bibitem[{{Kim} {et~al.}(2018){Kim}, {Kreiner}, {Zakrzewski}, {Og{\l}oza},
  {Kim}, \& {Jeong}}]{Kim18}
{Kim}, C.~H., {Kreiner}, J.~M., {Zakrzewski}, B., {et~al.} 2018, \apjs, 235, 41

\bibitem[{{Kiseleva} {et~al.}(1998){Kiseleva}, {Eggleton}, \&
  {Mikkola}}]{Kis98}
{Kiseleva}, L.~G., {Eggleton}, P.~P., \& {Mikkola}, S. 1998, \mnras, 300, 292

\bibitem[{{Kozai}(1962)}]{Koz62}
{Kozai}, Y. 1962, \aj, 67, 591

\bibitem[{{Kreiner} {et~al.}(2001){Kreiner}, {Kim}, \& {Nha}}]{Kre01}
{Kreiner}, J.~M., {Kim}, C.-H., \& {Nha}, I.-S. 2001, {An Atlas of O-C Diagrams
  of Eclipsing Binary Stars}

\bibitem[{{Kwee}(1958)}]{Kwe58}
{Kwee}, K.~K. 1958, \bain, 14, 131

\bibitem[{{Lehtinen} {et~al.}(2011){Lehtinen}, {Jetsu}, {Hackman}, {Kajatkari},
  \& {Henry}}]{Leh11}
{Lehtinen}, J., {Jetsu}, L., {Hackman}, T., {Kajatkari}, P., \& {Henry}, G.~W.
  2011, \aap, 527, A136

\bibitem[{{Li} {et~al.}(2018){Li}, {Rattenbury}, {Bond}, {Sumi}, {Bennett},
  {Koshimoto}, {Abe}, {Asakura}, {Barry}, {Bhattacharya}, {Donachie}, {Evans},
  {Fukui}, {Hirao}, {Itow}, {Masuda}, {Matsubara}, {Muraki}, {Nagakane},
  {Ohnishi}, {Saito}, {Sharan}, {Sullivan}, {Suzuki}, {Tristram}, \&
  {Yonehara}}]{Li18}
{Li}, M.~C.~A., {Rattenbury}, N.~J., {Bond}, I.~A., {et~al.} 2018, \mnras, 480,
  4557

\bibitem[{{Lohr} {et~al.}(2015){Lohr}, {Norton}, {Payne}, {West}, \&
  {Wheatley}}]{Loh15}
{Lohr}, M.~E., {Norton}, A.~J., {Payne}, S.~G., {West}, R.~G., \& {Wheatley},
  P.~J. 2015, \aap, 578, A136

\bibitem[{{Manzoori}(2016)}]{Man16}
{Manzoori}, D. 2016, Astronomy Letters, 42, 329

\bibitem[{{Miller}(1981)}]{Mil81}
{Miller}, R. 1981, Simultaneous Statistical Inference (Springer New York)

\bibitem[{{Mueller}(1995)}]{Mul95}
{Mueller}, M. 1995, Acta Phys. Pol., 88A, S49

\bibitem[{Porceddu {et~al.}(2018)Porceddu, Jetsu, Markkanen, Lyytinen,
  Kajatkari, Lehtinen, \& Toivari-Viitala}]{Por18}
Porceddu, S., Jetsu, L., Markkanen, T., {et~al.} 2018, Open Astronomy, 27, 232

\bibitem[{Porceddu {et~al.}(2008)Porceddu, Jetsu, Markkanen, \&
  Toivari-Viitala}]{Por08}
Porceddu, S., Jetsu, L., Markkanen, T., \& Toivari-Viitala, J. 2008, Cambridge
  Archaeological Journal, 18, 327

\bibitem[{{Powell} {et~al.}(2021){Powell}, {Kostov}, {Rappaport}, {Borkovits},
  {Zasche}, {Tokovinin}, {Kruse}, {Latham}, {Montet}, {Jensen}, {Jayaraman},
  {Collins}, {Masek}, {Hellier}, {Evans}, {Tan}, {Schlieder}, {Torres},
  {Smale}, {Friedman}, {Barclay}, {Gagliano}, {Quintana}, {Jacobs}, {Gilbert},
  {Kristiansen}, {Colon}, {LaCourse}, {Olmschenk}, {Omohundro}, {Schnittman},
  {Schwengeler}, {Barry}, {Terentev}, {Boyd}, {Schmitt}, {Quinn}, {Vanderburg},
  {Palle}, {Armstrong}, {Ricker}, {Vanderspek}, {Seager}, {Winn}, {Jenkins},
  {Caldwell}, {Wohler}, {Shiao}, {Burke}, {Daylan}, \& {Villasenor}}]{Pow21}
{Powell}, B.~P., {Kostov}, V.~B., {Rappaport}, S.~A., {et~al.} 2021, arXiv
  e-prints, arXiv:2101.03433

\bibitem[{{Reinhold} {et~al.}(2013){Reinhold}, {Reiners}, \& {Basri}}]{Rei13}
{Reinhold}, T., {Reiners}, A., \& {Basri}, G. 2013, \aap, 560, A4

\bibitem[{{Roy}(2005)}]{Roy05}
{Roy}, A.~E. 2005, {Orbital motion}

\bibitem[{{Soderhjelm}(1974)}]{Sod74}
{Soderhjelm}, S. 1974, Information Bulletin on Variable Stars, 885, 1

\bibitem[{{Soderhjelm}(1975)}]{Sod75}
---. 1975, \aap, 42, 229

\bibitem[{{Song} {et~al.}(2019){Song}, {Mai}, {Mutel}, {Pulley}, {Faillace}, \&
  {Watkins}}]{Son19}
{Song}, S., {Mai}, X., {Mutel}, R.~L., {et~al.} 2019, \aj, 157, 184

\bibitem[{{Tokovinin}(2021)}]{Tok21}
{Tokovinin}, A. 2021, \aj, 161, 144

\bibitem[{{Torra} {et~al.}(2020){Torra}, {Casta{\~n}eda}, {Fabricius},
  {Lindegren}, {Clotet}, {Gonz{\'a}lez-Vidal}, {Bartolom{\'e}}, {Bastian},
  {Bernet}, {Biermann}, {Garralda}, {G{\'u}rpide}, {Lammers}, {Portell}, \&
  {Torra}}]{Tor20}
{Torra}, F., {Casta{\~n}eda}, J., {Fabricius}, C., {et~al.} 2020, arXiv
  e-prints, arXiv:2012.06420

\bibitem[{{van Leeuwen}(2007)}]{Van07}
{van Leeuwen}, F. 2007, \aap, 474, 653

\bibitem[{{Wilson}(1953)}]{Wil53}
{Wilson}, R.~E. 1953, Carnegie Institute Washington D.C. Publication, 0

\bibitem[{{Wolf} {et~al.}(1999){Wolf}, {Diethelm}, \&
  {{\v{S}}arounov{\'a}}}]{Wol99}
{Wolf}, M., {Diethelm}, R., \& {{\v{S}}arounov{\'a}}, L. 1999, \aap, 345, 553

\bibitem[{{Zasche} \& {Uhla{\v{r}}}(2013)}]{Zas13}
{Zasche}, P., \& {Uhla{\v{r}}}, R. 2013, \mnras, 429, 3472

\bibitem[{{Zasche} \& {Wolf}(2007)}]{Zas07}
{Zasche}, P., \& {Wolf}, M. 2007, Astronomische Nachrichten, 328, 928

\bibitem[{{Zavala} {et~al.}(2010){Zavala}, {Hummel}, {Boboltz}, {Ojha},
  {Shaffer}, {Tycner}, {Richards}, \& {Hutter}}]{Zav10}
{Zavala}, R.~T., {Hummel}, C.~A., {Boboltz}, D.~A., {et~al.} 2010, \apjl, 715,
  L44

\end{thebibliography}

\clearpage
\appendix

\setcounter{table}{0}
\setcounter{figure}{0}
\renewcommand{\thetable}{A\arabic{table}}
\renewcommand{\thefigure}{A\arabic{figure}}

\begin{deluxetable*}{lccccccc}
   \tablecaption{Cases I, II and III.
    Cols 1-5 give O-C curve period $(p)$,
    peak to peak amplitude $(A)$,
    pericentre epoch $(t_p)$,
    eccentricity $(e)$ and
    periastron longitude $(\omega)$
    (Eqs. \ref{EqIrwOne}-\ref{EqIrwFive}). 
    Cols 6-7 give connected figures and tables.
    \label{TableCases}}
\addtolength{\tabcolsep}{-0.10cm}
\tablewidth{700pt}
\tabletypesize{\scriptsize}
\tablehead{
  & \colhead{Col 1}
  & \colhead{Col 2}
  & \colhead{Col 3}
  & \colhead{Col 4}
  & \colhead{Col 5}
  & \colhead{Col 6}
  & \colhead{Col 7}}
  \startdata
  & $\Dd$ & $\Dd$ & ${\mathrm{[HJD]}}$ & Dimensionless & $[\oo]$ & Fig. & Table \\
\hline
  Case I & $p=45976$    & $A=0.0994$  &  $t_p=2373019.94$ & $e=0.05, 0.10, 0.20, 0.30$ or 0.40 & $\omega=0, 45, 90, 135, 180, 225, 270$ or 315
 & \ref{FigOCone}    & \ref{TableOCone}, \ref{TableOCtwo} \\
  \hline
  Case II & $p_1=12295$ & $A_1=0.0174$ & $t_{p,1}=2375140.04$ & $e_1=0$                          & $\omega_1=0$
  & \ref{FigOCtwo}      &                     -                        \\
          & $p_2=46159$ & $A_2=0.1024$ & $t_{p,2}=2372653.76$ & $e_2=0$                          & $\omega_2=0$
  &                    &                                               \\
  \hline
  Case III& $p_1=12304$ & $A_1=0.0187$ & $t_{p,1}=2374760.75$ & $e_1=0$                          & $\omega_1=0$
  & \ref{FigOCthree}   &                     -                        \\
          & $p_2=25274$ & $A_2=0.020$  & $t_{p,2}=2380427.13$ & $e_2=0$                          & $\omega_2=0$
  &                     &                                              \\
\enddata
\addtolength{\tabcolsep}{+0.10cm}
\end{deluxetable*}

\begin{figure*}[ht!] 
\plotone{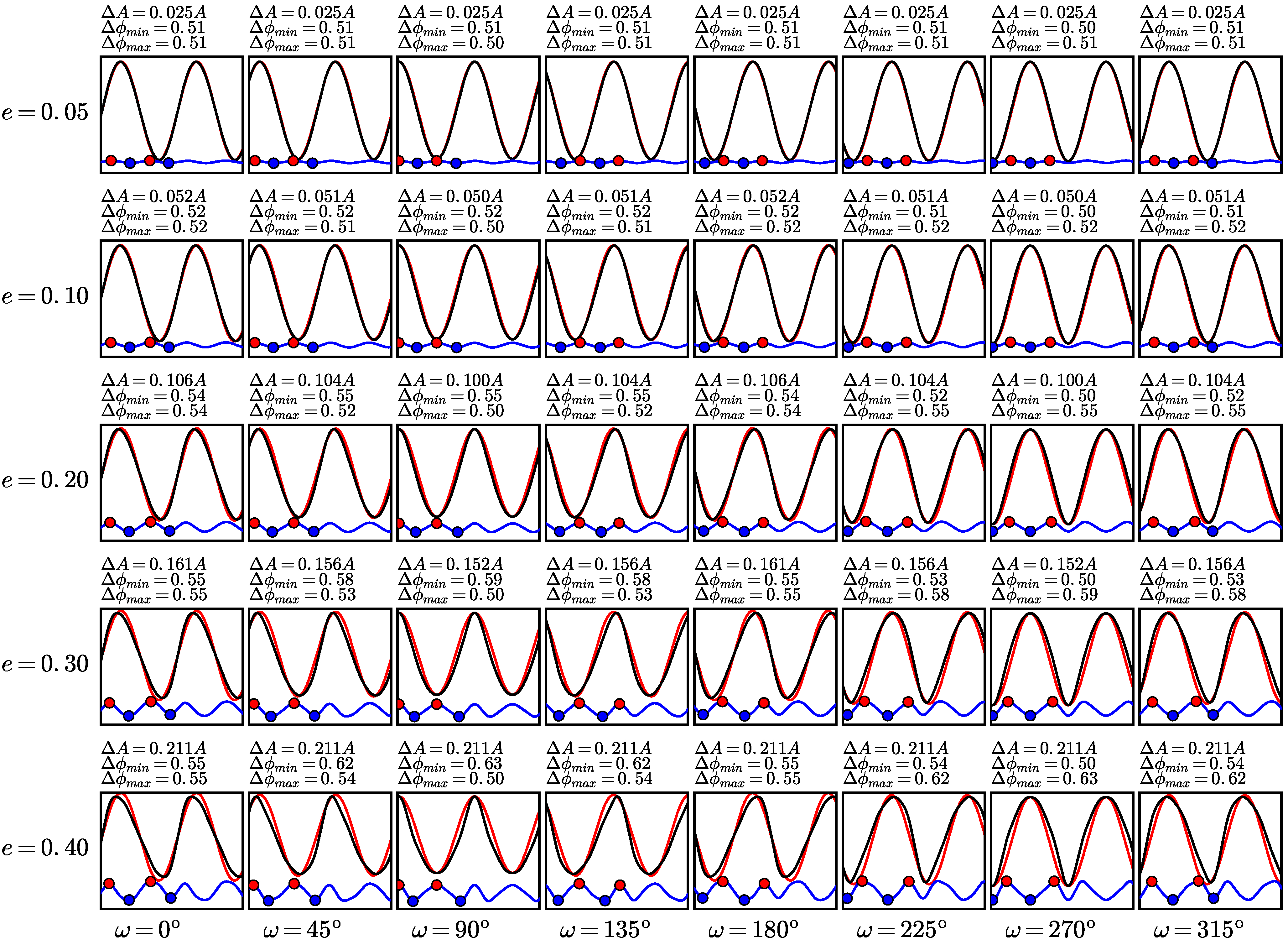}
\caption{Case I. 
  Black lines show forty eccentric orbit ${\mathrm{O-C}}_{e>0}$ curves
  (Eq. \ref{EqEcce}) having parameters 
  specified in Table \ref{TableCases} (Case I).
  Red lines show respective 
  circular orbit  ${\mathrm{O-C}}_{e=0}$ curves (Eq. \ref{EqCirc}). 
  Blue lines denote difference curves
  ${\mathrm{(O-C)}}_{\mathrm{diff}}$
  (Eq. \ref{EqDiff}). 
  Parameters
  $\Delta A$,
  $\Delta \phi_{\mathrm{min}}$
    and
    $\Delta \phi_{\mathrm{max}}$ (Eqs. \ref{EqDeltaAmp}-\ref{EqDeltaMax})
    are given above each panel.
    Blue and red circles 
    denote first two  ${\mathrm{(O-C)}}_{\mathrm{diff}}$
    curve minima and maxima.
    To save space, we show no quantitative xy-axis label values,
    and we offset ${\mathrm{(O-C)}}_{\mathrm{diff}}$ curve below
    ${\mathrm{O-C}}_{e>0}$ and  ${\mathrm{O-C}}_{e=0}$ curves.
    Units are $[t]={\mathrm{HJD}}$ (x-axis)
      and $[{\mathrm{O\!-\!C}}]=$ d (y-axis).
  \label{FigOCone}}
\end{figure*}

\section{DCM analysis of simulated O-C data
\label{SectSimulations}} 

If the third body orbit is circular $(e=0)$,
the suitable DCM model order is $K_2=1$,
because the O-C curve is a pure sinusoid
(Eq. \ref{EqIrwOne}: $e=0$).
If the third body orbit is not circular $(e>0)$,
the O-C curve is not a pure sinusoid.
In this case,
the suitable DCM model order for these eccentric orbits
is $K_2=2$ \citep[][]{Hof06}.
Our notations for circular $(e=0)$ and eccentric $(e>0)$ orbit O-C curves are
\begin{eqnarray}
  ({\mathrm{O-C}})_{e=0} & & 
\label{EqCirc} \\
  ({\mathrm{O-C}})_{e>0}. & &
\label{EqEcce}
\end{eqnarray}
These $({\mathrm{O-C}})_{e>0}$ and $({\mathrm{O-C}})_{e=0}$ curves
have the same peak to peak amplitude $A$
for any $p$, $t_p$, $e$ and $\omega$ combination
(Eqs. \ref{EqIrwOne} - \ref{EqIrwFive}).
Our notation for their difference curve is 
\begin{eqnarray}
  ({\mathrm{O-C}})_{\mathrm{diff}}
  = ({\mathrm{O-C}})_{e>0}-({\mathrm{O-C}})_{e=0}
\label{EqDiff}
\end{eqnarray}
having a peak to peak amplitude $A_{\mathrm{diff}}$.
The amplitude ratio is
\begin{eqnarray}
\Delta A=A_{\mathrm{diff}}/A.
\label{EqDeltaAmp}
\end{eqnarray}
We also determine the phase differences 
\begin{eqnarray}
  \Delta \phi_{\mathrm{min}} &=&(t_{\mathrm{2nd.min}}-t_{\mathrm{1st.min}})/p
  \label{EqDeltaMin} \\
\Delta \phi_{\mathrm{max}} &=&(t_{\mathrm{2nd.max}}-t_{\mathrm{1st.max}})/p,
 \label{EqDeltaMax}
\end{eqnarray}
of two first
 minimum $(t_{\mathrm{1st.min}},$$t_{\mathrm{2nd.min}})$
 and maximum $(t_{\mathrm{1st.max}},$$t_{\mathrm{2nd.max}})$ epochs
 of $({\mathrm{O-C}})_{\mathrm{diff}}$ curve.

We simulate three cases of artificial O-C data
(Table \ref{TableCases}: Cases I, II and III).
The simulated O-C values are computed
for the real data time points $t_i$
from Table \ref{TableData} $(n=2224)$.
We add $0.^{\mathrm{d}}005$ Gaussian random errors
to these simulated O-C values.
DCM period search for these simulated O-C data
is performed between 8000 and 80000 days.
We use the same period interval also
in our DCM analysis of real data
(Sects. \ref{SectAlldata}-\ref{SectExperiments})

\begin{deluxetable*}{cccccccccc}
  \tablecaption{Case I: Correct model results. 
    Simulated $(O-C)_{\mathrm{e>0}}$ data
    signal period is 
    $p=45976^{\mathrm{d}}$.
    Signal peak to peak amplitude is $A=0.^{\mathrm{d}}0994$.
    For different $e$ and $\omega$ combinations,
    one signal DCM \RModel{1,2,0} search detects
    periods $P_1$ and peak to peak amplitudes $A_1$.
    Abbreviation ``\Dan'' denotes
    \PD ~effect
    cases, where
    spurious period $P_1\sim2p$ may be detected, if 
    tested frequency grid is too sparse. 
 \label{TableOCone}}
\tablewidth{700pt}
\tabletypesize{\scriptsize}
\tablehead{
  & & \colhead{$\omega=0\oo$}
  &   \colhead{$\omega=45\oo$}
  & \colhead{$\omega=90\oo$}
  & \colhead{$\omega=135\oo$}
  & \colhead{$\omega=180\oo$}
  & \colhead{$\omega=225\oo$}
  & \colhead{$\omega=270\oo$}
  &  \colhead{$\omega=315\oo$} } 
  \startdata
  $e=0.05$ & $P_1$ &45978 \Dan&45933 \Dan&45996 \Dan&45968 \Dan &46082 \Dan&45946 \Dan&45976 \Dan&46001 \Dan \\
           & $A_1$ & 0.0995  & 0.0995  & 0.0996  & 0.0990   & 0.0990  & 0.0998  & 0.0993  & 0.0995   \\
           & $P_1/p$ &1.00   &1.00     & 1.00    & 1.00     & 1.00    &1.00     & 1.00    & 1.00     \\
           & $A_1/A$ &1.00   &1.00     & 1.00    & 1.00     & 1.00    &1.00     & 1.00    & 1.00     \\
  \hline
  $e=0.10$ & $P_1$ &46016 \Dan&45952 \Dan&45874 \Dan&45920 \Dan&45997 \Dan&45921 \Dan&45943 \Dan&45981 \Dan \\
           & $A_1$ & 0.0996  & 0.0995  & 0.0990  & 0.0986  & 0.0990  & 0.0994  & 0.0990  & 0.0991   \\
           & $P_1/p$ &1.00   &1.00     & 1.00    & 1.00    & 1.00    & 1.00    & 1.00    & 1.00     \\
           & $A_1/A$ &1.00   &1.00     & 1.00    & 0.99    & 1.00    & 1.00    & 1.00    & 1.00    \\
  \hline
  $e=0.20$ & $P_1$ & 46004   & 46057   & 46058   & 45990   & 46020   &46038    &46093    &45970     \\
           & $A_1$ & 0.0990  & 0.0984  & 0.0977  & 0.0985  & 0.0987  & 0.0983  & 0.0976  & 0.0984   \\
           & $P_1/p$ &1.00   & 1.00    & 1.00    & 1.00    & 1.00    & 1.00    & 1.00    & 1.00     \\
           & $A_1/A$ &1.00   & 0.99    & 0.98    & 0.99    & 0.99    & 0.99    & 0.98    & 0.99     \\
  \hline
  $e=0.30$ & $P_1$ & 46074   & 46056   & 46004   & 45937   & 45980   &46053    &45959    &46001    \\
           & $A_1$ & 0.0977  & 0.0964  & 0.0964  & 0.0975  & 0.0978  & 0.0968  & 0.0960  & 0.0981  \\
           & $P_1/p$ &1.00   & 1.00    & 1.00    & 1.00    & 1.00    & 1.00    & 1.00    & 1.00    \\
           & $A_1/A$ &0.98   & 0.97    & 0.97    & 0.98    & 0.98    & 0.97    & 0.96    & 0.99    \\
  \hline 
  $e=0.40$ & $P_1$ & 46084   & 46251   & 45930   & 45934   & 46079   & 46292   &45929    &45882    \\
           & $A_1$ & 0.0954  & 0.0939  & 0.0930  & 0.0960  & 0.0948  & 0.0932  &0.0935   & 0.0960  \\
           & $P_1/p$ &1.00   & 1.00    & 1.00    & 1.00    & 1.00    & 1.01    &1.00     & 1.00     \\
           & $A_1/A$ &0.96   & 0.94    & 0.94    & 0.96    & 0.95    & 0.94    &0.94     & 0.96    \\
\enddata
\end{deluxetable*}

\subsection{Case I: Simulated eccentric orbit data
 \label{SectEccentricCaseI}}

In this section,
we use $p$, $A$, $t_p$, $e$
and $\omega$ values of Case I
(Table \ref{TableCases}).
Our Fig. \ref{FigOCone} shows all forty
$({\mathrm{O-C}})_{e>0}$ and
$({\mathrm{O-C}})_{e=0}$ curve pairs,
as well as their $
({\mathrm{O-C}})_{\mathrm{diff}}$
difference curves.
We study only cases
$e \le 0.4$, because
our $\nu(t)$
Fourier expansion (Eq. \ref{EqIrwFour})
does not give
the exact quantitative
$\nu(t)$ values
for higher eccentricities.
However, our $\nu(t)$
estimates are sufficient
for illustrating how 
eccentric orbit $({\mathrm{O-C}})_{e>0}$
curves deviate from purely sinusoidal
circular orbit $({\mathrm{O-C}})_{e=0}$
curves.

The $\Delta A$,
$\Delta \phi_{\mathrm{min}}$
and
$\Delta \phi_{\mathrm{max}}$ values
for forty eccentric $({\mathrm{O-C}})_{e>0}$ curves
are given above each panel of  Fig. \ref{FigOCone}.
When eccentricity $e$ increases,
the amplitude ratio $\Delta A$ increases.
At the same time,
the $({\mathrm{O-C}})_{\mathrm{diff}}$ curve
symmetry decreases, because 
$\Delta \phi_{\mathrm{min}}$
and
$\Delta \phi_{\mathrm{max}}$
values deviate more from 0.5.
Both of these effects confirm that when eccentricity increases,
the $({\mathrm{O-C}})_{e>0}$ curve
deviates more from the pure  $({\mathrm{O-C}})_{e=0}$ sinusoid.
One symmetry remains:
adding $180^{\mathrm{o}}$ to
$\omega$ reverses the $\Delta \phi_{\mathrm{min}}$
and $\Delta \phi_{\mathrm{max}}$ pair values.

\subsubsection{Case I:  Correct model analysis
  \label{SectEccentricCaseI}} 

In Case I, the {\it correct} one signal
DCM model for simulated data
has an order $K_2=2 \equiv e>0$ (\RModel{1,2,0}).
The number of signals
$(K_1=1)$ and the signal order $(K_2=2)$
are both {\it correct}.
The results for DCM search with this
correct model are given in Table \ref{TableOCone}.
This table has the same structure as Fig. \ref{FigOCone}.
For example, the results for
combination $e=0.05$ and $\omega=0\oo$ 
are given in the upper left corner of both
Table \ref{TableOCone} and Fig. \ref{FigOCone}.

DCM always detects the correct period $p$,
because the ratio $P_1/p$ is close to unity for {\it all}
forty $e$ and $\omega$ combinations.
The amplitude ratio $A_1/A$ is close to unity for
lower eccentricities $e \le 0.2$.
This ratio decreases for higher eccentricities.
Yet, even in these cases
the amplitude ratio is $A_1/A \ge 0.95$.
The inaccuracy of our $\nu(t)$
Fourier expansion (Eq. \ref{EqIrwFour})
may partly explain this $A_1/A$ ratio decrease.
DCM can certainly detect the
correct simulated signal period $p=45976^{\mathrm{d}}$
and amplitude $A=0.^{\mathrm{d}}0994$.
Our abbreviation for this
correct $p$ period detection is 
\begin{itemize}
\item[] ``\PC'' effect.
\end{itemize}

For eccentricities close to $e=0$,
the $(O-C)_{\mathrm{e>0}}$ curves 
for $P_1=p$ and $P_1=2p$ periods are
nearly identical.
We use the abbreviation ``\Dan''
to highlight all $P_1$ values
for lower eccentricities $e\le 0.1$ 
(Table \ref{TableOCone}).
In these cases,
the spurious
double period $P_1=2p$
detection is possible,
if the grid of tested frequencies
is too sparse. The probability
for detecting this spurious $P_1=2p$
period would of course decrease,
if our chosen simulated data
error $0.^{\mathrm{d}}005$ were smaller.
We call this 
spurious $2p$ period detection
\begin{itemize}
\item[] ``\PD'' effect.
\end{itemize}

\begin{deluxetable*}{cccccccccc}
  \tablecaption{Case I: Wrong model results.
    Simulated ${\mathrm{(O-C)}}_{\mathrm{e>0}}$ 
    signal period is $p=45976^{\mathrm{d}}$.
    Signal peak to peak amplitude is $A=0.^{\mathrm{d}}0994$.
    For different $e$ and $\omega$ combinations,
    two signal DCM \RModel{2,1,0} search
    detects signals having periods $P_1$ and $P_2$,
    and peak to peak amplitudes $A_1$ and $A_2$.
    Abbreviation ``\Vv''
    highlights the \PH ~effect cases, where 
    detection of weaker $P_1\sim p/2$ signal requires a
    denser tested frequency grid.
 \label{TableOCtwo}}
\tablewidth{700pt}
\tabletypesize{\scriptsize}
\tablehead{
  & & \colhead{$\omega=0\oo$}
  &   \colhead{$\omega=45\oo$}
  & \colhead{$\omega=90\oo$}
  & \colhead{$\omega=135\oo$}
  & \colhead{$\omega=180\oo$}
  & \colhead{$\omega=225\oo$}
  & \colhead{$\omega=270\oo$}
  &  \colhead{$\omega=315\oo$} } 
  \startdata
$e=0.05$ &$P_1    $&24823 \Vv&23076 \Vv&24924 \Vv&24927 \Vv&24793 \Vv&23813 \Vv&24554 \Vv&24867 \Vv\\
         &$A_1    $&0.0021  &0.0025  &0.0026  &0.0021  &0.0024  &0.0024  &0.0028  &0.0021  \\
         &$P_2    $&46032   &46032   &46018   &45973   &46003   &46072   &46055   &45982   \\
         &$A_2    $&0.0990  &0.0994  &0.0994  &0.0991  &0.0995  &0.0990  &0.0995  &0.0994  \\
         &$P_2/P_1$&1.85    &1.99    &1.85    &1.84    &1.85    &1.93    &1.88    &1.85    \\
         &$A_2/A_1$&47.1    &39.8    &38.2    &47.2    &41.4    &41.2    &35.5    &47.3    \\
         &$P_2/p  $&1.00    &1.00    &1.00    &1.00    &1.00    &1.00    &1.00    &1.00    \\
         &$A_2/A  $&1.00    &1.00    &1.00    &1.00    &1.00    &1.00    &1.00    &1.00    \\
\hline
$e=0.10$ &$P_1    $&24993 \Vv&22788   &23172   &24840 \Vv&25051 \Vv&23170   &24418 \Vv&24900 \Vv\\
         &$A_1    $&0.0044  &0.0052  &0.0048  &0.0044  &0.0044  &0.0052  &0.0051  &0.0045  \\
         &$P_2    $&46043   &45996   &45937   &45998   &46032   &46025   &46121   &45993   \\
         &$A_2    $&0.0988  &0.0992  &0.0992  &0.0986  &0.0993  &0.0987  &0.0987  &0.0988  \\
         &$P_2/P_1$&1.84    &2.02    &1.98    &1.85    &1.84    &1.99    &1.89    &1.85    \\
         &$A_2/A_1$&22.4    &19.1    &20.7    &24.2    &22.6    &19.0    &19.4    &22.0    \\
         &$P_2/p  $&1.00    &1.00    &1.00    &1.00    &1.00    &1.00    &1.00    &1.00    \\
         &$A_2/A  $&0.99    &1.00    &1.00    &0.99    &1.00    &0.99    &0.99    &0.99    \\
\hline
$e=0.20$ &$P_1    $&23010   &23029   &22848   &22899   &23024   &23044   &22863   &22879  \\
         &$A_1    $&0.0098  &0.0089  &0.0092  &0.0095  &0.0096  &0.0096  &0.0097  &0.0094 \\
         &$P_2    $&45953   &46029   &45973   &45955   &46000   &46074   &45941   &45962  \\
         &$A_2    $&0.0974  &0.0975  &0.0981  &0.0974  &0.0973  &0.0974  &0.0978  &0.0973 \\
         &$P_2/P_1$&2.00    &2.00    &2.01    &2.01    &2.00    &2.00    &2.01    &2.01   \\
         &$A_2/A_1$&9.9     &11.0    &10.7    &10.2    &10.1    &10.1    &10.1    &10.4   \\
         &$P_2/p  $&1.00    &1.00    &1.00    &1.00    &1.00    &1.00    &1.00    &1.00   \\
         &$A_2/A  $&0.98    &0.98    &0.99    &0.98    &0.98    &0.98    &0.98    &0.98   \\
\hline
$e=0.30$ &$P_1    $&22977   &23116   &22833   &22996   &23098   &23048   &22806   &23085   \\
         &$A_1    $&0.0135  &0.0135  &0.0139  &0.0139  &0.0135  &0.0134  &0.0140  &0.0141  \\
         &$P_2    $&46069   &46073   &45987   &46018   &46060   &46030   &45942   &45976   \\
         &$A_2    $&0.0940  &0.0944  &0.0962  &0.0949  &0.0940  &0.0947  &0.0961  &0.0949  \\
         &$P_2/P_1$&2.00    &1.99    &2.01    &2.00    &1.99    &2.00    &2.01    &1.99    \\
         &$A_2/A_1$&7.0     &7.0     &6.9     &6.8     &7.0     &7.1     &6.9     &6.7     \\
         &$P_2/p  $&1.00    &1.00    &1.00    &1.00    &1.00    &1.00    &1.00    &1.00    \\
         &$A_2/A  $&0.94    &0.95    &0.97    &0.95    &0.94    &0.95    &0.97    &0.95    \\
\hline
$e=0.40$ &$P_1    $&22998   &23119   &22916   &22899   &23068   &23210   &22909   &22976   \\
         &$A_1    $&0.0166  &0.0166  &0.0167  &0.0178  &0.0165  &0.0164  &0.0170  &0.0174  \\
         &$P_2    $&45997   &46089   &45978   &45961   &46054   &46267   &45983   &45917   \\
         &$A_2    $&0.0901  &0.0909  &0.0938  &0.0922  &0.0898  &0.0904  &0.0934  &0.0920  \\
         &$P_2/P_1$&2.00    &1.99    &2.01    &2.01    &2.00    &1.99    &2.01    &2.00    \\
         &$A_2/A_1$&5.4     &5.5     &5.6     &5.2     &5.4     &5.5     &5.5     &5.3     \\
         &$P_2/p  $&1.00    &1.00    &1.00    &1.00    &1.00    &1.01    &1.00    &1.00   \\
         &$A_2/A  $&0.91    &0.91    &0.94    &0.93    &0.90    &0.91    &0.94    &0.92   \\
\enddata
\end{deluxetable*}

\begin{figure*}[ht!]
\plotone{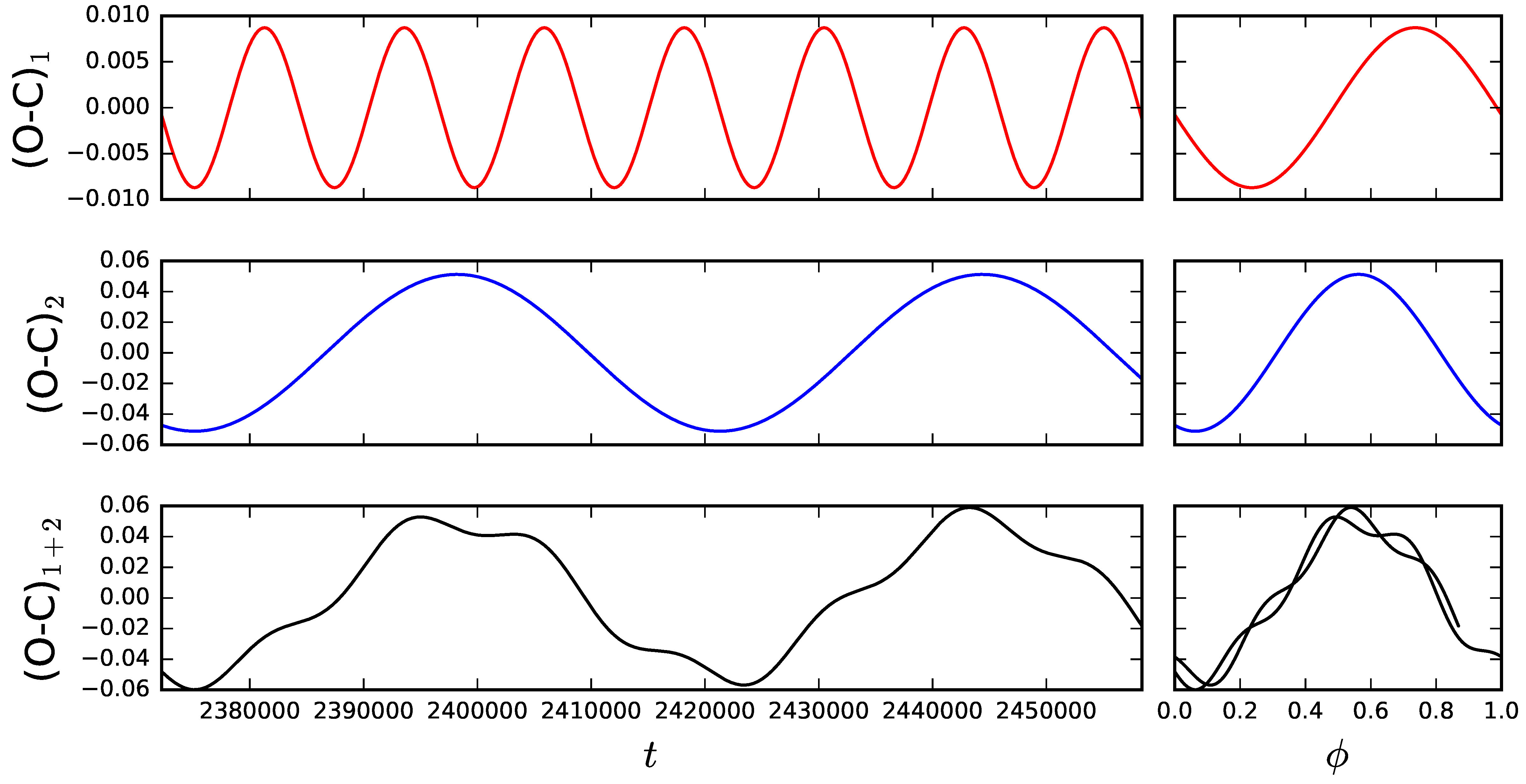}
\caption{Case II:
  Interference of two circular orbit
  ${\mathrm{(O-C)}}_{\mathrm{e=0}}$ curves.
  Red and blue curve periods are
  $p_1=12295^{\mathrm{d}}$
  and
  $p_1=46159^{\mathrm{d}}$, respectively.
  Other parameters are given in Table \ref{TableCases}.
  Black curve shows combined ${\mathrm{(O-C)_{1+2}}}$ effect
  having a period $P_1=46122^{\mathrm{d}}$.
  All curves are shown as a function of time
  (left-hand panels: $t$) and phase  (right-hand panels: $\phi$).
  Left-hand panel
    units are $[t]={\mathrm{HJD}}$
    and $[{\mathrm{O\!-\!C}}]={\mathrm{d}}$.
    Right-hand panel
    units are $[\phi]=$ dimensionless 
    and $[{\mathrm{O\!-\!C}}]={\mathrm{d}}$.
  \label{FigOCtwo}}
\end{figure*}

\begin{figure*}[ht!]
\plotone{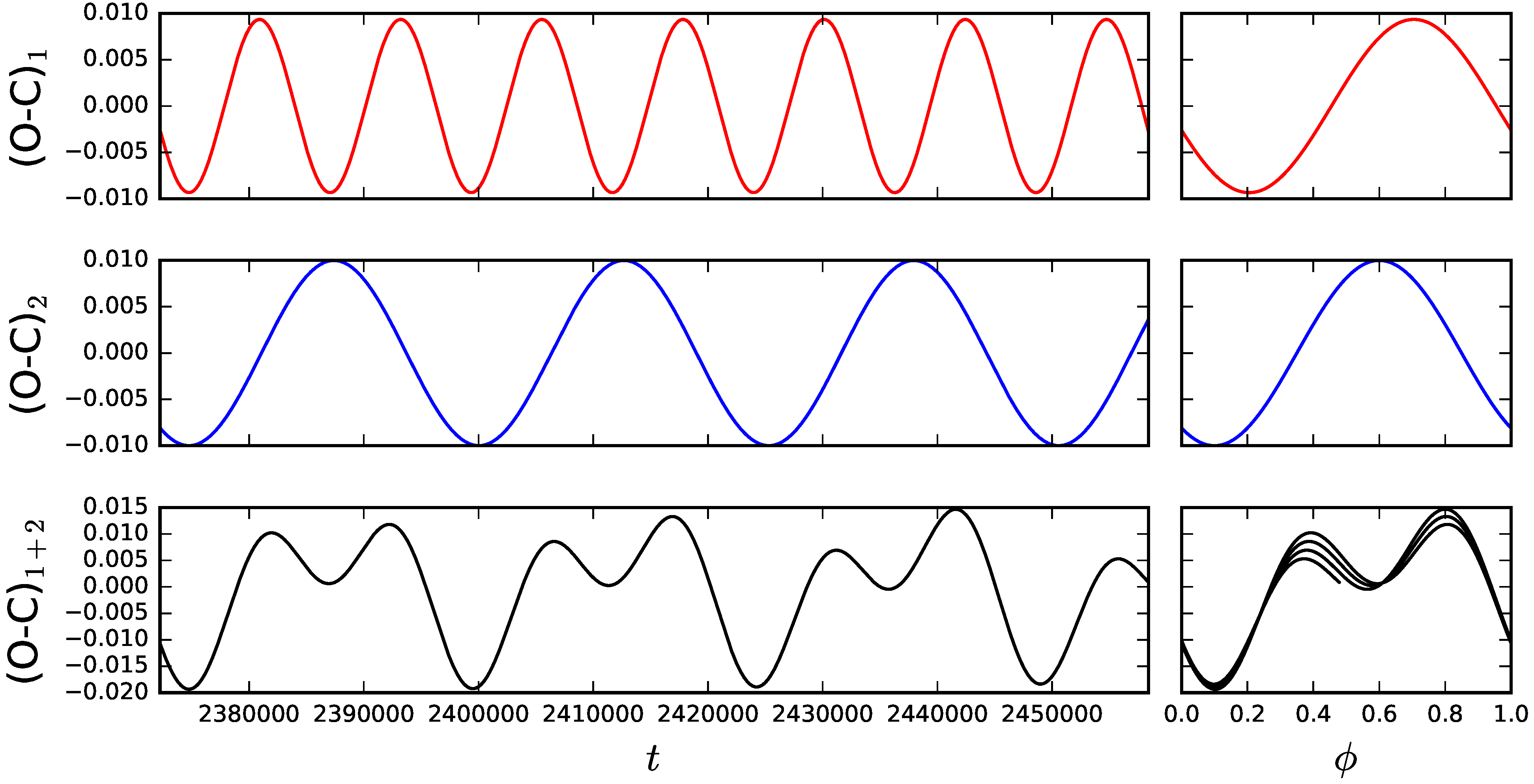}
\caption{Case III:
  Interference of two circular orbit
  ${\mathrm{(O-C)}}_{\mathrm{e=0}}$  curves.
  Red and blue curve periods are
  $p_1=12304^{\mathrm{d}}$ and $p_2=25274^{\mathrm{d}}$.
  Black curve shows combined ${\mathrm{(O-C)_{1+2}}}$  effect
  having a period $P_1=24771^{\mathrm{d}}$.
  Otherwise, as in Fig. \ref{FigOCtwo}.
  \label{FigOCthree}}
\end{figure*}

\subsubsection{Case I:  Wrong model analysis
  \label{SectCircularCaseI} } 

Here, we analyse again the same one signal simulated
eccentric orbit $({\mathrm{O-C}})_{e>0}$ data of Case~I,
but our two signal DCM \RModel{2,1,0} is {\it wrong}.
The number of signals $(K_1=2)$
and the signal order $(K_2=1)$
are both {\it wrong}.
In other words, we make the false assumption
that the one signal eccentric orbit  $({\mathrm{O-C}})_{e>0}$ curve
is a sum of two circular orbit  $({\mathrm{O-C}})_{e=0}$ curves.
The results for this wrong model analysis 
are given in Table \ref{TableOCtwo}.
Note that this table also
has the same structure as
Table \ref{TableOCone} and Fig. \ref{FigOCone}.

The correct period $P_2=p$ is always detected,
because the $P_2/p$ ratio is very close to unity for
{\it all} forty $e$ and $\omega$ combinations.
The  period of weaker detected signal is always 
$P_1 \approx P_2/2 \approx p/2$.
Furthermore,
the accuracy of this approximation increases when
$e$ increases!
Both of the $p$ and $p/2$ periods are certainly
detected at larger eccentricities $e \ge 0.2$.
The $A_2/A_1$ amplitude ratio of these
$p$ and $p/2$ signals decreases
for higher eccentricities.
This happens at the expense of $P_2$ signal,
because $A_2/A$ decreases about 10\%
when eccentricity
increases from $e=0.05$ and 0.40.
All these effects are also illustrated
in Fig. \ref{FigOCone}.

For nearly circular orbits $ e\le 0.10$,
the $A_2/A_1$ signal amplitude ratio 
is between 19 and 47.
We use the abbreviation ``\Vv'''
to highlight the cases,
where the detection of
weaker spurious $p/2$ period signal
requires
a denser tested frequency grid
(Table \ref{TableOCone}).
Our abbreviation for this
spurious $p/2$ period  detection is
\begin{itemize}
\item[] ``\PH'' effect.
\end{itemize}
Some $\omega$ values can eliminate
the symmetry of the
$({\mathrm{O-C}})_{e>0}$ curve
even at these low $e\le 0.10$ eccentricities,
like the $e=0.10$ and $\omega=45^{\mathrm{o}}$
combination  $({\mathrm{O-C}})_{e>0}$  curve that
shows no ``\Vv'' ~effect.

  It is important to realize that 
  every {\it real} eccentric orbit ${\mathrm{(O-C)_{\mathrm{e>0}}}}$
  curve can be presented as a sum of purely sinusoidal
  circular orbit ${\mathrm{(O-C)_{\mathrm{e=0}}}}$ curve and
  a nearly sinusoidal ${\mathrm{(O-C)_{\mathrm{diff}}}}$ curve.
  The respective periods of these curves are $p$, $p$ and $\sim p/2$.
  All these three curves are ``in-phase'',
  and therefore the eccentric orbit
  ${\mathrm{(O-C)_{\mathrm{e>0}}}}$ sum curve
  has only one minimum and one
  maximum.

\subsection{Case II: Correct model analysis
  \label{SectCircularCaseII}} 

In Case II, the simulated data
contains a sum of
two sinusoidal circular orbit  ${\mathrm{(O-C)}}_{\mathrm{e=0}}$  signals
having periods $p_1=12295^{\mathrm{d}}$ and $p_2=46159^{\mathrm{d}}$
(Fig. \ref{FigOCtwo}).
The other parameters can be found from Table \ref{TableCases}
(Case II).
The higher amplitude $p_2$ signal dominates over
the lower amplitude $p_1$ signal.
These red and blue $({\mathrm{O-C}})_{e=0}$ 
curves, and their black 
 $({\mathrm{O-C}})_{1+2}$ interference curve,
are shown in Fig. \ref{FigOCtwo}.

In this Case II, the {\it correct}
circular orbit model is DCM \RModel{2,1,0}.
This model has the correct number of signals $(K_1=2)$ and
the correct order $(K_2=1)$.
DCM detects the correct simulated
$P_1=12286^{\mathrm{d}} \pm 18^{\mathrm{d}}$ and
$P_2=46122^{\mathrm{d}} \pm 57^{\mathrm{d}}$ signal periods,
as well as the correct
amplitudes
$A_1=0.^{\mathrm{d}}0170 \pm 0.^{\mathrm{d}}0004$ and
$A_2=0.^{\mathrm{d}}1019 \pm 0.^{\mathrm{d}}0003$.
In short, DCM succeeds in detecting both simulated
circular orbit (O-C)$_{\mathrm{e=0}}$
signals.

\subsection{Case II: Wrong model analysis
  \label{SectEccentricCaseII}} 

Here, we analyse Case II simulated data 
using the {\it wrong} eccentric orbit
one signal DCM \RModel{1,2,0}.
Both the number of signals $(K_1=1)$ and the model
order $(K_2=2)$ are wrong.
We detect $P_1=46400^{\mathrm{d}}\pm 81^{\mathrm{d}}$ period signal
having a peak to peak
amplitude $A_1=0.^{\mathrm{d}}1015 \pm 0.^{\mathrm{d}}0005$.
Since the $p_2=46159^{\mathrm{d}}$
period of the stronger
signal dominates in the black
${\mathrm{(O-C)_{1+2}}}$
interference curve of Fig. \ref{FigOCtwo},
this detected $P_1$ period 
is close to, but slightly larger than,
the $p_2$ period.
Our DCM search result for $P_1$ is confirmed by
the distance between the black
${\mathrm{(O-C)}_{1+2}}$
interference curve minima,
which is indeed
longer than the distance
between the dominating blue ${\mathrm{(O-C)_{2}}}$
curve minima (Fig. \ref{FigOCtwo}).

\subsection{Case III: Correct model analysis
  \label{SectCircularCaseIII}} 

In Case III, the simulated sinusoidal
${\mathrm{(O-C)}}_{\mathrm{e=0}}$ 
signal periods are
$p_1=12304^{\mathrm{d}}$ and
$p_2=25274^{\mathrm{d}}$
(Fig. \ref{FigOCthree}).
The signal amplitudes are nearly equal
(Table \ref{TableCases}: Case~II).
The {\it correct} model for these simulated data
is the DCM \RModel{2,1,0}, which searches for the
sum of two circular orbit  ${\mathrm{(O-C)}}_{\mathrm{e=0}}$ curves
$(K_1=2, K_2=1)$.
DCM detects the correct 
$P_1=12322^{\mathrm{d}} \pm 20^{\mathrm{d}}$ and
$P_2=25259^{\mathrm{d}} \pm 89^{\mathrm{d}}$ signals,
as well as the correct 
amplitudes 
$A_1=0.^{\mathrm{d}}0190 \pm 0.^{\mathrm{d}}0004$ and
$A_2=0.^{\mathrm{d}}020 \pm 0.^{\mathrm{d}}0003$.
Again, DCM succeeds in detecting both simulated
circular orbit ${\mathrm{(O-C)}}_{\mathrm{e=0}}$ signals.

\subsection{Case III: Wrong model analysis
  \label{SectEccentricCaseIII}} 

Finally, the same simulated Case III data is
analysed by using the {\it wrong} eccentric DCM \RModel{1,2,0}.
In other words, we search for only one eccentric orbit
${\mathrm{(O-C)}}_{\mathrm{e>0}}$ signal
when the data contains two circular orbit 
${\mathrm{(O-C)}}_{\mathrm{e=0}}$ signals.
DCM detects a signal having
$P_1=24771^{\mathrm{d}} \pm 34^{\mathrm{d}}$ and
$A_1=0.^{\mathrm{d}}0319 \pm 0.^{\mathrm{d}}0005$.
The simulated
$p_1$ and $p_2$ signals' interference period is
\begin{eqnarray}
{p'}=k (p_1^{-1}-p_2^{-1})^{-1},
\label{EqSpurious}
\end{eqnarray}
where $k= \pm 1, \pm 2, ...$ is the phase
  difference during $p'$.
  In this particular case, $k=1$ gives $p'=23976^{\mathrm{d}}$. 
  This black double wave
${\mathrm{(O-C)}}_{1+2}$
  curve is shown in Fig. \ref{FigOCthree}.
DCM detects this ``correct''
interference signal period $p'$,
which is repeated through out the whole data.
We call this spurious interference period $p'$ detection
\begin{itemize}
\item[] ``\PI'' effect.
\end{itemize}
\noindent
The black $p'$ interference
${\mathrm{(O-C)}}_{1+2}$
curve shows two minima
and two maxima, because the red $p_1$ period
and the blue $p_2$ period sinusoids
are ``off-phase''(Fig. \ref{FigOCthree}).
Therefore, this black ${\mathrm{(O-C)}}_{1+2}$
curve can not represent
a {\it real} eccentric
orbit ${\mathrm{(O-C)}}_{\mathrm{e>0}}$ curve.

\newpage

\begin{deluxetable*}{ccc}

  \tablecaption{O-C data.
  Only first three
  of all $n=2224$ values are shown.
 \label{TableData}}
\tablewidth{700pt}
\tabletypesize{\scriptsize}
\tablehead{ 
\colhead{$t$}   & \colhead{$y$}   & \colhead{$\sigma_y$} \\
\colhead{$\Dd$} & \colhead{$\Dd$} & \colhead{$\Dd$}     }
\startdata
       2372238.35100   & -0.17216    &    0.00010 \\
       2372284.23700   & -0.16303    &    0.00010 \\
       2372301.39700   & -0.20686    &    0.00010 \\
       \enddata
       \tablecomments{The arbitrary
         $0.^{\mathrm{d}}00010$ errors give the correct
  format for our data file (file1),
  because
  dcm.py
requires that the
  third column of file1
  contains some numerical values for the errors.
  Since these errors are unknown,
  we use equal weights
  for all observations, and the numerical
  values of these errors are irrelevant
  (see Appendix of \paperone: TestStat$\neq 1$).}
     \end{deluxetable*}

\begin{deluxetable*}{lrllrrc}
  \tablecaption{Samples. 
    Number of observations $(n)$,
    first observing time $(t_1)$,
    last observing time $(t_n)$,
    time span $(\Delta T)$
    and
    data file (file1)
    \label{TableSamples}}
\tablewidth{700pt}
\tabletypesize{\scriptsize}
\tablehead{
  \colhead{}  & \colhead{$n$~~} & \colhead{~~~$t_1$}  & \colhead{~~$t_n$} & \multicolumn{2}{c}{$\Delta T$}  & \colhead{}  \\
  \cline{5-6}
  \colhead{Sample} & -        & \colhead{[HJD]}  & \colhead{[HJD]}  & \colhead{$\Dd$} & \colhead{[y]} & \colhead{file1}}
\startdata
All data   & 2224  & 2372238.351   &  2458409.7612  &  86171.4102  &  235.9 & 1hjdAlgol.dat \\
\Adataone  & 2174  & 2372238.351   &  2454839.9189  &  82601.5679  &  226.2 & 2hjdAlgol.dat \\
\Adatatwo  & 50    & 2455063.566   &  2458409.7612  &  3346.1952   &  9.2   & - \\
\Bdataone  & 1731  & 2372238.351   &  2439918.358   &  67680.007   &  185.3 & 3hjdAlgol.dat  \\
\Bdatatwo  & 493   & 2440144.8771  &  2458409.7612  &  18264.8841  &  50.0  & - \\
\enddata
\end{deluxetable*}

\begin{deluxetable*}{cll}
  \tablecaption{Abbreviations.
    These abbreviations are used in Tables
    \ref{TableOCone} - \ref{TableOCtwo} and
    Tables \ref{TableAllFisher}-\ref{TableCompare}.
    Second column gives their meaning. Third column gives section
    where this meaning is defined.
\label{TableSymbols}}
\tablewidth{700pt}
\tabletypesize{\scriptsize}
\tablehead{\colhead{Symbol}  & \colhead{Meaning} & \colhead{Definition} }
\startdata
 ``\RD''  & Unstable model               & Sect. \ref{SectMethod} \\
 ``\HD''  & Intersecting frequencies     & Sect. \ref{SectMethod} \\
 ``\AD''  & Dispersing amplitudes        & Sect. \ref{SectMethod} \\
 ``\LK''  & Leaking period               & Sect. \ref{SectTrend}  \\
 ``\SP''  & Spurious period              & Sect. \ref{SectAllDataEccentric}  \\
 ``\Vv''  & \PH ~effect                  & Sect. \ref{SectEccentricCaseI} \\
 ``\Dan'' & \PD ~effect                      & Sect. \ref{SectCircularCaseI} \\
\enddata
\end{deluxetable*}

\begin{deluxetable*}{cccccccccccccc}
  \tablecaption{
    All data trend (Sect. \ref{SectTrend}):
    One, two and three signal
    ($K_1=1, 2$ and 3) models (\M) have eccentric
    third body orbits ($K_2=2 \equiv e>0)$.
    Trend orders $K_3=0, 1,2$ and 3 for $p(t)$ are compared.
    Fisher-test is used to compare DCM
    search results between 8000 and 80000 days.
    Notations are
    ``$\uparrow$ $" \equiv$ complex model above is
    better than left side simple model,
    and
    ``$\leftarrow$" $\equiv$ left side simple model is better
    than complex model above.
    Parameters are
    $F=$ Fisher test statistic and
    $Q_F=$ critical level.
    Control file is dcm.dat.
    Unstable models are denoted with ``\RD''.
    They have dispersing amplitudes ``\AD'',
    or
    intersecting frequencies ``\HD''.
    Some models have
    leaking periods  ``\LK'' larger than $\Delta T=86171^{\mathrm{d}}$.
    Best DCM model for all data is linear trend
    $K_3=1$ order \RModel{3,2,1}.
     \label{TableAllFisher}}
  \tablewidth{700pt}
    \addtolength{\tabcolsep}{-0.16cm}
\tabletypesize{\scriptsize}
\tablehead{  
   & & & & & & & & & & & & & \\ 
   & & \multicolumn{10}{c}{Period analysis: All original data = 1hjdAlgol.dat} & \\
  \colhead{}&
  \colhead{}&  
  \colhead{\M=2}&
  \colhead{\M=3}&
  \colhead{\M=4}&
  \colhead{\M=5}&
  \colhead{\M=6}&
  \colhead{\M=7}&
  \colhead{\M=8}&
  \colhead{\M=9}&
  \colhead{\M=10}&
  \colhead{\M=11}&
  \colhead{\M=12}&
               \\
  \colhead{\M}&
  \colhead{Model}&
  \colhead{\RModel{1,2,1}}&
  \colhead{\RModel{1,2,2}}&
  \colhead{\RModel{1,2,3}}&
  \colhead{\RModel{2,2,0}}&
  \colhead{\RModel{2,2,1}}&
  \colhead{\RModel{2,2,2}}&
  \colhead{\RModel{2,2,3}}&
  \colhead{\RModel{3,2,0}}&
  \colhead{\RModel{3,2,1}}&
  \colhead{\RModel{3,2,2}}&
  \colhead{\RModel{3,2,3}}&
  \colhead{dcm.dat}
}
\startdata
1  & \RModel{1,2,0} &$\uparrow$&$\uparrow$&$\uparrow$&$\uparrow$&$\uparrow  $&$\uparrow$&$\uparrow$&$\uparrow$&$\uparrow$&$\uparrow$&$\uparrow$& 1hjd13R120S.dat \\
   &  \p$=6        $&$F=2821  $&$F=1542  $&$F=1154  $&$F=684   $&$F=815      $&$F=702  $&$F=623   $&$F=498   $&$F=503   $&$F=461   $&$F=427   $&                     \\
   &  $R=1.0678    $&\REJ      &\REJ      &\REJ      &\REJ      &\REJ        &\REJ      &\REJ      &\REJ      &\REJ      &\REJ      &\REJ      &                     \\
\hline
2  & \RModel{1,2,1} &-         &$\uparrow$&$\uparrow$&$\uparrow $&$\uparrow  $&$\uparrow$&$\uparrow$&$\uparrow$&$\uparrow$&$\uparrow$&$\uparrow$& 1hjd13R121S.dat \\
   &  \p$=7        $&-         &$F=116   $&$F=142   $&$F=67     $&$F=183      $&$F=154   $&$F=137  $&$F=106   $&$F=120  $&$F=109   $&$F=101   $&                     \\
\LK&  $R=0.4698    $&-         &\REJ      &\REJ      &\REJ       &\REJ        &\REJ      &\REJ      &\REJ     &\REJ      &\REJ      &\REJ      &                     \\
\hline
3  &  \RModel{1,2,2}&-         & -        &$\uparrow$&$\uparrow$&$\uparrow  $&$\uparrow$&$\uparrow$&$\uparrow$&$\uparrow$&$\uparrow$&$\uparrow$& 1hjd13R122S.dat \\
\RD   &\p$=8       $&-         & -        &$F=161   $&$F=34    $&$F=190     $&$F=154   $&$F=134   $&$F=100   $&$F=115   $&$F=103   $&$F=95    $&                     \\
\enddata
  \addtolength{\tabcolsep}{+0.16cm},
\end{deluxetable*}

\begin{deluxetable*}{clccccccccl}
  \tablecaption{All data eccentric orbits (Sect. \ref{SectAllDataEccentric}).
    Col 1. Model number \M.
    Col 2. \RModel{K_1,K_2,K_3},
    \p= number of free parameters and $R=$ sum of squared residuals.
    Cols 3-6. Period analysis results:
    Detected periods $P_1, ..., P_4$ and amplitudes $A_1, ..., A_4.$
    Cols 7-9. Fisher-test results.
    Col 10. Control file is dcm.dat.
    Model \M=3+2 is the best one for all data.
    \label{Table1hjd14R}}
  \tablewidth{700pt}
  \addtolength{\tabcolsep}{-0.12cm}
\tabletypesize{\scriptsize}
\tablehead{
  \colhead{Col 1} &
  \colhead{Col 2} &
  \colhead{Col 3} &
  \colhead{Col 4} &
  \colhead{Col 5} &
  \colhead{Col 6} &
  \colhead{     } &
  \colhead{Col 7} &
  \colhead{Col 8} &
  \colhead{Col 9} &
  \colhead{Col 10}\\
  & & \multicolumn{8}{c}{Search between 8000 and 80000 days} & \\
 \cline{3-10}
  & & \multicolumn{4}{c}{Period analysis: All original data =  1hjdAlgol.dat} & & \multicolumn{3}{c}{Fisher-test} \\
  \cline{3-6} \cline{8-10}
  \colhead{\M}&   \colhead{Model} & \colhead{$P_1$\& $A_1 \Dd$} & \colhead{$P_2$\& $A_2\Dd$} &\colhead{$P_3$\& $A_3\Dd$} &\colhead{$P_4$\& $A_4\Dd$} & \colhead{} &
  \colhead{\RModel{2,2,1}} & \colhead{\RModel{3,2,1}} & \colhead{\RModel{4,2,1}} & \colhead{dcm.dat}}
\startdata
1 & \RModel{1,2,1}&$88183\pm816$ \LK &           -    & -                    & -                    & &$\uparrow             $&$\uparrow         $&$\uparrow$& 1hjd14R121S.dat \\
  & \p$=7        $&$0.313 \pm0.004 $&            -    & -                    & -                    & &$F=183                $&$F=120            $&$F=90    $&                    \\
\LK&$R=0.4698    $&                 &                 &                      &                      & &\REJ                   &\REJ               &\REJ      &                    \\
  \hline
2 &\RModel{2,2,1}&$24984\pm  99   $&$80087\pm744     $& -                    & -                    & & -                     &$\uparrow$         &$\uparrow$& 1hjd14R221S.dat\\
  &\p$=12       $&$0.0316\pm0.0008$&$0.288\pm 0.004  $& -                    & -                    & & -                     &$F=41             $&$F=31    $&                    \\
  &$R=0.3324    $&                 &                  &                      &                      & & -                     &\REJ               &\REJ      &                    \\
\hline
 3&\RModel{3,2,1}&$20358\pm 128   $&$24742\pm142     $&$79999\pm1216        $& -                    & & -                     & -                 &$\uparrow $& 1hjd14R321S.dat \\
  &\p$=17       $&$0.013\pm0.001  $&$0.029\pm 0.001  $&$0.287\pm0.005       $& -                    & & -                     & -                 &$F=19     $&                    \\
\enddata
\addtolength{\tabcolsep}{+0.15cm}
\end{deluxetable*}

\begin{deluxetable*}{clccccccccl}
  \tablecaption{All data circular orbits (Sect. \ref{SectAllDataCircular}: First alternative).
    Otherwise as in Table \ref{Table1hjd14R}.
    \label{Table1hjd14C}}
  \tablewidth{700pt}
  \addtolength{\tabcolsep}{-0.12cm}
\tabletypesize{\scriptsize}
\tablehead{
  \colhead{} & 
  \colhead{} &
  \colhead{} &
  \colhead{} &
  \colhead{} &
  \colhead{} &
  \colhead{} &
  \colhead{} &
  \colhead{} &
  \colhead{} &
  \colhead{} \\
  & & \multicolumn{8}{c}{Search between 8000 and 80000 days} & \\
 \cline{3-10}
  & & \multicolumn{4}{c}{Period analysis: All original data = 1hjdAlgol.dat} & & \multicolumn{3}{c}{Fisher-test} \\
  \cline{3-6} \cline{8-10}
  \colhead{\M}&   \colhead{Model} & \colhead{$P_1$\& $A_1 \Dd$} & \colhead{$P_2$\& $A_2\Dd$} &\colhead{$P_3$\& $A_3\Dd$} &\colhead{$P_4$\& $A_4\Dd$} & \colhead{} &
  \colhead{\RModel{2,1,1}} & \colhead{\RModel{3,1,1}} & \colhead{\RModel{4,1,1}} & \colhead{dcm.dat}}
\startdata
1 & \RModel{1,1,1}&$59474\pm255    $&           -     & -                     & -                    & &$\uparrow             $&$\uparrow         $&$\uparrow$& 1hjd14R111S.dat \\
  & \p$=5        $&$0.199 \pm0.001 $&            -    & -                     & -                    & &$F=863                $&$F=598            $&$F=503   $&                    \\
  &$R=0.9998     $&                 &                 &                       &                      & &\REJ                   &\REJ               &\REJ      &                    \\
  \hline
2 &\RModel{2,1,1}&$45204\pm 350    $&$145483\pm13700$ \LK& -                  & -                    & & -                     &$\uparrow$         &$\uparrow$& 1hjd14R211S.dat \\
\RD&\p$=8       $&$0.091\pm0.003  $&$0.8\pm 0.2$ \AD  & -                     & -                    & & -                     &$F=154            $&$F=149   $&                    \\
\AD,\LK&$R=0.4610$&               &                   &                       &                      & & -                     &\REJ               &\REJ      &                    \\
  \hline
 3&\RModel{3,1,1}&$12338\pm 55    $&$45423\pm424     $&$145456\pm15326$ \LK   & -                    & & -                     & -                 &$\uparrow $& 1hjd14R311S.dat \\
\RD&  \p$=11    $&$0.0177\pm0.0006$&$0.094\pm 0.004  $&$0.8 \pm0.2$     \AD   & -                    & & -                     & -                 &$F=120    $&                    \\
\enddata
\addtolength{\tabcolsep}{+0.15cm}
\end{deluxetable*}

\begin{deluxetable*}{clcccccccccl}
  \tablecaption{All data circular orbits (Sect. \ref{SectAllDataCircular}: Second alternative).
    Otherwise as in Table \ref{Table1hjd14R}.
   \label{Table1hjd16C}}
  \tablewidth{700pt}
  \addtolength{\tabcolsep}{-0.15cm}
\tabletypesize{\scriptsize}
\tablehead{
  \colhead{} &
  \colhead{} & 
  \colhead{} &
  \colhead{} &
  \colhead{} &
  \colhead{} &
  \colhead{} &
  \colhead{} &
  \colhead{} &
  \colhead{} &
  \colhead{} &
  \colhead{} \\
  & & \multicolumn{9}{c}{Search between 8000 and 80000 days} & \\
 \cline{3-11}
  & & \multicolumn{6}{c}{Period analysis: All original data = 1hjdAlgol.dat} &  \multicolumn{2}{c}{Fisher-test} \\
  \cline{3-8}  \cline{9-10}
  \colhead{\M}&   \colhead{Model} & \colhead{$P_1$\& $A_1 \Dd$} & \colhead{$P_2$\& $A_2\Dd$} &\colhead{$P_3$\& $A_3\Dd$}
  &\colhead{$P_4$\& $A_4\Dd$} & \colhead{$P_5$\& $A_5\Dd$} & \colhead{$P_6$\& $A_6\Dd$} & 
  \colhead{\RModel{5,1,1}} & \colhead{\RModel{6,1,1}} & \colhead{dcm.dat}}
\startdata
1 & \RModel{4,1,1}&$12377\pm50     $&$24633\pm578    $&$42318\pm 240 $&$120817\pm31482$ \LK& -          & -             &$\uparrow         $&$\uparrow  $& 1hjd16R411S.dat \\
\RD  &\p$=14     $&$0.0188\pm0.0008$&$0.018 \pm0.004 $&$0.085\pm0.004 $&$0.6\pm0.7$ \AD   &-            & -             &$F=57             $&$F=31      $&                    \\
\AD,\LK&$R=0.3284$&                 &                 &                &                  & -           & -             &\REJ               &\REJ        &                    \\
\hline
2 & \RModel{5,1,1}&$10144\pm30     $&$12294\pm109    $&$24247\pm872    $&$42422\pm640    $&$120740\pm41002$ \LK& -      &                  &$\leftarrow$& 1hjd16R511S.dat \\
\RD  &\p$=17     $&$0.0097\pm0.0004$&$0.018\pm0.001  $&$0.018\pm0.002  $&$0.08\pm0.01    $&$0.6\pm0.5$ \AD& -           &                  &$F=4.6     $&                    \\
\AD,\LK&$R=0.3047$&                 &                 &                 &                 &              & -            &                  &$Q_F=0.0032$&                    \\
\hline
3 & \RModel{6,1,1}&$10778\pm256$ \HD&$11219\pm206$ \HD&$11368\pm182$ \HD&$23475\pm625   $&$42617\pm267 $&$96213\pm20664$ \LK&              &  -         & 1hjd16R611S.dat \\
\RD  &\p$=20   $&$0.08\pm0.02$   \AD&$0.3\pm0.1$  \AD&$0.28\pm0.08$ \AD&$0.016\pm0.001 $&$0.078\pm0.006$&$0.4\pm0.2$ \AD  &                &  -         &                    \\
\enddata
\addtolength{\tabcolsep}{+0.15cm}
\end{deluxetable*}


\begin{deluxetable*}{clccccccccl}
  \tablecaption{\Adataone ~eccentric orbits
    (Sect. \ref{SectAdata}).
    Otherwise as in Table \ref{Table1hjd14R}.
    \label{Table2hjd14R}}
  \tablewidth{700pt}
  \addtolength{\tabcolsep}{-0.12cm}
\tabletypesize{\scriptsize}
\tablehead{
  \colhead{Col 1} &
  \colhead{Col 2} &
  \colhead{Col 3} &
  \colhead{Col 4} &
  \colhead{Col 5} &
  \colhead{Col 6} &
  \colhead{     } &
  \colhead{Col 7} &
  \colhead{Col 8} &
  \colhead{Col 9} &
  \colhead{Col 10}\\
  & & \multicolumn{8}{c}{Search between 8000 and 80000 days} & \\
 \cline{3-10}
  & & \multicolumn{4}{c}{Period analysis: All original data = 2hjdAlgol.dat} & & \multicolumn{3}{c}{Fisher-test} \\
  \cline{3-6} \cline{8-10}
  \colhead{\M}&   \colhead{Model} & \colhead{$P_1$\& $A_1 \Dd$} & \colhead{$P_2$\& $A_2\Dd$} &\colhead{$P_3$\& $A_3\Dd$} &\colhead{$P_4$\& $A_4\Dd$} & \colhead{} &
  \colhead{\RModel{2,2,1}} & \colhead{\RModel{3,2,1}} & \colhead{\RModel{4,2,1}} & \colhead{dcm.dat}}
\startdata
1 & \RModel{1,2,1}&$92489\pm1016$ \LK&           -    & -                    & -                    & &$\uparrow             $&$\uparrow         $&$\uparrow $& 2hjd14R121S.dat \\
  & \p$=7        $&$0.320 \pm0.004 $&            -    & -                    & -                    & &$F=160                $&$F=104            $&$F=77     $&                    \\
\LK&$R=0.4443    $&                 &                 &                      &                      & &\REJ                   &\REJ               &\REJ         &                    \\
\hline
2 &\RModel{2,2,1}&$26623\pm  107  $&$84215\pm1466$ \LK & -                    & -                    & & -                     &$\uparrow$         &$\uparrow $& 2hjd14R221S.dat\\
  &\p$=12       $&$0.028\pm0.001  $&$0.300\pm 0.005  $& -                    & -                    & & -                     &$F=35             $&$F=26     $&                    \\
\LK  &$R=0.3243 $&                 &                  &                      &                      & & -                     &\REJ               &\REJ       &                    \\
\hline
 3&\RModel{3,2,1}&$20592\pm 172   $&$24870\pm190     $&$78589\pm2004        $& -                    & & -                     & -                 &$\uparrow  $& 2hjd14R321S.dat \\
  &\p$=17       $&$0.014\pm0.002  $&$0.030\pm 0.008  $&$0.282\pm0.007       $& -                    & & -                     & -                 &$F=17      $&                    \\
\enddata
\addtolength{\tabcolsep}{+0.12cm}
\end{deluxetable*}

\begin{deluxetable*}{clccccccccl}
  \tablecaption{\Bdataone
    ~eccentric orbits (Sect. \ref{SectBdata}).
    Otherwise as in Table \ref{Table1hjd14R}.
    \label{Table3hjd14R}}
  \tablewidth{700pt}
  \addtolength{\tabcolsep}{-0.12cm}
\tabletypesize{\scriptsize}
\tablehead{
  \colhead{Col 1} &
  \colhead{Col 2} &
  \colhead{Col 3} &
  \colhead{Col 4} &
  \colhead{Col 5} &
  \colhead{Col 6} &
  \colhead{     } &
  \colhead{Col 7} &
  \colhead{Col 8} &
  \colhead{Col 9} &
  \colhead{Col 10}\\
  & & \multicolumn{8}{c}{Search between 8000 and 80000 days} & \\
 \cline{3-10}
  & & \multicolumn{4}{c}{Period analysis: All original data = 3hjdAlgol.dat} & & \multicolumn{3}{c}{Fisher-test} \\
  \cline{3-6} \cline{8-10}
  \colhead{\M}&   \colhead{Model} & \colhead{$P_1$\& $A_1 \Dd$} & \colhead{$P_2$\& $A_2\Dd$} &\colhead{$P_3$\& $A_3\Dd$} &\colhead{$P_4$\& $A_4\Dd$} & \colhead{} &
  \colhead{\RModel{2,2,1}} & \colhead{\RModel{3,2,1}} & \colhead{\RModel{4,2,1}} & \colhead{dcm.dat}}
\startdata
1 & \RModel{1,2,1}&$64454\pm 880   $&            -    & -                    & -                    & &$\uparrow             $&$\uparrow         $&$\uparrow $& 3hjd14R121S.dat \\
  & \p$=7        $&$0.258 \pm0.005 $&            -    & -                    & -                    & &$F=  118              $&$F=69             $&$F=50     $&                    \\
  &$R=0.3407     $&                 &                 &                      &                      & &\REJ                   &\REJ               &\REJ       &                    \\
\hline
2 &\RModel{2,2,1}&$23477\pm  144  $&$120755\pm24037$ \LK& -                  & -                    & & -                     &$\uparrow$         &$\uparrow  $& 3hjd14R221S.dat \\
\RD  &\p$=12    $&$0.031\pm0.002  $&$0.5 \pm 0.9$ \AD & -                    & -                    & & -                     &$F=16             $&$F=12      $&                    \\
\AD,\LK&$R=0.2534 $&               &                  &                      &                      & & -                  &$Q_F=5.6\times 10^{-15}$&\REJ       &                    \\
\hline
 3&\RModel{3,2,1}&$14197\pm 134   $&$22873\pm289     $&$97561\pm23777$ \LK   & -                    & & -                     & -                 &$\uparrow$& 3hjd14R321S.dat \\
\RD &\p$=17    $&$0.010\pm0.001  $&$0.029\pm0.002   $&$0.3\pm0.8$   \AD      & -                    & & -                     & -                  &$F=7.9   $&                    \\
\enddata
\addtolength{\tabcolsep}{+0.12cm}
\end{deluxetable*}

  \begin{deluxetable*}{lllllll}
  \tablecaption{All data: Comparison of circular orbit results.
    Cols 1-3.
    Periods and amplitudes of best model \M=4+5+9 from Table \ref{Table1hjd14C}.
    Cols 4-6. 
    Periods and amplitudes of best model \M=2+6 from Table \ref{Table1hjd16C}.    
  \label{TableCompareCircular}}
\tablewidth{700pt}
\tabletypesize{\scriptsize}
\tablehead{\colhead{Col 1} & \colhead{Col 2} & \colhead{Col 3} & \colhead{} & \colhead{Col 4} & \colhead{Col 5} & \colhead{Col 6}} 
\startdata
\multicolumn{3}{c}{Table \ref{Table1hjd14C}: Circular $e=0 \equiv K_2=1$} & & \multicolumn{3}{c}{Table \ref{Table1hjd16C}: Circular $e=0 \equiv K_2=1$}  \\
\multicolumn{3}{c}{Best model \M=4+5+9}                                   & & \multicolumn{3}{c}{Best model \M=2+5}   \\   
\cline{1-3} \cline{5-7}
   & ~~~~~~~~ $\Dd$            & ~~~~~~~~ $\Dd$         &  &    & ~~~~~~~~ $\Dd$          & ~~~~~~~~ $\Dd$      \\
   \hline
\M=4&$P_4=145456\pm 16180$ \LK  &$A_4=0.9\pm 0.2$ \AD     & &\M=2&$P_5=120740\pm41002$ \LK&$A_5=0.6\pm0.5$  \AD \\
\M=4&$P_3=42610 \pm 556        $&$A_3=0.088\pm 0.004     $& &\M=2&$P_4=42422 \pm640      $&$A_4=0.08 \pm 0.01  $\\
\M=4&$P_2=24773 \pm 526        $&$A_2=0.018\pm 0.001     $& &\M=2&$P_3=24247 \pm872      $&$A_3=0.018\pm 0.002 $\\
\M=4&$P_1=12352 \pm 58         $&$A_1=0.0188\pm 0.0006   $& &\M=2&$P_2=12294 \pm 109     $&$A_2=0.018\pm 0.002 $\\
\M=5&$P_1=10175 \pm 83         $&$A_1=0.0087\pm0.0005    $& &\M=2&$P_1=10144 \pm 30      $&$A_1=0.0097\pm0.0004$\\ 
\M=9&$P_3=7360  \pm 55         $&$A_1=0.0056\pm0.0006    $& &\M=5&$P_3=7395  \pm 37      $&$A_3=0.0061\pm0.0006$\\
\M=9&$P_1=680.7 \pm 0.4        $&$A_1=0.0056\pm0.0006    $& &\M=5&$P_3=680.7 \pm 0.5     $&$A_1=0.0057\pm0.0009$\\
\M=9&$P_2=2986  \pm 39         $&$A_2=0.0031\pm0.0008    $& &   &                        &                    \\
\enddata
\end{deluxetable*}

\begin{deluxetable*}{llllllllll}
  \tablecaption{All data: comparison of eccentric and circular orbit results.
    Cols 1-3. Eccentric orbit results (Table \ref{Table1hjd14R}).
    Cols 4-6. Circular orbit results (Table \ref{Table1hjd16C}).
    Col 7. Connection between eccentric and circular orbit periods.
    Col 8. Effects are explained in Sect. \ref{SectAllDataSignals}.
    Eccentric and circular orbit periods are denoted with
    subscripts ``e'' and ``c'', respectively.
    \label{TableCompare}}
\tablewidth{700pt}
\tabletypesize{\scriptsize}
\tablehead{\colhead{Col 1} & \colhead{Col 2} & \colhead{Col 3} && \colhead{Col 4} & \colhead{Col 5} & \colhead{Col 6} & \colhead{} & \colhead{Col 7} & \colhead{Col 8} }
\startdata
  \multicolumn{3}{c}{Table \ref{Table1hjd14R}: Eccentric $e>0 \equiv K_2=2 $}&& \multicolumn{3}{c}{Table \ref{Table1hjd16C}: Circular  $e =0 \equiv K_2=1 $} &&  Connection &Effect \\
\cline{1-3}   \cline{5-7} \cline{9-10}
\M=3&$P_{e,5}=79999 \pm 1216$&$A_{e,5}=0.287\pm0.005$&&\M=2&$P_{c,7}=120740\pm 41002$ \LK&$A_{c,7}=0.6\pm 0.5$ \AD &&$P_{e,5}\approx 1 \times P_{c,7}$  & \PC \\
    &                       &                       &&\M=2&$P_{c,6}=42422  \pm 640   $&$A_{c,6}=0.08\pm 0.01   $&&$P_{e,5}\approx 2 \times  P_{c,6}$ & \PH \\
\M=3&$P_{e,4}=24742 \pm 141 $&$A_{e,4}=0.029\pm0.001 $&&\M=2&$P_{c,5}=24247 \pm 872     $&$A_{c,5}=0.018\pm0.002  $&&$P_{e,4}\approx 1 \times P_{c,5}$  & \PC\\
    &                       &                       &&\M=2&$P_{c,4}=12294 \pm109     $&$A_{c,4}=0.018 \pm0.001 $&&$P_{e,4}\approx 2 \times P_{c,4}$  & \PH \\
\M=3&$P_{e,3}=20358\pm128   $&$A_{e,3}=0.013\pm0.001 $&&\M=2&$P_{c,3}=10144\pm 91       $&$A_{c,3}=0.0097\pm0.0004$&&$P_{e,3} \approx 2\times P_{c,3}$ & \PH \\
\M=6&$P_{e,2}=7269\pm 29    $&$A_{e,2}=0.007\pm0.001 $&&\M=5&$P_{c,2}=7395\pm37         $&$A_{c,2}=0.0061\pm0.0006$&&$P_{e,2} \approx 1\times P_{c,2}$ & \PC \\  
\M=6&$P_{e,1}=680.4\pm 0.4  $&$A_{e,1}=0.0064\pm0.0007$&&\M=5&$P_{c,1}=680.7\pm0.5       $&$A_{c,1}=0.0057\pm0.0009$&&$P_{e,1} \approx 1\times P_{c,1}$ & \PC \\
\enddata
     \end{deluxetable*}

\begin{deluxetable*}{llllllllllll}
  \tablecaption{Eccentric orbit results for three samples.
    Cols 1-3. All data eccentric orbit results (Table \ref{Table1hjd14R}).
    Cols 4-6. \Adataone ~eccentric orbit results (Table \ref{Table2hjd14R}).
    Cols 7-9. \Bdataone ~eccentric orbit results (Table \ref{Table3hjd14R}).
    Col 10. \Bdataone ~period effect explained in Sect. \ref{SectAllDataSignals}.
    \label{TableThreeSamples}}
\tablewidth{700pt}
\addtolength{\tabcolsep}{-0.10cm}
\tabletypesize{\scriptsize}
\tablehead{\colhead{Col 1} &  \colhead{Col 2}  & \colhead{Col 3} & & \colhead{Col 4}   & \colhead{Col 5}  & \colhead{Col 6} & &\colhead{Col 7} & \colhead{Col8} & \colhead{Col 9}&
  \colhead{Col 10}}
\startdata
\multicolumn{3}{c}{All data: $\Delta T=86171^{\mathrm{d}},n=2224$}            & & \multicolumn{3}{c}{\Adataone: $\Delta T=82602^{\mathrm{d}}, n=2174$} & &   \multicolumn{4}{c}{\Bdataone: $\Delta T=67680^{\mathrm{d}}, n=1731$}\\
\multicolumn{3}{c}{Table \ref{Table1hjd14R}}      & & \multicolumn{3}{c}{Table \ref{Table2hjd14R}} &&  \multicolumn{4}{c}{Table \ref{Table3hjd14R}} \\
\cline{1-3} \cline{5-7} \cline{9-12}
     & ~~~~~~~~ $\Dd$     & ~~~~~~~~ $\Dd$     &  &    & ~~~~~~~~ $\Dd$    & ~~~~~~~~ $\Dd$    &  &   &   ~~~~~~~~ $\Dd$ & ~~~~~~~~ $\Dd$      & Effect    \\
\M=3 & $P_3=79999\pm 1216$&$A_3=0.287\pm0.005  $& &\M=3&$P_3=78589\pm2004$&$A_3=0.282\pm0.007  $& &\M=4& $P_4=62992\pm2499$&$A_4=0.25 \pm0.01   $& \PC \\
\M=3 & $P_2=24742\pm142  $&$A_2=0.029\pm0.001  $& &\M=3&$P_2=24870\pm190 $&$A_2=0.030\pm0.008  $& &\M=4&$P_1=12370\pm296  $&$A_1=0.018 \pm0.002 $& \PH \\
\M=3 & $P_1=20358\pm128  $&$A_3=0.013\pm0.001  $& &\M=3&$P_1=20592\pm172 $&$A_1=0.014\pm0.002  $& &\M=4&$P_2=20037\pm420  $&$A_2=0.015 \pm0.002 $& \PC \\
\M=6 & $P_2=7290\pm29    $&$A_2=0.007\pm0.001  $& &\M=6&$P_2=7287\pm40   $&$A_2=0.007\pm0.001  $& &\M=4&$P_3=15429\pm222  $&$A_3=0.008 \pm0.002 $& \PD \\
\M=6 & $P_1=680.4\pm0.4  $&$A_2=0.0064\pm0.0007$& &\M=6&$P_1=680.3\pm0.3 $&$A_1=0.0063\pm0.0008$& &\M=6&$P_1=679.6\pm0.4  $&$A_1=0.0074\pm0.0007 $& \PC \\
\enddata
\addtolength{\tabcolsep}{+0.10cm}
\end{deluxetable*}

 \begin{deluxetable*}{lrrrrrrrrrrrrrrrr}
   \tablecaption{Detection limits (Sect. \ref{SectDetection}).
     Col 1. Candidate. Cols 2-5. Mass, period and
     semi-major axis $(m_3,p_3,a_3)$.
     Cols 6-7. Radial velocity limits during $\Delta t=p_3$
     (Eqs. \ref{EqVradMax}-\ref{EqVradMin}: $v_{\mathrm{min}},v_{\mathrm{max}})$. 
     Cols 8-11. Distance change limits during $\Delta t=5^{\mathrm{y}}$
     and $\Delta t=20^{\mathrm{y}}$
     (Eqs. \ref{EqMoveMax}-\ref{EqMoveMin}:
     $\Delta a_{\mathrm{min}}$, $\Delta a_{\mathrm{max}}$).
     Col 12-13. Proper motion limits 
     (Eqs. \ref{EqMuMin}-\ref{EqMuMax}: $\mu_{\mathrm{min}},\mu_{\mathrm{max}}$).
     \label{TableDetection}
     }
\tablewidth{700pt}
\tabletypesize{\scriptsize}
\tablehead{
  \colhead{Col 1}                    & \colhead{Col 2}              & \colhead{Col 3}              & \colhead{Col 4} &
  \colhead{Col 5}                    & \colhead{     }              & \colhead{Col 6}              & \colhead{Col 7} &
  \colhead{     }                    & \colhead{Col 8}              & \colhead{Col 9}              & \colhead{     } &
  \colhead{Col 10}                   & \colhead{Col 11}             & \colhead{     }              & \colhead{Col 12} &
  \colhead{Col 13}                   \\
  \colhead{}                         & \colhead{$m_3$}              & \colhead{$p_3$}              & \colhead{$a_3$}             &
  \colhead{$a_3$}                    & \colhead{}                          & \colhead{$v_{\mathrm{min}}$}       & \colhead{$v_{\mathrm{max}}$}    &
  \colhead{}                         &  \colhead{$\Delta a_{\mathrm{min}}$}  & \colhead{$\Delta a_{\mathrm{max}}$} & \colhead{}                   & \colhead{$\Delta a_{\mathrm{min}}$}   & \colhead{$\Delta a_{\mathrm{max}}$}   & &
  \colhead{$\mu_{\mathrm{min}}$} & \colhead{$\mu_{\mathrm{max}}$}  \\
  & \multicolumn{4}{c}{Table \ref{TableOrbits}}
  &  \multicolumn{3}{c}{During $\Delta t=p_3$} & & 
  \multicolumn{2}{c}{During $\Delta t=5^{\mathrm{y}}$} & &
  \multicolumn{2}{c}{During $\Delta t=20^{\mathrm{y}}$} & &
  \multicolumn{2}{c}{During $\Delta t=1^{\mathrm{y}}$}\\
  \cline{2-5} \cline{7-8} \cline{10-11} \cline{13-14} \cline{16-17}
\colhead{Candidate}&\colhead{[$m_{\odot}$]}  & \colhead{[y]}        & \colhead{[AU]}    &
  \colhead{[mas]}  & \colhead{}             & \colhead{[km/s]}     & \colhead{[km/s]}  &
  \colhead{}       & \colhead{[mas]}      & \colhead{[mas]}    & \colhead{}        &
  \colhead{[mas]}  & \colhead{[mas]}        & \colhead{}           & \colhead{[mas/y]}   &
  \colhead{[mas/y]}}
\startdata
   Algol H& 2.50& 219.0& 44.7&  1569&&    -2&    10&&     4&   225&&    64&   888&&     0&    47 \\
   Algol G& 0.27&  66.4& 26.1&   911&&    -8&    16&&    25&   427&&   379&  1479&&     0&    89 \\
   Algol F& 0.43&  33.7& 16.2&   572&&   -10&    18&&    61&   515&&   739&  1144&&     0&   109 \\
   Algol E& 0.26&  27.8& 14.6&   510&&   -12&    20&&    79&   547&&   835&  1020&&     0&   118 \\
   Algol D& 0.24&  20.0& 11.7&   416&&   -14&    22&&   123&   591&&   832&   832&&     0&   133 \\
   Algol C& 1.2 &   1.9&  2.1&    75&&   -30&    38&&   151&   151&&   151&   151&&     0&   153 \\
\enddata
\end{deluxetable*}

\clearpage 

\begin{figure*}[ht!]
\plotone{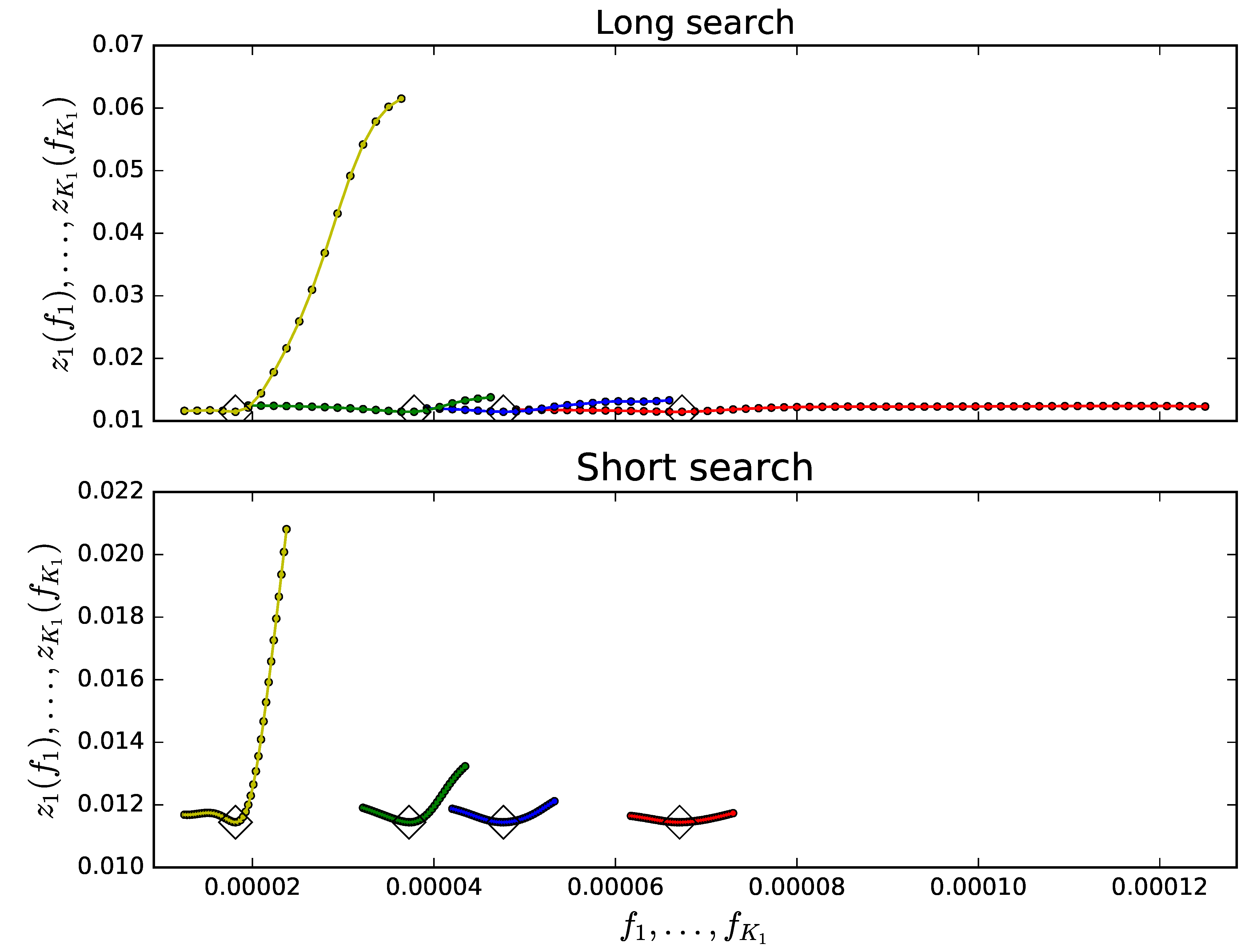}
\caption{All data: Unstable four signal eccentric
  orbit model periodograms
  (Table \ref{Table1hjd14R}: \M=4).
  Upper and lower panels show long and short search
  periodograms (Eq. \ref{EqSlices}).
  Their colours are
  red $(z_1(f_1))$,
  blue $(z_2(f_2))$,
  green $(z_3(f_3))$
  and
  yellow $(z_4(f_4))$.  
  Open diamonds denote locations of best frequencies.
  Their corresponding periods are
      $P_1=14912^{\mathrm{d}}$,
      $P_2=20984^{\mathrm{d}}$,
      $P_3=26846^{\mathrm{d}}$
    and
    $P_4=55172^{\mathrm{d}}$
    (Table \ref{Table1hjd14R}: \M=4).
  Units are frequencies
    $[f_1]=, ..., =[f_{4}]={\mathrm{d^{-1}}}$ 
    and periodogram slices
    $[z_1(f_1)]=, ..., =[z_4(f_{4})]={\mathrm{d}}$.
      \label{1hjd14R421Sz}}
\end{figure*}

\begin{figure*}[ht!]
\plotone{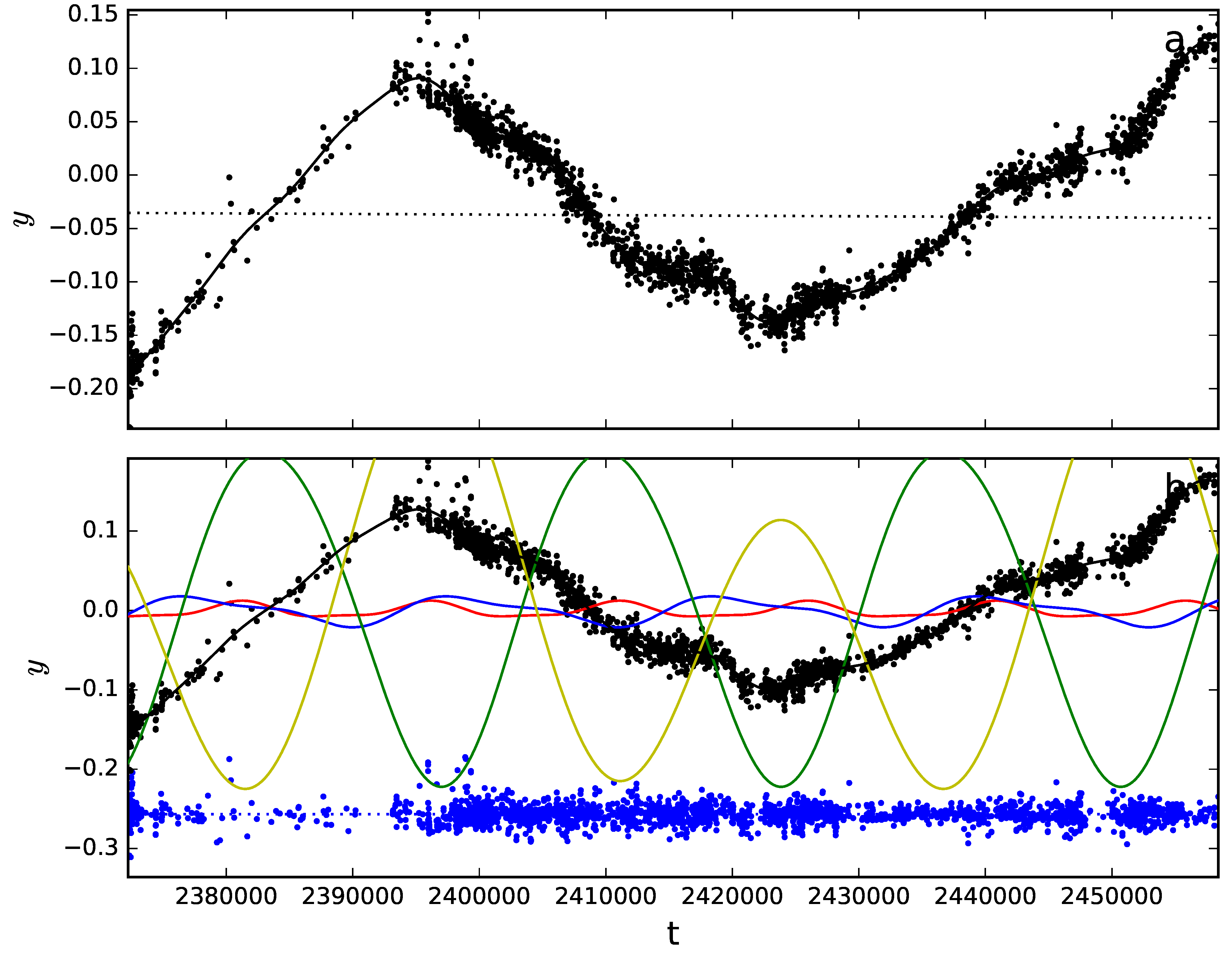}
\caption{All data: Unstable four signal eccentric orbit model
  (Table \ref{Table1hjd14R}: \M=4).
  (a) Data (black dots),
  model $g(t)$ (continuous black line)
  and $p(t)$ trend (dotted black line).
  (b) Data minus $p(t)$ trend (black dots),
  $g(t)$ minus $p(t)$ (black line),
  $g_1(t)$ signal (red line),
  $g_2(t)$ signal (blue line),
  $g_3(t)$ signal (green line)
  and
  $g_4(t)$ signal (yellow line).
  Signal periods are
      $P_1=14912^{\mathrm{d}}$,
      $P_2=20984^{\mathrm{d}}$,
      $P_3=26846^{\mathrm{d}}$
    and
    $P_4=55172^{\mathrm{d}}$.
  Residuals (blue dots) are offset
  to -0.15 (dotted blue line).
  Units are
    $[t]={\mathrm{d}}$
    and
    $[y]={\mathrm{d}}$.
  \label{1hjd14R421Sgdet}}
\end{figure*}


\begin{figure*}[ht!]
\plotone{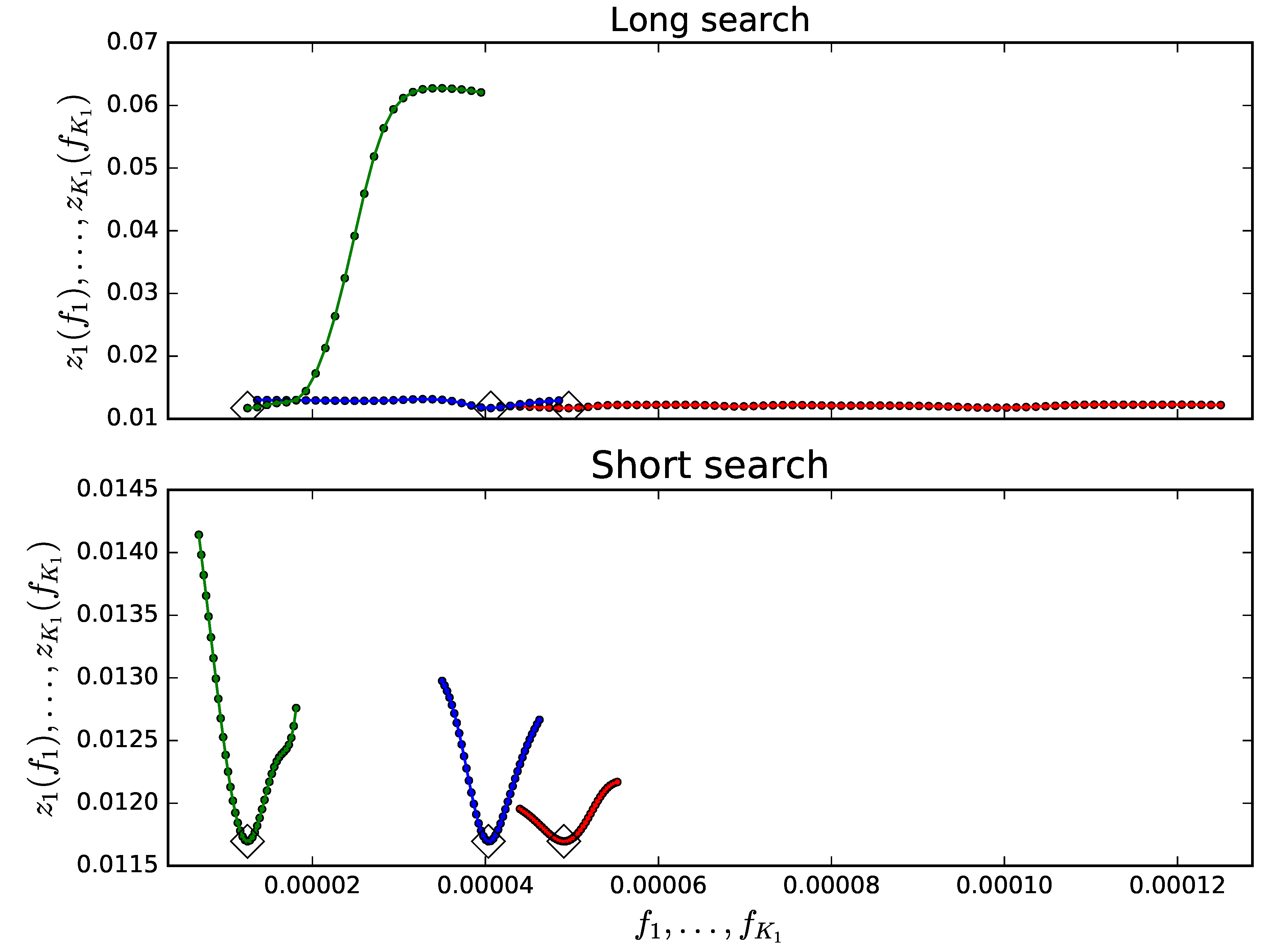}
\caption{All data: Stable three signal eccentric orbit periodograms
  (Table \ref{Table1hjd14R}: \M=3).
    Best periods are at
      $P_1=20358^{\mathrm{d}}$,
      $P_2=24742^{\mathrm{d}}$
    and
    $P_3=79999^{\mathrm{d}}$.
  Otherwise as in Fig. \ref{1hjd14R421Sz}.
  \label{1hjd14R321Sz}}
\end{figure*}

\begin{figure*}[ht!]
\plotone{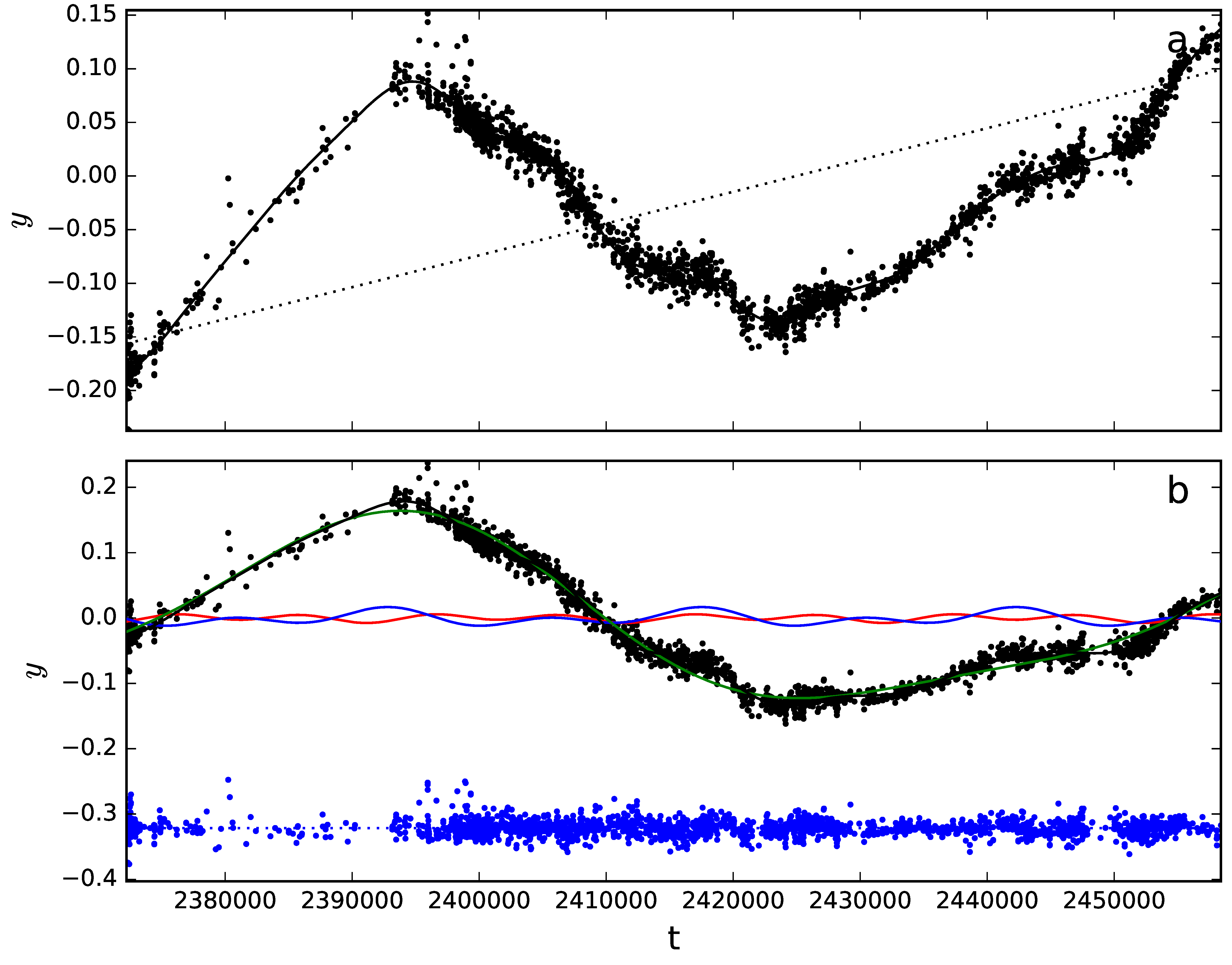}
\caption{All data: Stable three signal eccentric orbit model
  (Table \ref{Table1hjd14R}: \M=3).
  Signal periods are
      $P_1=20358^{\mathrm{d}}$,
      $P_2=24742^{\mathrm{d}}$
    and
    $P_3=79999^{\mathrm{d}}$.
  Otherwise as in Fig. \ref{1hjd14R421Sgdet}.
 \label{1hjd14R321Sgdet}}
\end{figure*}

\begin{figure*}[ht!]
\plotone{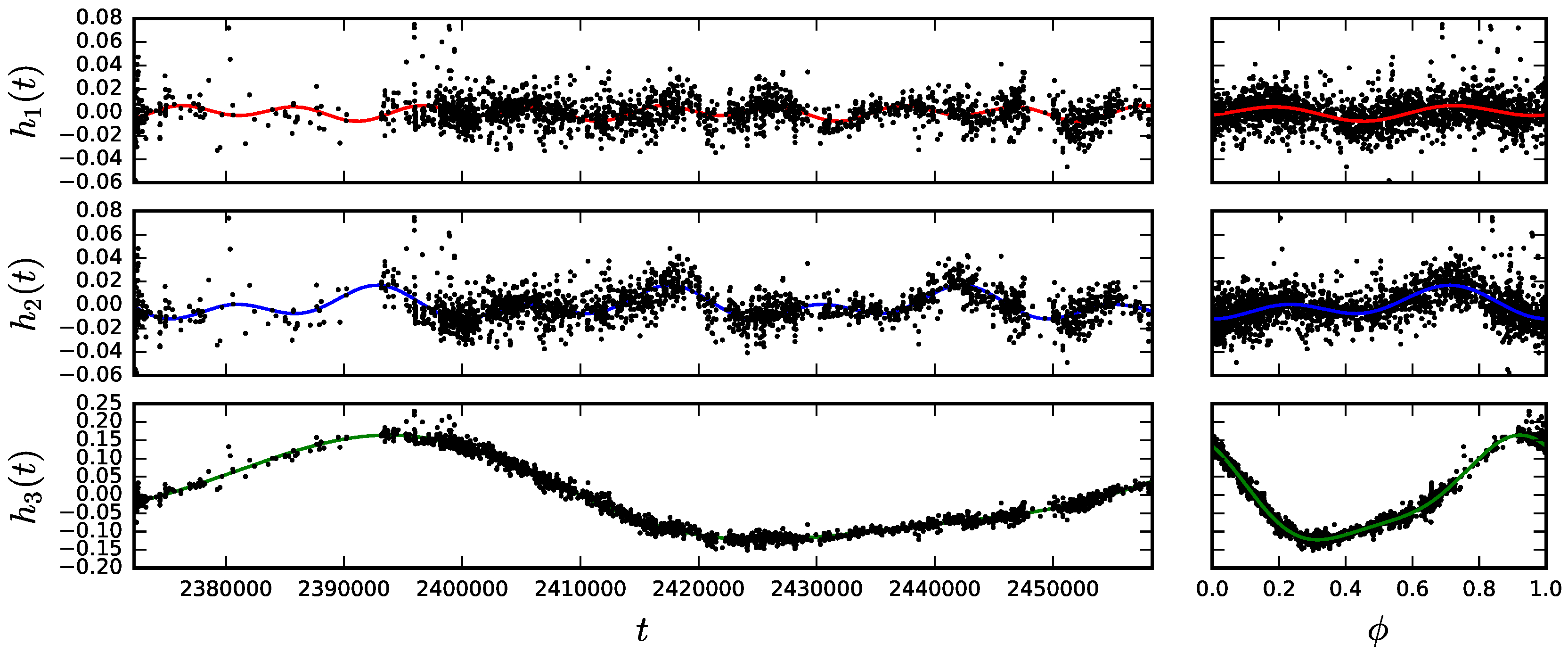}
\caption{All data: Three signals $y_{i,j}$ (Eq. \ref{EqSignals})
  of stable eccentric orbit model (Table \ref{Table1hjd14R}: \M=3).
  Each signal is plotted as a function
  of time $(t)$ and phase $(\phi)$.
  Signal curve colours are as in Fig. \ref{1hjd14R321Sgdet}.
  Signal periods are
      $P_1=20358^{\mathrm{d}}$,
      $P_2=24742^{\mathrm{d}}$
    and
    $P_3=79999^{\mathrm{d}}$.
  Left-hand panel units are
    $[t]={\mathrm{d}}$
    and
    $[h_1(t)]=[h_3(t)]=[h_3(t)]={\mathrm{d}}$.
    Right-hand panel units are
    $[\phi]=$ dimensionless
    and
    $[h_1(t)]=[h_3(t)]=[h_3(t)]={\mathrm{d}}$.
\label{1hjd14R321SSignals}}
\end{figure*}

\begin{figure*}[ht!]
\plotone{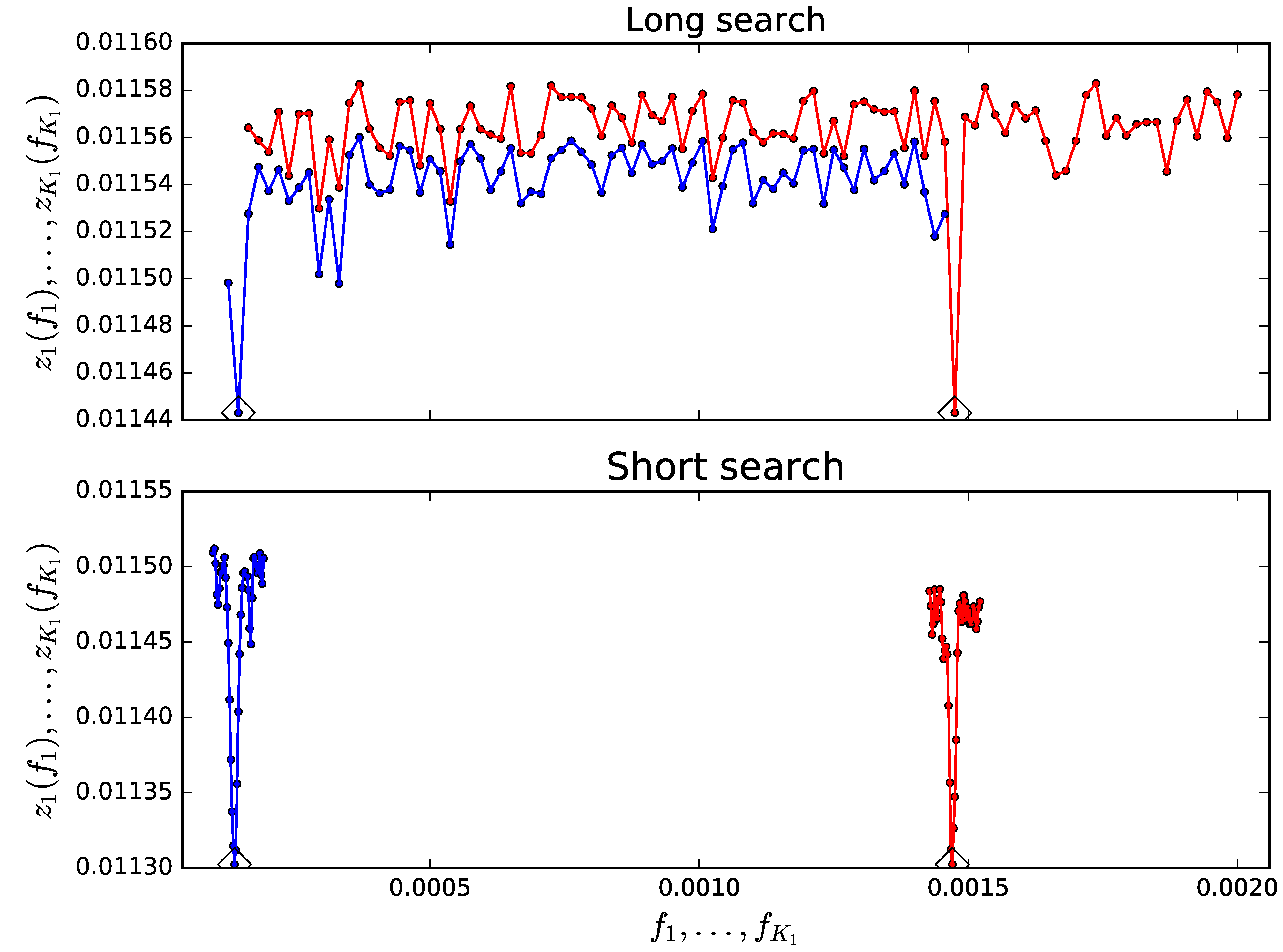}
\caption{Periodograms for residuals of \M=3 model of all data
  (Fig. \ref{1hjd14R321Sgdet}, blue dots):
  Two signal model periodograms (Table \ref{Table1hjd14R}: \M=6).
 Best periods are at
  $P_1=680.^{\mathrm{d}}4$
  and
  $P_2=7290^{\mathrm{d}}$. 
  Otherwise as in
  Fig. \ref{1hjd14R421Sz}.
  \label{1hjd58R220Sz}}
\end{figure*}

\begin{figure*}[ht!]
\plotone{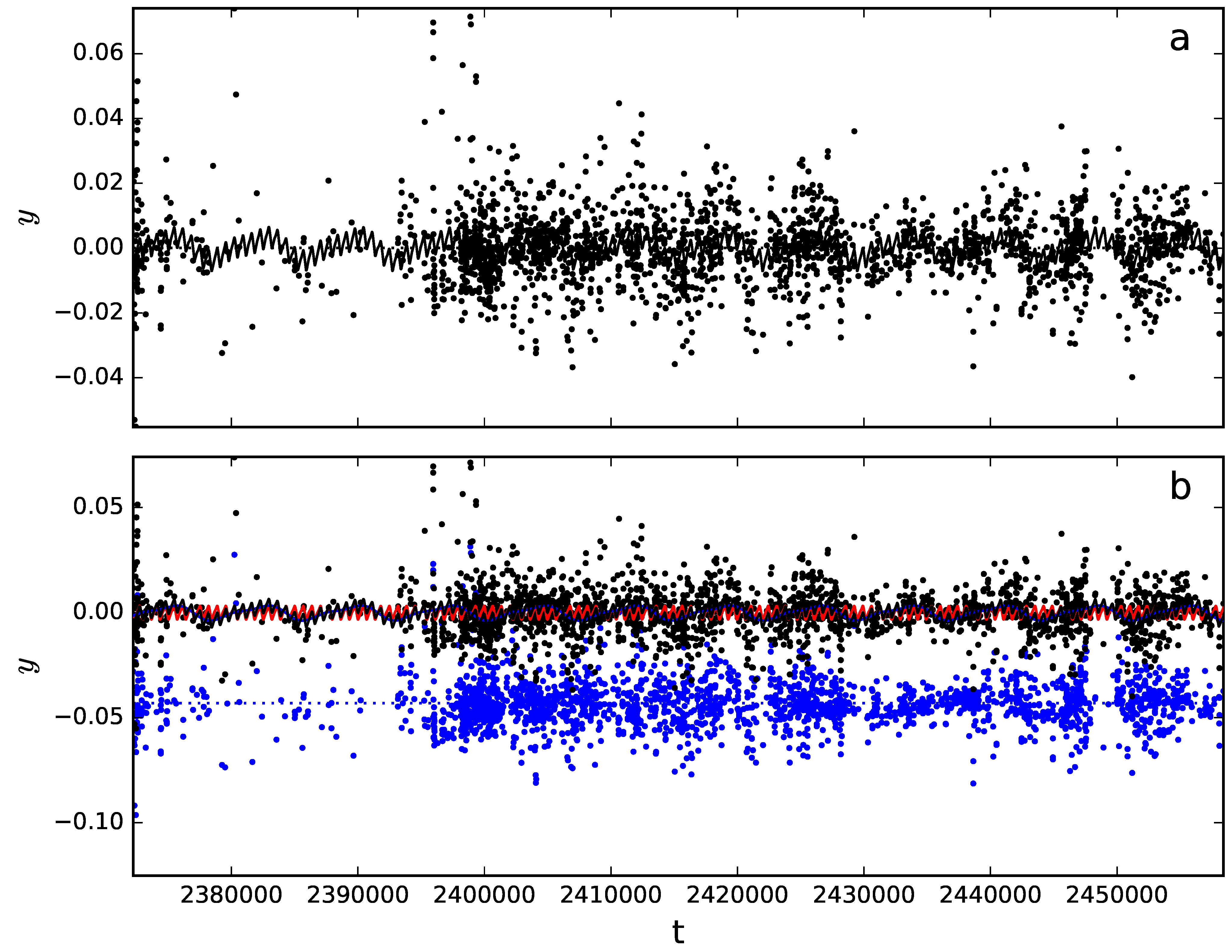}
\caption{Model for residuals of \M=3 model for all data
  (Fig. \ref{1hjd14R321Sgdet}, blue dots):
  Two signals have
  periods
  $P_1=680.^{\mathrm{d}}4$
  and
  $P_2=7290^{\mathrm{d}}$
 (Table \ref{Table1hjd14R}, \M=6 model).
  Otherwise as in
  Fig. \ref{1hjd14R421Sgdet}.
  \label{1hjd58R220Sgdet}}
\end{figure*}

\begin{figure*}[ht!]
\plotone{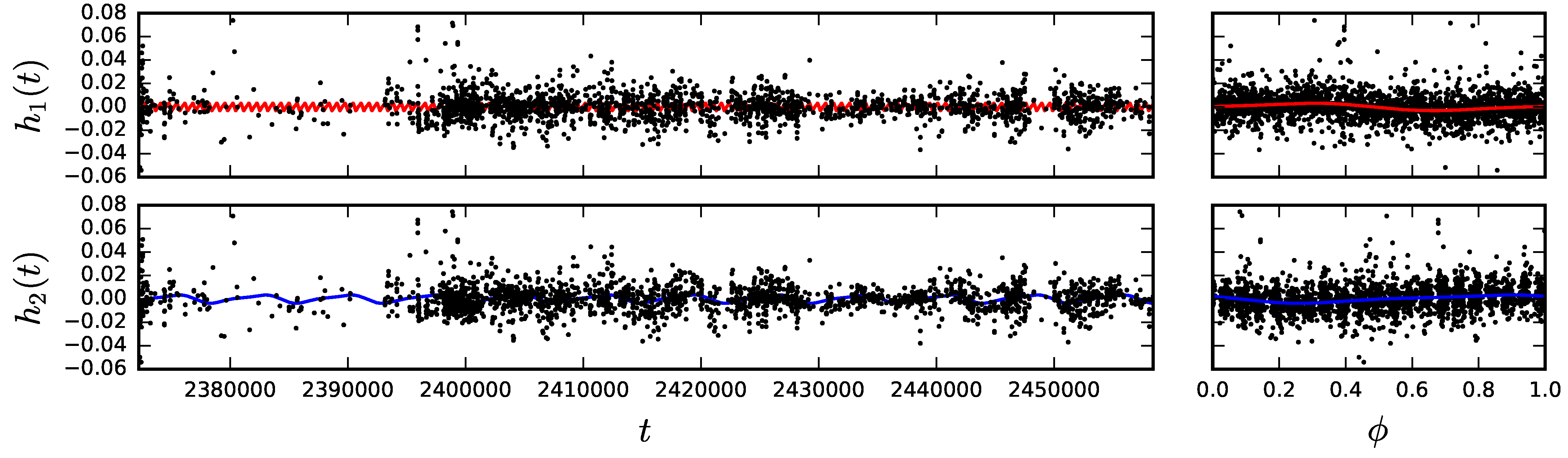}
\caption{Signals in residuals of \M=3 model
  for all data
  (Fig. \ref{1hjd14R321Sgdet}, blue dots):
  Two signals $y_{i,j}$ (Eq. \ref{EqSignals})
  have  periods
  $P_1=680.^{\mathrm{d}}4$
  and
  $P_2=7290^{\mathrm{d}}$
  (Table \ref{Table1hjd14R}, \M=6 model).
  Otherwise as in Fig. \ref{1hjd14R321SSignals}.
\label{1hjd58R220SSignals}}
\end{figure*}

\begin{figure*}[ht!]
\plotone{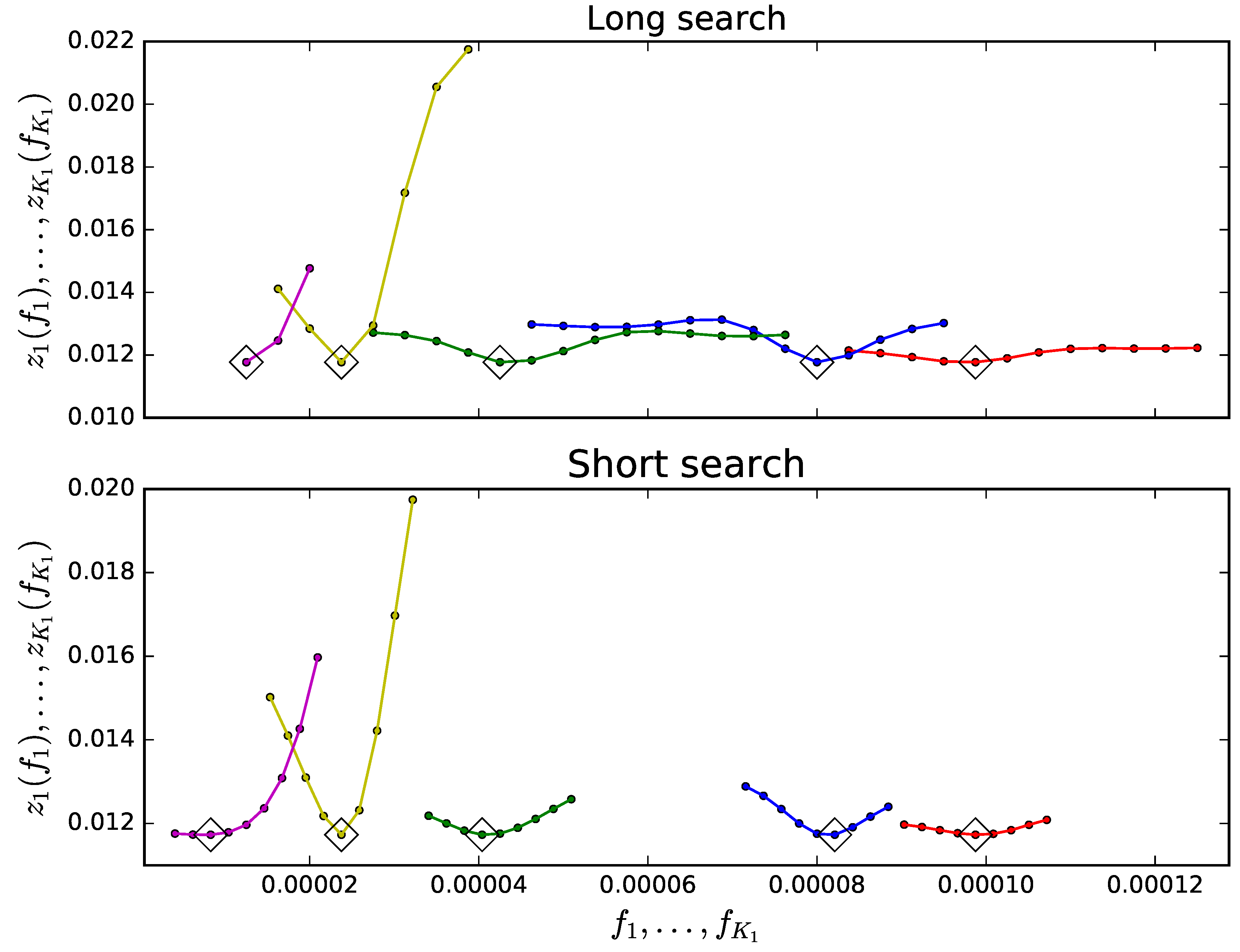}
\caption{All data: Five signal circular orbit model periodograms
  (Table \ref{Table1hjd16C}: \M=2).
  Best periods are at
      $P_1=10144^{\mathrm{d}}$,
      $P_2=12294^{\mathrm{d}}$,
      $P_3=24247^{\mathrm{d}}$,
      $P_4=42422^{\mathrm{d}}$
    and
    $P_5=120740^{\mathrm{d}}$.
    Otherwise as in Fig. \ref{1hjd14R421Sz}.
  \label{1hjd16R511Sz}}
\end{figure*}

\begin{figure*}[ht!]
\plotone{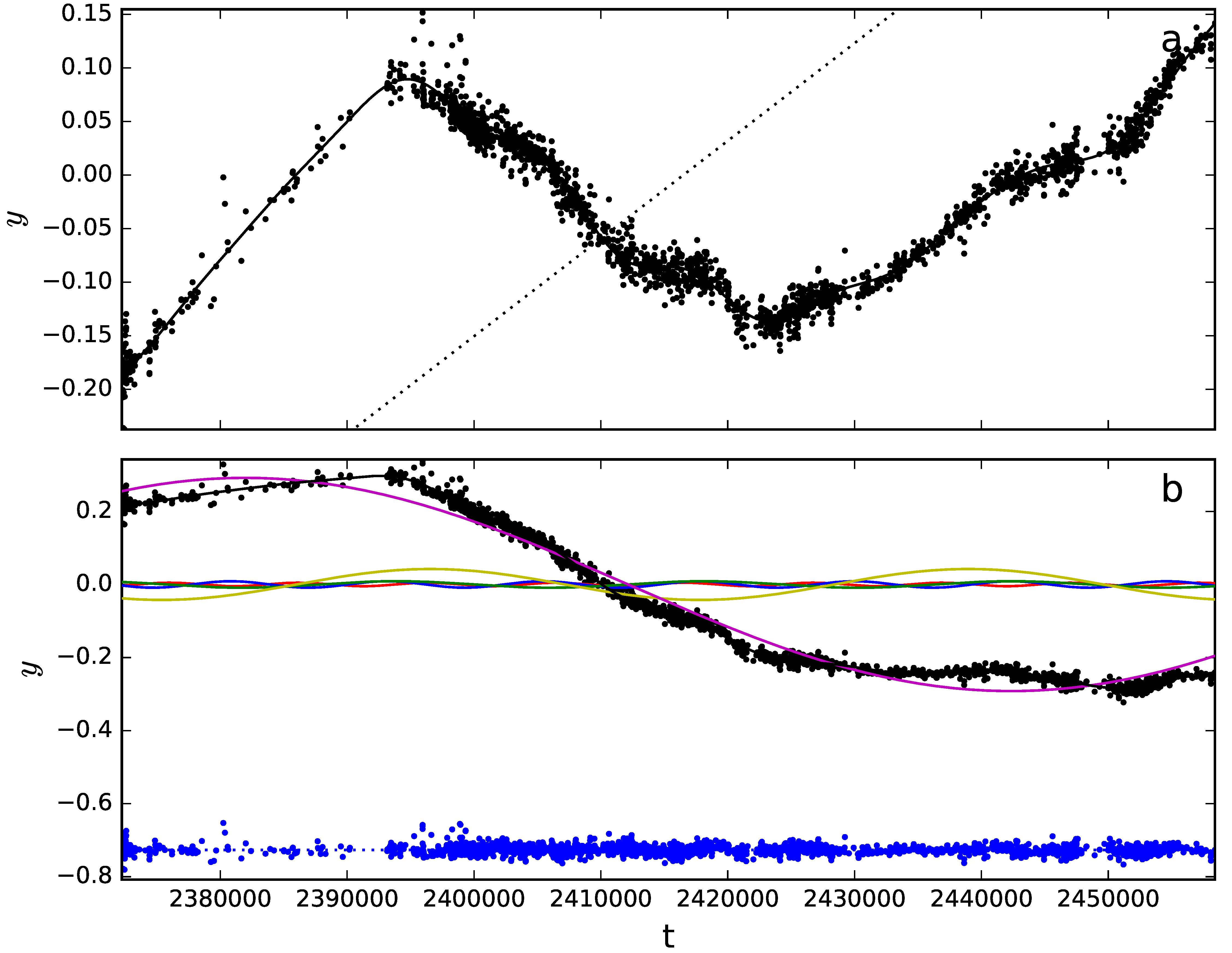}
\caption{All data: Five signal circular orbit model
  (Table \ref{Table1hjd16C}: \M=2).
    Signal periods are
      $P_1=10144^{\mathrm{d}}$,
      $P_2=12294^{\mathrm{d}}$,
      $P_3=24247^{\mathrm{d}}$,
      $P_4=42422^{\mathrm{d}}$
    and
    $P_5=120740^{\mathrm{d}}$.
  Otherwise as in Fig. \ref{1hjd14R421Sgdet}.
  \label{1hjd16R511Sgdet}}
\end{figure*}

\begin{figure*}[ht!]
\plotone{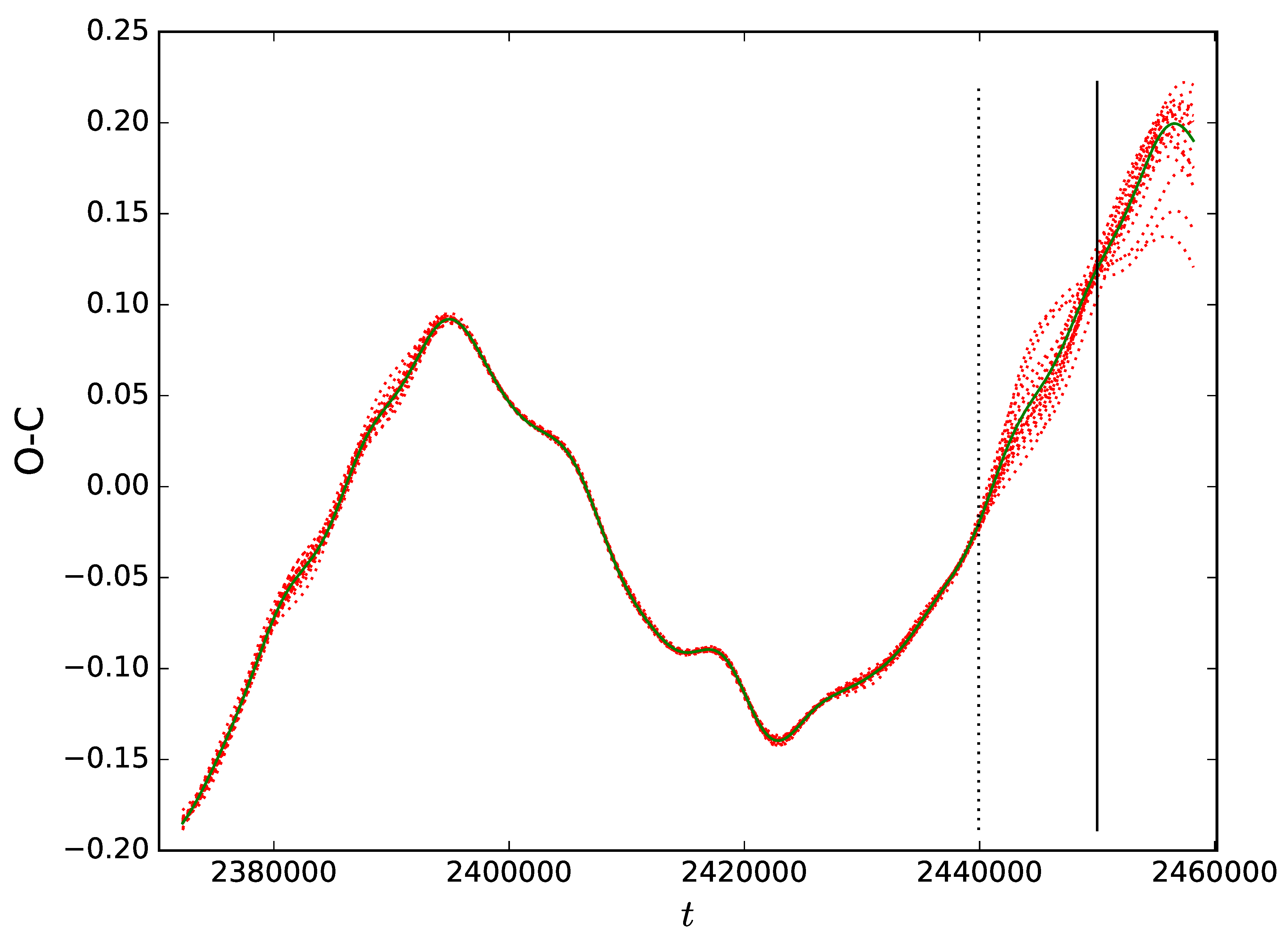}
\caption{\Bdataone
  ~eccentric orbit analysis (Sect. \ref{SectBdata}).
  Green line denotes  $g(t)$ model \M=4+6 (Fig. \ref{Prediction3}).
  Dotted red lines show models for 20 bootstrap samples.
  \Bdatatwo ~prediction begins
  from dotted vertical line.
  Continuous vertical line is
  data turning point in Fig. \ref{Prediction3}b.
  Units are
    $[t]={\mathrm{HJD}}$
    and
    $[{\mathrm{O\!-\!C}}]={\mathrm{d}}$.
  \label{AllSolutions}}
\end{figure*}

\begin{figure*}[ht!]
\plotone{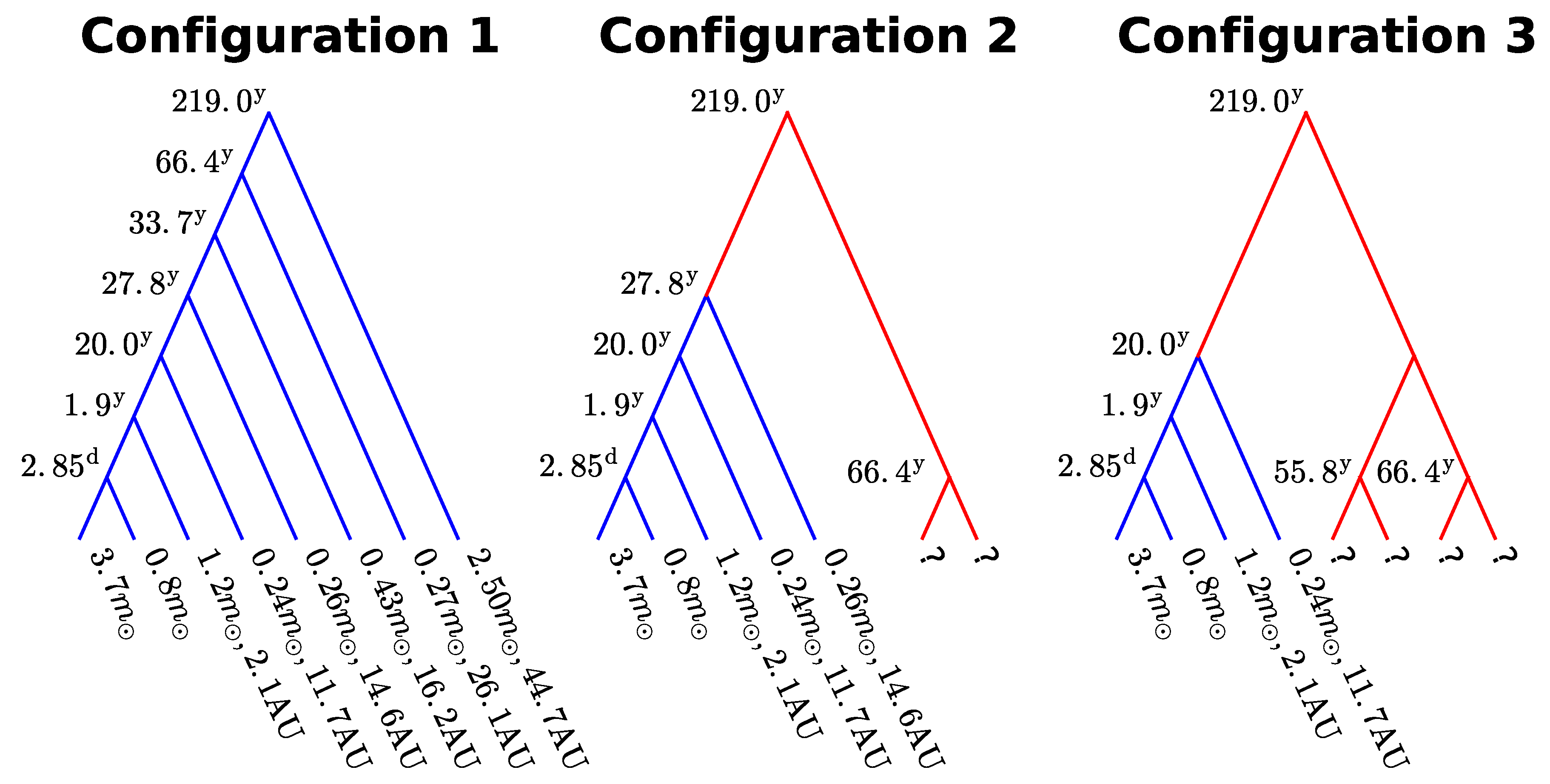}
\caption{Alternative configurations.
  Configuration 1.
  Eight members are cEB and six WOSs
  on circular orbits
  (Table \ref{TableOrbits}: $i=90^{\mathrm{o}}$).
  In next configurations,
  red lines highlight 
  differences from this first configuration.
  Configuration 2.
  Seven members are cEB and five WOSs.
  Red lines illustrate a long-period
  $66.^{\mathrm{y}}4$ binary on a distant
  $219.^{\mathrm{y}}0$ orbit.
  Configuration 3.
  Eight members are cEB and six WOSs.
  Red lines illustarate two long-period
  $55.^{\mathrm{y}}8$ and $66.^{\mathrm{y}}4$ binaries
  on a distant
  $219.^{\mathrm{y}}0$ orbit.
  \label{Configurations}}
\end{figure*}

\end{document}